\documentclass[preprintnumbers,amsmath,amssymb]{revtex4}
\usepackage{graphicx}
\usepackage{dcolumn}
\usepackage{bm}

\begin{document}

\title{The Two-Measure Theory  and an Overview of Some of its Manifestations
}




\author
{ A.  B. Kaganovich
\thanks{alexk@bgu.ac.il}}
\address{Physics Department, Ben Gurion University of the Negev, Beer
Sheva 84105, Israel \\
and Sami Shamoon College of Engineering, \\Beer Sheva, Israel}
\date{\today}



\begin{abstract}

The Two-Measure theory (TMT) has been developing since 1998 and has yielded a number of highly interesting results, including those not realized in traditional field theory models. The most important advantage of TMT as an alternative theory is that, under the conditions under which all classical tests of general relativity are performed, TMT models are able to accurately reproduce Einstein's general relativity. 
Despite this, TMT is still often perceived as something too exotic to be relevant to reality. In fact, the fundamental idea underlying TMT seems undeniable: if we truly believe in the effectiveness of mathematics in studying nature, we must agree that there must be a correspondence between the fundamental laws of nature and the structure of the mathematical apparatus necessary to adequately describe them. It then turns out that there is no reason to ignore the volume measure existing on the differentiable manifold on which the theory of gravity and matter fields is built. This idea has far-reaching implications. The goals of this paper are: 1) to provide a clear mathematical and conceptual justification for TMT; 2) to collect in a single article some of the main results of TMT obtained over the past 25 years.

\vspace{0.5cm}

\hspace{1.5cm} K-essence, phantom dark energy,
neutrino dark energy, initial conditions for inflation.

\end{abstract}

\maketitle

\section{Introduction}
\label{Introduction}

What are these two measures? Why the theory of two measures?
To answer these questions, let's begin by analyzing the generally accepted description of gravity.
The mathematical structure on the basis of which the action integral is constructed in Einstein's General Relativity (GR) can be symbolically represented as
\[ \boxed{\text{Riemannian  geometry  $\oplus$  matter  fields} =  \text{Einstein's GR}}\]

In  {\em the Palatini formulation of gravity} (P'GR),  the metric tensor and the (affine) connection are treated as independent variables. As is well known, in the absence of a non-minimal coupling of matter with curvature, the theory of gravity obtained by variation of the original action coincides with Einstein's GR. However, in the presence of, for example, a non-minimal coupling of a scalar field with a curvature scalar, the solution of the equations obtained by variation with respect to the affine connection leads to connection coefficients with which the covariant derivative of the metric tensor is non-zero.  Violation of the metricity condition means that the space-time is non-Riemannian, and to formulate the theory in a Riemannian space-time, it is necessary to perform the corresponding Weyl transformation of the metric tensor. Typically, such a transformation is performed in the original action in such a way as to eliminate the non-minimal coupling of the scalar field with the curvature scalar, and the set of variables used in this case is called the Einstein frame. 
The mathematical structure on the basis of which the action integral  is constructed  in P'GR can be symbolically represented as
\[ \boxed{\text{(affine connection  $+$   metric)  $+$  matter  fields} =  \text{P'GR}}\]

Considering the formulation of P'GR from the point of view of the applied mathematical apparatus, we inevitably come to the conclusion that we are dealing with a 4-dimensional differentiable manifold at the stages of equipping it with  an affine connection and a metric structure (see, for example, \cite{Hawking}). 
\[ \boxed{\text{(paracompact topological space $+$ Differential structure }=\text{4D differentiable manifold}}\]
\[ \boxed{\text{($4D$ differentiable manifold $+$ affine connection  $+$  metric)  $+$ matter  fields} = 
 \text{P'GR}}\]

If we really believe in the effectiveness of mathematics in studying nature, we can go further and agree that there must be a correspondence between the fundamental laws of nature and the structure of the mathematical apparatus necessary for their adequate description. At the same time, we note that among the mathematical objects commonly used in field theories in 4D space-time, there is no the volume form, which can be defined as a 4-form on  the 4D differentiable manifold even before equipping it with an affine connection and a metric structure. 
The appropriate volume element can be constructed  by means of 4 scalar  functions $\varphi_a$ ($a =1 ,..,4$) defined on the 4D differentiable manifold, as following
\begin{eqnarray}
dV_{\Upsilon}=\Upsilon d^4x\equiv
\varepsilon^{\mu\nu\gamma\beta}\varepsilon_{abcd}\partial_{\mu}\varphi_{a}
\partial_{\nu}\varphi_{b}\partial_{\gamma}\varphi_{c}
\partial_{\beta}\varphi_{d}d^4x
=4!d\varphi_1\wedge d\varphi_2\wedge d\varphi_3\wedge d\varphi_4.
\label{Phi}
\end{eqnarray}
 After replacing Einstein's GR with Palatini's formulation of gravity, the construction of the Two-Measure theory can be seen as the next step in the realization of a clearly formulated  idea:
\underline{all possible degrees of freedom}
 contained in the  action should be considered as independent of each other, and all relationships between them appear as a result of applying the principle of least action. For brevity, we will call this approach to constructing a theory "the principle of maximum dynamic independence (PMDI)". 
The key to this formulation is the word "all" in the expression "all possible degrees of freedom", and we will now discuss what is meant by this. 
Using $dV_{\Upsilon}$ as a volume measure in the action integral, according to PMDI, we must consider the functions $\varphi_a$ as degrees of freedom and, therefore, when finding the action extremum, we must  perform variations with respect to $\varphi_a$ as well.

Apparently, it is necessary to more carefully analyze the extent to which it is justified to ignore the volume measure of type (\ref{Phi}) in the generally accepted approach.
 After the metric structure is defined on the manifold, if the latter is orientable, $dV_g=\sqrt{-g}d^4x$ can also be chosen as a 4-form to describe the volume measure in the action integral. The possibility of such two alternative choices is based on the fact that under general coordinate transformations with a positive Jacobian (i.e. without changing the orientation of the manifold), the scalar densities $\sqrt{-g}$ and $\Upsilon$ are transformed according to the same rule. 
The standard approach (see, for example, section 2.8 in ref. \cite{Hawking})   is to use only the volume element $dV_g$, {\em by default assuming that the volume element $dV_{\Upsilon}$ is redundant}. 
A more detailed argument for this conclusion is given by R. Wald in appendix B2 in ref. \cite{Wald}. Considering on n-dimensional manifold M the volume element 
$f(x^1,...,x^n) dx^1\wedge ...\wedge dx^n$, where the function $f$ is non-vanishing, Wald reasonably notes that "{\em if one is given only the structute of a manifold $M$,  there is no natural choice of volume element}". Wald then shows that {\em if a metric is defined on M, then there exists a natural volume element, given by the n-form}
$\sqrt{|g|}dx^1\wedge ...\wedge dx^n$. Continuing Wald's analysis, we note that the possibility of treating the volume element 
$\sqrt{|g|}dx^1\wedge ...\wedge dx^n$ as natural exists only because in Einstein's GR the metric tensor is a dynamic variable, whose variation
 in the Einstein-Hilbert action leads to Einstein's equations.
 The original idea of the TMT might be that including the volume element $f(x^1,...,x^n) dx^1\wedge ...\wedge dx^n$ in the action integral allows us to consider the function $f$ as another variable subject to variation in the action, along with the metric. But this modification of the theory leads to a trivial result: the function $f$ plays the role of a Lagrange multiplier, and the Lagrangian term entering the action with the volume element 
 $f(x^1,...,x^n) dx^1\wedge ...\wedge dx^n$ turns out to be equal to zero. The idea of the TMT makes sense if, following PMDI, we fully exploit the capabilities of a differentiable manifold M. Returning from the n-dimensional manifold to the 4-dimensional one and using four arbitrary functions $\varphi_i$ ($i=1,2,3,4$) (differentiable a sufficient number of times), we can construct a 4-form (\ref{Phi}) and consider it as a volume element in the action integral together with the volume element $dV_g=\sqrt{-g}d^4x$. In this formulation of the theory, varying the functions $\varphi_i$ leads to an equation with a significant dynamic effect. This is the essence of the TMT, and, returning to the analysis of Wald's argumentation, we see that {\em the volume elements $dV_g$ and $dV_{\Upsilon}$ turn out to be  equally natural}.

A closer look at the problem when the $\varphi_a$ functions are added to the set of degrees of freedom in action shows that
 as a result of varying the  action,  the scalar function in the form of the ratio of volume measures 
\begin{equation}
 \zeta(x)\stackrel{\mathrm{def}}{=} \frac{dV_{\Upsilon}}{dV_g}\equiv\frac{\Upsilon}{\sqrt{-g}}   
 \label{zeta}
\end{equation}
 appears in all equations of motion and plays a decisive role in all new effects of TMT. Bearing in mind that gravity in GR
 has a geometric nature described by the metric tensor, we will use the term \underline{pregeometry} for effects
 caused by the scalar $\zeta(x)$.

The organization of the paper is as follows. In Section \ref{Two-Measure Theory in empty space}, we show that in empty space  the solution of the equations of the gravity model with two volume measures leads to the result $\Upsilon=const.\cdot \sqrt{-g}$, i.e., there is no fundamental difference between the volume measures $dV_{\Upsilon}$ and $dV_g$. 
However, even in empty space the gravity model with two measures acquires new interesting properties compared to the gravity model with only one measure $dV_g$. In Section \ref{Some additional mathematical aspects taken into account in TMT}, we show that in TMT, the physical results depend on orientation of the space-time manifold and, moreover, that the latter arises spontaneously. By analyzing the  mathematical differences between the volume measures $dV_g$ and $dV_{\Upsilon}$, we arrive at a formulation of possible structural properties of the action integral in TMT. An important note regarding the terminology used in this article is also provided at the end of this section.
In Section \ref{toy model}, using a toy model as an example, we demonstrate the main features of the TMT and describe in detail
 the computational procedure in TMT.
In Section \ref{quint with additional symm}, a class of TMT models
 that provide quintessence without the need for fine-tuning is described. 
In Section \ref{Scale invar models}, four TMT models with global scale invariance are considered. These models demonstrate the possibility of solving a number of fundamental problems in gravity and cosmology.
A model of dark energy generated by neutrinos without a scalar field is presented in Section
\ref{Neutrino genereted dark energy}.
The TMT effect, related to a possible connection between the Borde--Guth--Vilenkin theorem and initial conditions for inflation, is discussed in 
Section \ref{BGV and initial conditions for inflation}.

\section{The Two-Measure theory in empty space}
\label{Two-Measure Theory in empty space}

When constructing models with two volume measures, we assume that the total volume measures in the action integral  have the form of linear combinations 
$(\alpha\sqrt{-g}+\beta\Upsilon)d^4x$, in which the constant real coefficients $\alpha$, $\beta$ can be different depending in which term of the Lagrangian density they are present with in the  original action. 

According to PMDI, the affine connection $\Gamma^{\lambda}_{\mu\nu}$ and metric tensor $g_{\alpha\beta}$ should be treated as independent degrees of freedom, i.e., one should proceed in the Palatini formulation. Using the definitions $R(\Gamma,g)=g^{\mu\nu}R_{\mu\nu}(\Gamma)$,
$R_{\mu\nu}(\Gamma)=R^{\lambda}_{\mu\nu\lambda}(\Gamma)$ and
$R^{\lambda}_{\mu\nu\sigma}(\Gamma)\equiv \Gamma^{\lambda}_{\mu\nu
,\sigma}+ \Gamma^{\lambda}_{\gamma\sigma}\Gamma^{\gamma}_{\mu\nu}-
(\nu\leftrightarrow\sigma)$, 
 let us consider the TMT model in the empty space with the following   action 
\begin{equation}
S_{empty}=\int\Biggl[(\alpha\sqrt{-g}+\beta\tilde{\Upsilon})\left( -\frac{\tilde{M}_P^2}{2}\right)g^{\mu\nu}R_{\mu\nu}(\Gamma)-\sqrt{-g}V_1-
\frac{\tilde{\Upsilon}^2}{\sqrt{-g}}\tilde{V}_2\Biggr]d^4x.
\label{S-empty}
\end{equation} 
 Bearing in mind that the further description of this model will also use redefinitions of the parameters $\tilde{M}_P$, $\tilde{V}_2$ and the functions 
$\tilde{\varphi}_a$ (and hence $\tilde{\Upsilon}$), we use the tilde to denote them here.
Last two terms in (\ref{S-empty})  have the nature of the  vacuum-like terms. 
If the first term, $-\sqrt{-g}V_1$   was present in Einstein's GR,  $V_1$ would be a cosmological constant. 
The term with $V_2$ was first introduced in ref. \cite{GK2}. The  reason for adding the term with $V_2$ is that the term
 $\propto \tilde{\Upsilon}=\tilde{\zeta}(x)\sqrt{-g}$, which can be expected as the contribution of quantum-gravitational effects
 to a vacuum-like action, does not contribute to the equations of motion. Therefore, the next  in powers of $\tilde{\zeta}$ 
vacuum-like term can be of the form $\propto \tilde{\zeta}^2\sqrt{-g}= \tilde{\Upsilon}^2/\sqrt{-g}$. 
It is assumed that $\alpha>0$, $\beta>0$ and  $V_1<0$, $\tilde{V}_2<0$.

The action  includes the following independent variables (and takes into account the effect of pre-geometry): the affine connection $\Gamma^{\lambda}_{\mu\nu}$, the metric tensor $g_{\mu\nu}$ and four scalar functions  $\tilde{\varphi}_a$, (a=1...4), with the help of which the measure density  $\tilde{\Upsilon}$ is constructed.
By varying the action with respect to scalar functions $\tilde{\varphi}_{a}$ we get 
\begin{equation}
B^{\mu}_{a}\partial_{\mu}\left[\beta\left( -\frac{\tilde{M}_P^2}{2}\right)R(\Gamma,g)-
2\tilde{\zeta} \tilde{V}_2\right]=0 \quad \textsf{where} \quad
B^{\mu}_{a}=\varepsilon^{\mu\nu\alpha\beta}\varepsilon_{abcd}
\partial_{\nu}\tilde{\varphi}_{b}\partial_{\alpha}\tilde{\varphi}_{c}
\partial_{\beta}\tilde{\varphi}_{d}.
\label{varphiB}
\end{equation} 
Since $Det (B^{\mu}_{a}) \propto\tilde{\Upsilon}^{3}$ it follows
that if
 \begin{equation}
\quad \textsf{everywhere} \quad \tilde{\Upsilon}(x)\neq 0, 
\label{Phi neq 0}
\end{equation}
the equality
\begin{equation}
-\frac{\tilde{M}_P^2}{2}g^{\mu\nu}R_{\mu\nu}(\Gamma)=\frac{1}{\beta}(\tilde{{\mathcal M}}+2\tilde{\zeta} \tilde{V}_2)
\label{var varphi empty}
\end{equation} 
must be satisfied, where  $\tilde{{\mathcal M}}$ is a constant of integration with the dimension of $(mass)^4$.

 Variation with respect to $g^{\mu\nu}$ yields  the equation
\begin{equation}
(\alpha+\beta\zeta)\left(-\frac{\tilde{M}_P^2}{2}\right)R_{\mu\nu}(\Gamma)
- \frac{\alpha}{2}g_{\mu\nu}\left(-\frac{\tilde{M}_P^2}{2}\right)g^{\mu\nu}R_{\mu\nu}(\Gamma)
+\frac{1}{2}g_{\mu\nu}(V_1-\tilde{\zeta}^2\tilde{V}_2)=0.
\label{Grav.eq empty}
\end{equation}
The condition for compatibility of eqs.(\ref{var varphi empty}) and  (\ref{Grav.eq empty}) is a constraint in the form of the following expression for a constant scalar $\tilde{\zeta}$
 \begin{equation}
\tilde{\zeta}=\frac{\alpha\tilde{{\mathcal M}}-2\beta V_1}{\beta\tilde{{\mathcal M}}-2\alpha \tilde{V}_2}.
\label{zeta empty}
\end{equation}
As noted above in assertion 1, the latter means that the solution in empty space reveals that the volume measures $dV_{\Upsilon}$ and $dV_g$ differ only by a constant factor. 
From the constancy of $\tilde{\zeta}$ it follows that the variation of $\Gamma^{\lambda}_{\mu\nu}$ leads to an equation whose solution is the equality
$\Gamma^{\lambda}_{\mu\nu}=\{^{\lambda}_{\mu\nu}\}$,
where $\{^{\lambda}_{\mu\nu}\}$  are the Christoffel's connection
coefficients of the metric $g_{\mu\nu}$. This means that spacetime is Riemannian, there is no need for a Weyl transformation and
$R_{\mu\nu}(\Gamma)=R_{\mu\nu}(g)$ is the Ricci tensor of the spacetime with the metric $g_{\mu\nu}$. Substituting the value of $\tilde{\zeta}$ into eq.(\ref{Grav.eq empty}), we obtain Einstein's equations in empty space
\begin{equation}
R_{\mu\nu}(g)=
- \frac{1}{2\tilde{M}_P^2}\cdot\frac{\tilde{{\mathcal M}}^2-4V_1 \tilde{V}_2}{\beta\tilde{{\mathcal M}}-2\alpha \tilde{V}_2}g_{\mu\nu}.
\label{Grav.final eq empty}
\end{equation}
An important novelty of the solution of the TMT model in empty space  compared to the convensional theory is that by choosing the constant of integration 
$\tilde{{\mathcal M}}$, it is possible to realize a solution with any TMT-effective cosmological constant  $\Lambda$. In particular, by choosing $\tilde{{\mathcal M}}=2\sqrt{V_1\tilde{V_2}}$, we obtain a solution with $\Lambda =0$ and
\begin{equation}
\tilde{\zeta}|_{\Lambda =0}=\sqrt{\frac{V_1}{\tilde{V_2}}}.
\label{zeta empty lambda 0}
\end{equation}

It is worth paying special attention to the fact that the relations $\tilde{{\mathcal M}}=2\sqrt{V_1\tilde{V_2}}$ and (\ref{zeta empty lambda 0}), which play a very important role, do not contain the parameters $\alpha$ and $\beta$. It can be shown that this fact is not accidental. To do this, we rewrite the action in empty space (\ref{S-empty}) using the following redefinitions:
 $\alpha\tilde{M}_P^2=M_P^2$, \,
$(\frac{\beta}{\alpha})^{1/4}\cdot\tilde{\varphi}_a=\varphi_a$, \, $\frac{\beta}{\alpha}\cdot\tilde{\Upsilon}=\Upsilon$, \,
$\left(\frac{\alpha}{\beta}\right)^2\cdot\tilde{V}_2=V_2$. As a result, the action (\ref{S-empty}) takes the form
\begin{equation}
S_{empty}=\int\Biggl[(\sqrt{-g}+\Upsilon)\left( -\frac{M_P^2}{2}\right)g^{\mu\nu}R_{\mu\nu}(\Gamma)-\sqrt{-g}V_1-
\frac{\Upsilon^2}{\sqrt{-g}}V_2\Biggr]d^4x.
\label{S-empty redef}
\end{equation} 
 It is worth noting that {\em the presence of vacuum-like terms violates the symmetry $\Upsilon \longleftrightarrow \sqrt{-g}$}.
By acting with (\ref{S-empty redef}) in the same way as we did with (\ref{S-empty}), instead of (\ref{var varphi empty}) we first get 
\begin{equation}
-\frac{M_P^2}{2}g^{\mu\nu}R_{\mu\nu}(\Gamma)={\mathcal M}+2\zeta V_2,
\label{var varphi empty after redef}
\end{equation}
 where $\zeta\equiv\frac{\Upsilon}{\sqrt{-g}}$ and ${\mathcal M}$ is the integration constant. Then it follows that
\begin{equation}
\zeta=\frac{{\mathcal M}-2 V_1}{{\mathcal M}-2V_2},
\label{zeta empty redef}
\end{equation}
\begin{equation}
R_{\mu\nu}(g)=
- \frac{1}{2M_P^2}\cdot\frac{{\mathcal M}^2-4V_1 V_2}{{\mathcal M}-2 V_2}g_{\mu\nu}.
\label{Grav.final eq empty redef}
\end{equation}
Again, by choosing the integration constant
${\mathcal M}$, we can implement a solution with any TMT-effective  cosmological constant $\Lambda$. In particular, by choosing
 ${\mathcal M}=2\sqrt{V_1V_2}$, we obtain a solution with $\Lambda =0$ and $\zeta|_{\Lambda =0}=\sqrt{\frac{V_1}{V_2}}$.

\section{Some additional mathematical aspects taken into account in TMT}
\label{Some additional mathematical aspects taken into account in TMT}

Before moving on to the study of specific models, it is necessary to return to the mathematical aspects of the existence of the two volume measures $dV_g$ and $dV_{\Upsilon}$ and take into account the important differences between them.
The density $\sqrt{-g}$ of the volume measure
$dV_g$  is positive-definite and the volume measure $dV_g$ is defined  if the manifold is oriented. But $\Upsilon$, which plays the role of the density of the volume measure $dV_{\Upsilon}$, is generally sign-indefinite. Moreover, the sign of the 4-form
$dV_{\Upsilon}$ determines the sign of the orientation of the manifold (for more details see, for example, \cite{Lee}). Therefore, using linear combinations  
$(\alpha_i\sqrt{-g}+\beta_i\Upsilon)d^4x$  as a total  volume measures in constructing the  terms of the primordial action, {\em we leave the sign of the space-time orientation undefined}. And only as a result of choosing a solution to the equations of motion, the sign of $\Upsilon =\zeta\sqrt{-g}$ is fixed, and, therefore, the sign of the orientation. To understand how this effect occurs, let's return to the simplest example in empty space described by eqs. (\ref{zeta empty redef}) and (\ref{Grav.final eq empty redef}) (how this happens in the presence of matter we will see later in this  paper). To obtain a solution with $\Lambda=0$, we chose the integration constant ${\mathcal M}=2\sqrt{V_1V_2}$ and as a result found that $\zeta=\sqrt{\frac{V_1}{V_2}}>0$ and, therefore, $\Upsilon>0$. However, we could have chosen ${\mathcal M}=-2\sqrt{V_1V_2}$, which also ensures $\Lambda=0$. But with such a choice, the constraint (\ref{zeta empty redef}) leads to
 $\zeta=-\sqrt{\frac{V_1}{V_2}}<0$ and, therefore, $\Upsilon<0$. Hence, changing the sign of the integration constant ${\mathcal M}$, we obtain solutions with opposite sign of the space-time orientation.  In essence, we are discovering {\em the TMT-effect of  spontaneous restoration of the sign of the space-time orientation}, which is one of the effects of pregeometry.

The conclusion that follows from the described TMT effect of spontaneous restoration of space-time orientation is so fundamental that it allows us to formulate in the most concentrated form the main difference between TMT and conventional field theories. In conventional field theories, it is assumed by default that the sign of space-time orientation can be chosen arbitrarily. But in TMT, as we have seen, one of the two possible signs of orientation of the space-time manifold of our Universe is fixed by the choice of the solution of cosmological equations. And since the sign of orientation coincides with the sign of the scalar function  $\zeta$, which plays a decisive role in the dynamics of fields, 
{\em the physical results predicted by TMT turn out to be different for different signs of  the space-time orientation}.

To construct the  action in the presence of matter, we take the action in empty space (\ref{S-empty redef}) as the gravitational term, to which the matter terms are added. As matter, we can consider different types of models, and, compared to conventional models, the whole novelty in the context of
TMT is the possibility of choosing the coefficients $\alpha_i$ and  $\beta_i$ ($i=1,2,...n$) in the volume elements $(\alpha_i\sqrt{-g}+\beta_i\Upsilon)d^4x$  in each $i$-th term $(\alpha_i\sqrt{-g}+\beta_i\Upsilon)\tilde{L}_i$ of the Lagrangian density.

Since the original action does not imply the existence of a definite space-time orientation, and also because of the the sign-indefinitness of $\Upsilon$, when constructing the original action terms using linear combinations of $\sqrt{-g}$ and $\Upsilon$, it should be assumed that the coefficients in front of 
$\Upsilon$ in the total volume measures can be positive  as well as negative. That is, assuming  $\alpha_i>0$,  $\beta_i>0$, the total volume measures can have the form 
$(\alpha_i\sqrt{-g}\pm\beta_i\Upsilon)d^4x$. This finding is a direct consequence of the fact that the  original action structure takes pre-geometry into account to a large extent, and this is important for achieving the main results that will be presented in this review.
In the following sections, using examples of specific models, we will consider how the above-described features of TMT manifest themselves.
  But before that, one more remark, more of a technical nature, needs to be made.
As follows from the above analysis, the general form of the i-th term of the Lagrangian density is $(\alpha_i\sqrt{-g}\pm\beta_i\Upsilon)\tilde{L}_i$.
 Since the $\varphi_a$ functions have already been redefined to obtain the gravitational term (\ref{S-empty redef}) of the action, to eliminate the coefficient $\beta_i$, we can factor it out of the brackets and absorb it by redefining either the model parameters or the fields in $\tilde{L}_i$. As a result, the i-th term of
 the Lagrangian density takes the form $(b_i\sqrt{-g}\pm \Upsilon)L_i$, 
where $b_i=\frac{\alpha_i}{\beta_i}$ and $L_i$ is obtained from  the mentioned redefinitions in $\tilde{L}_i$. 
This is the form in which all Lagrangian density terms in TMT models will be used in what follows. 

An important note regarding the terminology used in what follows. In models of the "conventional" theory, the original action contains a single volume element $dV_g=\sqrt{-g}d^4x$. In conventional alternative gravity models, the original action differs from the Einstein-Hilbert action only in the form of the Lagrangian, for example, the presence of a non-minimal coupling with curvature.
The transformation to the Einstein  frame, which in conventional theories  is carried out  in the original action, simply changes the set of variables used to describe the theory, but does not change the theory itself. In TMT, the main difference is that, along with the volume measure $dV_g$, the original action contains the volume measure $dV_{\Upsilon}$, although a modification of the Lagrangian is also possible. Therefore, the original TMT action differs from the Einstein-Hilbert action in the form of the Lagrangian {\em density}, and to bring the TMT action to the Einstein frame, the conformal transformation must exclude from the gravitational term of the action not only the non-minimal coupling, but also $\Upsilon$. In a theory obtained in this way, the constraint determining $\zeta$ cannot arise, that is, by acting in this way we would be dealing with a different theory. It is therefore incorrect to use the term "Jordan frame" in the usual sense for the set of variables in the original action of  TMT. {\em To emphasize this distinction}, in what follows we will use  the terms "primordial variables", "primordial model parameters",  "primordial potential", "primordial Lagrangian" and "primordial action" instead of "variables in the original frame",  "original model parameters", "original potential", "original Lagrangian" and "original action".

In conclusion of this section, it should be noted that in addition to the ideas for constructing TMT described above, there is a wide class of models that use more than one non-canonical form of space-time volume. For an overview of these modified gravity theories and their applications in cosmology, see article \cite{more than one non-canonical form} and the references therein.

\section{TMT procedure using a toy model as an example}
\label{toy model}

The first scalar field TMT models in which the primordial action contains both the volume element $\sqrt{-g}d^4x$ and $\Upsilon d^4x$ were studied  in refs. \cite{GK2,G1,G2}.
To become familiar with the procedure for deriving equations and their initial analysis in TMT, it is useful to start with a toy model as a simple example. To simplify the TMT structure as much as possible, we restrict ourselves to a model in which the Lagrangian of the $\phi$ field enters the initial TMT action with the same volume element $(\sqrt{-g}+\Upsilon)d^4x$ as in the gravitational term in eq.(\ref{S-empty redef}). Thus, we choose the primordial action in the form
\begin{equation}
S_{toy}=\int\Biggl[(\sqrt{-g}+\Upsilon)\left(-\frac{M_P^2}{2}g^{\mu\nu}R_{\mu\nu}(\Gamma)+
\frac{1}{2}g^{\mu\nu}\phi_{,\mu}\phi_{,\nu}-V(\phi)\right)
-\sqrt{-g}V_1-
\frac{\Upsilon^2}{\sqrt{-g}}V_2\Biggr]d^4x,
\label{S-toy}
\end{equation} 
where the choice of the potential $V(\phi)$ will be discussed later.

Let's start by varying the action with respect to scalar functions $\varphi_{a}$. Similar to what was in section \ref{Two-Measure Theory in empty space}, we get 
\begin{equation}
B^{\mu}_{a}\partial_{\mu}\left( -\frac{\tilde{M}_P^2}{2}R(\Gamma,g)+\frac{1}{2}g^{\mu\nu}\phi_{,\mu}\phi_{,\nu}-V(\phi)-
2\zeta V_2\right)=0.
\label{varphiB toy}
\end{equation} 
where 
$B^{\mu}_{a}=\varepsilon^{\mu\nu\alpha\beta}\varepsilon_{abcd}
\partial_{\nu}\varphi_{b}\partial_{\alpha}\varphi_{c}
\partial_{\beta}\varphi_{d}$.
Since $Det (B^{\mu}_{a}) \propto\Upsilon^{3}$ it follows
that if
 \begin{equation}
\quad \textsf{everywhere} \quad \Upsilon(x)\neq 0, 
\label{Phi neq 0 toy}
\end{equation}
the equality
\begin{equation}
-\frac{M_P^2}{2}g^{\mu\nu}R_{\mu\nu}(\Gamma)=\Bigl({\mathcal M}-\frac{1}{2}g^{\mu\nu}\phi_{,\mu}\phi_{,\nu}+V(\phi)+2\zeta V_2\Bigr)
\label{var varphi toy}
\end{equation} 
must be satisfied, where  ${\mathcal M}$ is a constant of integration with the dimension of $(mass)^4$.

Variation of the action with respect to $g^{\mu\nu}$ yields  the equation
\begin{eqnarray}
&&-\frac{M_P^2}{2} R_{\mu\nu}(\Gamma)+\frac{1}{2}\phi_{,\mu}\phi_{,\nu}
\nonumber
\\
&&- \frac{g_{\mu\nu}}{2(1+\zeta)}\left[-\frac{M_P^2}{2} R(\Gamma, g) 
 +\frac{1}{2}g^{\alpha\beta}\phi_{,\alpha}\phi_{,\beta}-V(\phi) -V_1+ \zeta^2 V_2 \right]=0.
 \label{Grav.eq toy}
\end{eqnarray}
Using the trace of eq.(\ref{Grav.eq toy}) one can eliminate  the term  $-\frac{1}{2}M_P^2 R(\Gamma, g)$ from eq.(\ref{var varphi toy}). As a result, we come to the conclusion that for these equations to be consistent, it is necessary that an algebraic constraint be satisfied that defines scalar $\zeta$ as a function locally dependent on $\phi(x)$:
\begin{equation}
\zeta(\phi(x);{\mathcal M})=\frac{{\mathcal M}-2 V_1}{{\mathcal M}-2V_2+V(\phi)}.
\label{zeta toy}
\end{equation}

Variation of the affine connection yields the equations solution of which has the following form
\begin{equation}
\Gamma^{\lambda}_{\mu\nu}=\{^{\lambda}_{\mu\nu}\}+ (\delta^{\lambda}_{\mu}\chi,_{\nu}
+\delta^{\lambda}_{\nu}\chi,_{\mu}-
\chi,_{\beta}g_{\mu\nu}g^{\lambda\beta}),
 \label{GAM2 toy}
\end{equation}
where $\{^{\lambda}_{\mu\nu}\}$  are the Christoffel's connection
coefficients of the metric $g_{\mu\nu}$ and
\begin{equation}
\chi(x)=\frac{1}{2}\ln(1+\zeta).
\label{ln in Gamma toy}
\end{equation}
If $\chi(x)\neq const.$ the metricity condition does not hold and consequently
geometry of the space-time with the metric $g_{\mu\nu}$ is generically non-Riemannian. In what follows we will completely ignore a possibility to incorporate the torsion tensor, which could be an additional source for the space-time  to be different from Riemannian. 
It is easy to see that  the transformation of the metric
\begin{equation}
\tilde{g}_{\mu\nu}=(1+\zeta) g_{\mu\nu}
 \label{Ein toy}
\end{equation}
turns  the affine connection $\Gamma^{\lambda}_{\mu\nu}$ into
the Christoffel connection coefficients of the metric
$\tilde{g}_{\mu\nu}$ and the space-time turns into (pseudo)
Riemannian. 

Gravitational equations
(\ref{Grav.eq toy}) expressed in terms of the metric $\tilde{g}_{\mu\nu}$ take the
canonical GR  form 
\begin{equation}
R_{\mu\nu}(\tilde{g})-\frac{1}{2}\tilde{g}_{\mu\nu}R(\tilde{g})=\frac{1}{M_P^2}T_{\mu\nu}^{(eff)}
\label{grav eq Ein toy}
\end{equation}
with the same Newton constant as in the original frame.
Here $R_{\mu\nu}(\tilde{g})$ and $R(\tilde{g})$ are the Ricci tensor and the scalar curvature of the metric $\tilde{g}_{\mu\nu}$, respectively.
 Therefore, gravity becomes canonical, and the set of dynamical variables using the metric $\tilde{g}_{\mu\nu}$ can be called the Einstein frame. 
$T_{\mu\nu}^{(eff)}$ on the right side of the Einstein equations (\ref{grav eq Ein toy})  is {\em the  TMT-effective energy-momentum tensor}
\begin{equation}
T_{\mu\nu}^{(eff)}=\phi_{,\mu}\phi_{,\nu}-\tilde{g}_{\mu\nu}\frac{1}{2}\tilde{g}^{\alpha\beta}\phi_{,\alpha}\phi_{,\beta}
+\tilde{g}_{\mu\nu}U_{eff}(\phi,\zeta(\phi);{\mathcal M}),
 \label{Tmn toy}
\end{equation}
 and
\begin{equation}
U_{eff}(\phi,\zeta;{\mathcal M})=\frac
{{\mathcal M}-V_1-V_2}{(1+\zeta)^2}+V_2
 \label{U zeta toy}
\end{equation}
is the $\zeta$-dependent form of {\em the TMT-effective potential}.
To avoid confusion, it should be noted that the use of the term "TMT-effective potential" has nothing to do with the effective potential obtained taking into account quantum corrections. The terms "TMT-effective energy-momentum tensor" and "TMT-effective potential" are used to denote the energy-momentum tensor and potential that appear in the  equations of motion  after  performing all the steps of the TMT procedure described above, which begins with varying the primordial TMT action and ends with the transition to the Einstein frame in the equations of motion.
It is very important to note that the application of the  least action principle in the Palatini formalism does not lead to a differential equation for $\zeta$ and therefore {\em $\zeta$ is not a dynamical variable}. Substituting $\zeta$ given by the constraint (\ref{zeta toy}) into (\ref{U zeta toy}) and assuming, as in section \ref{Two-Measure Theory in empty space}, $V_1<0$ and $V_2<0$, we find the TMT-effective potential  of the field $\phi$ in the following form 
\begin{equation}
U_{eff}(\phi)=
\frac{{\mathcal M}^2-4V_1V_2+V(\phi)\bigl[2{\mathcal M}+4|V_2|+V(\phi)\bigr]}
{4(|V_1|+|V_2|+{\mathcal M})},
 \label{U with M toy}
\end{equation}
which still contains the integration constant ${\mathcal M}$.

Just as was done in the section \ref{Two-Measure Theory in empty space}, the desired value of  the TMT-effective cosmological constant (CC) can be found by fixing a certain value of ${\mathcal M}$. To do this, we first obtain  the $\phi$ field equation. By varying the action (\ref{S-toy}) with respect to $\phi$ with subsequent transition (\ref{Ein toy}) to the Einstein frame, we obtain
\begin{equation}
\frac{1}{\sqrt{-\tilde{g}}}\partial_{\mu}\bigl(\sqrt{-\tilde{g}}\tilde{g}^{\mu\nu}\partial_{\nu}\phi\bigr)+V^\prime (\phi)=0,
\label{phi eq toy}
\end{equation}
where $V^\prime (\phi)=\frac{dV}{d\phi}$.

Note that instead of the system of the Einstein equation (\ref{grav eq Ein toy}) with the energy-momentum tensor
(\ref{Tmn toy})   and the field $\phi$-equation (\ref{phi eq toy}), it is much more convenient to work with the
TMT-effective action, the variation of which gives these equations. As usual, if $\tilde{g}^{\alpha\beta}\phi_{,\alpha}\phi_{,\beta}>0$,  the energy-momentum tensor $T_{\mu\nu}^{(eff)}$ can be rewritten in the form of a perfect fluid. Then the pressure density plays the role of the matter Lagrangian  in the TMT-effective action, and we arrive
at the {\em TMT-effective action}
\begin{equation}
S_{eff}^{(TMT)}=\int \left(-\frac{M_P^2}{2}R(\tilde{g})+L_{eff}(\phi)\right)\sqrt{-\tilde{g}}d^4x,
\label{Seff toy}
\end{equation}
with the following  {\em TMT-effective Lagrangian}
\begin{equation}
L_{eff}(\phi, X_{\phi})=\frac{1}{2}\tilde{g}^{\alpha\beta}\phi_{,\alpha}\phi_{,\beta}-U_{eff}(\phi).
\label{Leff toy}
\end{equation}
The idea of working with TMT-effective action and TMT-effective Lagrangian will be useful in this paper when exploring more complex TMT models.

Proceeding with the simplest toy model, we choose the  original potential $V(\phi)=\frac{\lambda}{4}\phi^4$.
Then from eq.(\ref{phi eq toy}) it follows that the vacuum of the classical field $\phi$ is achieved at $\phi=0$. The value of $U_{eff}(\phi)$ at $\phi=0$ gives us  the following equation describing the dependence of the CC $\Lambda$ on the integration constant ${\mathcal M}$:
\begin{equation}
\Lambda =U_{eff}(0)=
\frac{{\mathcal M}^2-4V_1V_2}
{4(|V_1|+|V_2|+{\mathcal M})}.
 \label{U with M toy 0}
\end{equation}
 When choosing $\Lambda =0$, the solutions are ${\mathcal M}_{+}=2\sqrt{V_1V_2}$ and ${\mathcal M}_{-}=-2\sqrt{V_1V_2}$. Using (\ref{zeta toy}) we find the  corresponding values of $\zeta$:
\begin{equation}
\zeta_{v+}=\zeta(\phi;{\mathcal M})|_{(\phi=0, {\mathcal M}={\mathcal M}_{+})}=\sqrt{\frac{V_1}{V_2}}
\label{zeta toy vac Mplus}
\end{equation}
and 
\begin{equation}
\zeta_{v-}=\zeta(\phi;{\mathcal M})|_{(\phi=0, {\mathcal M}={\mathcal M}_{-})}=-\sqrt{\frac{V_1}{V_2}}.
\label{zeta toy vac Mminus}
\end{equation}
Thus, the model predicts two vacuum states with $\Lambda =0$ that differ in the sign of $\zeta$ in the vacuum and, consequently, in the sign of the corresponding values of $\Upsilon$.

A solution (\ref{var varphi toy}) of eq.(\ref{varphiB toy}) exists under the condition $\Upsilon(x)\neq 0$, eq.(\ref{Phi neq 0 toy}), i.e, 
only if $\Upsilon(x)>0$ or only if $\Upsilon(x)<0$. Therefore,  only those solutions of the system of equations  are valid for which {\em the corresponding $\Upsilon(x)$ is sign-definite}, since eq.(\ref{varphiB toy}) 
 is one of the equations of the system.
It should be noted here that, as usual, by default we assume that the original metric $g_{\mu\nu}$ in the primordial action is regular, that is  $g=\det(g_{\mu\nu})<0$. 
Therefore, the validity of the solution regarding the fulfillment of the condition on the sign of $\Upsilon$ can be controlled by checking the sign of the scalar $\zeta=\Upsilon/\sqrt{-g}$.  In other words, {\bf only those cosmological solutions (together with their initial conditions and the vacuum states)  are valid for which 
$\zeta(x)$ does not change the sign throughout the evolution of the universe.} This is another one of the effects of pregeometry.

We found that in the toy model under study there are two vacuum states with $\zeta_{v+}$ and $\zeta_{v-}$ differing in the signs. Therefore, in accordance with the analysis in the previous paragraph, there are two classes of solutions differing in the sign of $\zeta(x)$.
 Considering that the sign of $\zeta$ determines the sign of the orientation of the space-time manifold, we come to the conclusion that the difference between these two classes of solutions is that they are realized on space-time manifolds with opposite orientations.

It seems interesting to find the TMT-effective potentials  for the two classes of solutions indicated above. To do this, it is sufficient to substitute the corresponding values ${\mathcal M}_{+}$ and ${\mathcal M}_{-}$ of the  integration constant into formula (\ref{U with M toy}). As a result, we find
\begin{equation}
U_{eff}(\phi)\big|_{{\mathcal M}_{+}}=
\frac{1+\sqrt{\frac{V_1}{V_2}}+\frac{\lambda \phi^4}{16|V_2|}}
{\Bigl(1+\sqrt{\frac{V_1}{V_2}}\Bigr)^2}\cdot\frac{\lambda}{4}\phi^4,
 \label{U with Mplus toy}
\end{equation}
\begin{equation}
U_{eff}(\phi)\big|_{{\mathcal M}_{-}}=
\frac{1-\sqrt{\frac{V_1}{V_2}}+\frac{\lambda \phi^4}{16|V_2|}}
{\Bigl(1-\sqrt{\frac{V_1}{V_2}}\Bigr)^2}\cdot\frac{\lambda}{4}\phi^4.
 \label{U with Mminus toy}
\end{equation}
Thus, as was the case in the empty space model, we conclude that {\em the physical results predicted by TMT turn out to be different for different signs of spacetime orientation}.

\section{TMT with additional, internal, symmetry: a class of TMT models
 that provide quintessence without the need for fine-tuning}
\label{quint with additional symm}

In this section we will show that in the case where the TMT has additional symmetry, there is a broad class of TMT models that provide the quintessence without the need for fine-tuning. This will be done based on the ideas and results of ref. \cite{K}.
The primordial action can be represented in the form
\begin{equation}
    S = \int\left[ L_{1}\sqrt{-g}+L_{2}\Upsilon\right]d^{4}x, 
\label{S quint} 
\end{equation}
where
\begin{equation}
L_1=-\frac{M_P^2}{2}g^{\mu\nu}R_{\mu\nu}(\Gamma)+
\frac{1}{2}g^{\mu\nu}\phi_{,\mu}\phi_{,\nu}-v_1(\phi)-V_1,
\label{L1 quint}
\end{equation} 
\begin{equation}
L_2=-\frac{M_P^2}{2}g^{\mu\nu}R_{\mu\nu}(\Gamma)+
\frac{1}{2}g^{\mu\nu}\phi_{,\mu}\phi_{,\nu}-v_2(\phi).
\label{L2 quint}
\end{equation}
There are two differences in the structure of the actions (\ref{S quint}) and (\ref{S-toy}). First, instead of a single primordial potential $V(\phi)$ entering (\ref{S-toy}) with both volume element $\sqrt{-g}d^4x$ and $\Upsilon d^4x$, in the action (\ref{S quint}) the primordial potential is split into two: $v_1(\phi)$ entering with the volume element $\sqrt{-g}d^4x$, and $v_2(\phi)$ entering with the volume element $\Upsilon d^4x$. Second, the vacuum-like term
 \begin{equation}
-V_2\int\frac{\Upsilon^2}{\sqrt{-g}}d^4x
\label{V2 term}
\end{equation}
 is missing from action (\ref{S quint}). This choice may be justified if, for some as yet unknown reason (possibly due to a more fundamental theory), the Lagrangian $L_2$ does not depend on the $\varphi_a$ functions from which the volume measure density $\Upsilon$ is constructed. For example, as discussed in paper \cite {GK2}, the presence of such a term in the primordial action may be forbidden if a more fundamental theory requires that the primordial TMT action be invariant,  up to a total divergence, under transformations 
 \begin{equation}
\varphi_a\to \varphi_a+f_a(L_2), \quad \text{where} \quad f_a, \, a=1...4, \quad \text{are arbitrary differentiable functions of}\,  L_2.
\label{f(L2) transform}
\end{equation}

By performing with action (\ref{S quint})-(\ref{L2 quint}) all the operations that make up the TMT procedure described in detail in section \ref{toy model}, we arrive at the following results:
\begin{equation}
\zeta=\frac{\mathcal{M}-2V_1-2v_1(\phi)+v_2(\phi)}{\mathcal{M}+v_2(\phi)},
\label{zeta quint}
\end{equation}
where  ${\mathcal M}$ is the integration constant;

the metric tensor in the Einstein frame is defined as in eq.(\ref{Ein toy});

the Einstein equations (\ref{grav eq Ein toy}) with  the  TMT-effective energy-momentum tensor $T_{\mu\nu}^{(eff)}$ in the form as in eq.(\ref{Tmn toy}) and with the TMT-effective potential  
\begin{equation}
U_{eff}(\phi)=
\frac{\bigl[{\mathcal M}+v_2(\phi)\bigr]^2}
{4\bigl[{\mathcal M}-V_1+v_2(\phi)-v_1(\phi)\bigr]}.
 \label{U with M quint}
\end{equation}

In contrast to  convensional gravitational theories where the quintessential 
potential must be
slow {\em decreasing} function as $\phi\rightarrow\infty$, in TMT
we have an {\em absolutely new option}: the quintessential behaviour of the 
TMT-effective potential $U_{eff}(\phi)$ for $\phi$ large enough may be achieved 
with {\em increasing} primordial potentials $v_{1}(\phi)$ and $v_{2}(\phi)$. This
circumstance enables to avoid both the cosmological constant problem and
the problem of the flatness of the quintessential potential.

For illustration of these statements, following ref. \cite{K},  we consider two types of models. 

\paragraph{The inverse power low quintessential potential}

We start from the primordial potentials
  $v_{1}(\phi)$ and $v_{2}(\phi)$ which for $\phi$ large enough
are dominated by the positive power low functions
\begin{equation} 
v_{1}(\phi)  = -\frac{1}{4} m_{1}^{(4-2n_{1})} \phi^{2n_{1}}+\text{subdominant terms (as  $\phi$ large enough)},
\label{V1poz quint}
\end{equation}
\begin{equation} 
v_{2}(\phi)  = m_{2}^{(4-n_{2})} \phi^{n_{2}}+\text{subdominant terms (as  $\phi$ large enough)}
\label{V2poz quint}
\end{equation}
with $n_1>n_2$, we obtain the TMT-effective potential which for $\phi$ 
large enough has the inverse power low form
\begin{equation}
U_{eff}(\phi) \approx \frac{m_{2}^{2(4-n_{2})}}{m_{1}^{2(2-n_{1})}}
\frac{1}{\phi^{2(n_1 -n_2)}}.
\label{Upower}
\end{equation}
and does not depend on the integration constant ${\mathcal M}$. It should be especially emphasized
 that adding to the primordial action the term $-V_{1}\int\sqrt{-g}d^{4}x$, $V_1\equiv const$, which in the convensional  gravity models would have the meaning of a cosmological constant,  does not affect $U_{eff}(\phi)$ for sufficiently large $\phi$.
  Another particularly interesting case is $v_{2}(\phi)\equiv 0$ and, for example,
$v_{1}(\phi)\equiv -\frac{\lambda}{4}\phi^4$. Then $U_{eff}(\phi)(\phi)={\mathcal M}^{2}/\lambda\phi^{4}$.

 Thus starting from the polynomial
form of the  primordial potentials  $v_{1}(\phi)$ and $v_{2}(\phi)$ with an appropriate choice
of the powers $n_1$ and $n_2$ of the leading terms, one can in fact provide a
generation of the inverse power low quintessential potential in such a way
that neither the cosmological constant problem nor the the problem of the 
flatness of the quintessential potential do not appear at all.

\paragraph{The  quintessential potential of the exponential form.} If the primordial potentials $v_{1}(\phi)$ and $v_{2}(\phi)$ are chosen to be exponentially growing as  $\phi$ is large enough 
\begin{equation} 
v_{1}(\phi)  = -\frac{1}{4} m_{1}^4e^{2\alpha\phi/M_p}+\text{subdominant terms (as  $\phi$ large enough)},
\label{V1exp quint}
\end{equation}
\begin{equation} 
v_{2}(\phi)  = m_{2}^4 e^{\beta\phi/M_p}+\text{subdominant terms (as  $\phi$ large enough)},
\label{V2exp quint}
\end{equation}
where  $\alpha$ and $\beta$ are positive constants, then for $\phi$ large enough  the TMT-effective potential (\ref{U with M quint}) takes the form
\begin{equation}
    U_{eff}(\phi)=\frac{1}{m_{1}^{4}}\left( 
m_{2}^{4}e^{-(\alpha-\beta )\phi/M_{p}}+ 
\mathcal{M}^{4}e^{-\alpha \phi/M_{p}}\right)^{2}
\label{Ueff exp 1} 
\end{equation} 
The case $\alpha=\beta$ corresponds to a kind of scale-invariant theory studied by Gundelman \cite{G1,G2,G4}
 and will be considered in the next section.
The case $\alpha>\beta >0$  is the most interesting  from the viewpoint of the quintessence:
for $\alpha \phi\gg M_{p}$,
the TMT-effective potential  behaves as a decaying exponent
\begin{equation}
   U_{eff}(\phi) \simeq \frac{m_{2}^{8}}{m_{1}^{4}} 
e^{-2(\alpha-\beta)\phi /M_{p}}.
\label{U2}
\end{equation}
It is noteworthy that, as in case of the inverse power low quintessential potential (\ref{Upower}), the asymptotic behavior of $U_{eff}(\phi)$ does not depend on the integration constant $\mathcal{M}$.

It should be noted that if symmetry (\ref{f(L2) transform}) is broken, then in both types of models considered, adding the term (\ref{V2 term}) to the action (\ref{S quint}) leads to a TMT-effective  potential that tends to a constant $V_2$ as $\phi\to \infty$.

In \cite{K}, by choosing the primordial potentials $v_{1}(\phi)$ and $v_{2}(\phi)$ as linear combinations of the power and exponential functions of
$\phi$, it was demonstrated that it is possible to obtain the TMT-effective potential $U_{eff}(\phi)$, which allows realizing the idea 
of quintessential inflation \cite{PV}, and also without the  cosmological constant problem.
 In the range of $\phi$ values  corresponding to inflation, 
$U_{eff}(\phi)$ is a power function of $\phi$, which ensures a scenario of chaitic inflation.
 However, this contradicts the requirement of flatness of the inflationary potential imposed by modern CMB data \cite{Planck},\cite{{BICEP}}.

\section{Brief review  of the TMT models with global scale invariance.
 Some results in cosmology and gravity}
\label{Scale invar models}

This section provides an opportunity to get acquainted with the main results of a large series of articles investigating models
in which the primordial TMT action is invariant with respect to global scale transformations. A common feature 
of all these models is the absence of the term (\ref{V2 term}) in the primordial action.

\subsection{Scale invarint model originally developed by E. Guendelman}
\label{Guend mod}

In series of papers  \cite{G1,G2,G4}, E. Guendelman suggested and studied the scale invariant TMT model with the primordial action
\begin{equation}
S=\int d^{4}x 
\Bigl[\Upsilon\Bigl(-\frac{M_P^2}{2}R(\Gamma ,g)+\frac{1}{2}g^{\mu\nu}\phi_{,\mu}\phi_{,\nu}-V(\phi)\Bigr) +\sqrt{-g}U(\phi)\Bigr],
 \label{Ed primord action}
\end{equation}
where the primordial potentials are defined by the formulas
\begin{equation}
V(\phi)= f_1e^{\alpha\phi}, \qquad    U(\phi)= f_2e^{2\alpha\phi}.
 \label{Ed V U}
\end{equation}
 The action is invariant under the global scale transformations:
\begin{equation}
    \varphi_{a}\rightarrow \lambda_a\varphi_{a} \,
\text{(no sum on a), where} \, \prod_{a}\lambda_a=e^\theta, \text{which means} \quad \Upsilon
\rightarrow e^{\theta}\Upsilon,
 \label{st1}
\end{equation}
\begin{equation}
 g_{\mu\nu}\rightarrow e^{\theta }g_{\mu\nu}, \quad
\Gamma^{\mu}_{\alpha\beta}\rightarrow \Gamma^{\mu}_{\alpha\beta},
\quad \phi\rightarrow \phi-\frac{\theta}{\alpha}.
\label{st2}
\end{equation}

Variation of the primordial action (\ref{Ed primord action}) with respect to $\varphi_a$ under the condition $\Upsilon\neq 0$ yields the
equation
\begin{equation}
-\frac{M_P^2}{2}R(\Gamma ,g)+\frac{1}{2}g^{\mu\nu}\phi_{,\mu}\phi_{,\nu}-V(\phi)={\mathcal M}. 
\label{varying varphi a guend}
\end{equation}
The appearance of a nonzero integration constant ${\mathcal M}$ {\em spontaneously breaks the scale invariance}.
Continuing to execute all the operations that make up the TMT procedure using the primordial action (\ref{Ed primord action}) results in the following:
\begin{equation}
\zeta=\frac{2U(\phi)}{\mathcal{M}+V(\phi)};
\label{zeta quint}
\end{equation}

the metric tensor in the Einstein frame is defined by 
\begin{equation}
\tilde{g}_{\mu\nu}=\zeta\, g_{\mu\nu};
 \label{Ein frame Ed}
\end{equation}

the Einstein equations (\ref{grav eq Ein toy}) with  the  TMT-effective energy-momentum tensor $T_{\mu\nu}^{(eff)}$ in the form as in eq.(\ref{Tmn toy}) and with the TMT-effective potential  
\begin{equation}
U_{eff}(\phi)=\frac{1}{4f_2}
\bigl(f_1+{\mathcal M}e^{-\alpha\phi}\bigr)^2.
 \label{U eff Ed}
\end{equation}
This potential has remarkable properties. Firstly, if $f_1/{\mathcal M}<0$ and $f_2>0$, a minimum is achieved at $\phi_{min}=-\frac{1}{\alpha}\ln|\frac{f_1}{{\mathcal M}}|$ and $U_{eff}(\phi_{min})=0$. This means that {\em the model provides  zero cosmological
constant without fine tuning}. Second, for $\alpha\phi\gg 1$, $U_{eff}(\phi)$ has an infinite plateau with height $\frac{f_1^2}{4f_2}$. Therefore, the model can also describe  inflation in the slow-roll regime.

Note that since $\tilde{g}_{\mu\nu}$ is invariant under the scale transformations (\ref{st1}), (\ref{st2}), in the Einstein frame the latter are reduced to only a shift of the scalar field $\phi\rightarrow \phi-\frac{\theta}{\alpha}$. In ref.\cite{G2} Guendelman showed that Goldstone's theorem is generally inapplicable in this theory, and, consequently, the Goldstone boson associated with the spontaneous breaking of this symmetry is absent.

It should be noted that, unlike action (\ref{S-toy}), action (\ref{Ed primord action}) does not contain terms
\begin{equation}
\int\Bigl[-\frac{M_P^2}{2}g^{\mu\nu}R_{\mu\nu}(\Gamma)+
\frac{1}{2}g^{\mu\nu}\phi_{,\mu}\phi_{,\nu}\Bigr]\sqrt{-g}d^4x
\label{non invar terms Ed}
\end{equation}
 These terms cannot be added because they are not invariant with respect to scale transformations (\ref{st2}).
But in the model presented in the next subsection, these terms are included in a scale invariant manner.

\subsection{K-Essence and inflation in the scale invariant TMT model}
\label{K-Essence and inflation in the scale invariant}

 Here a brief review of refs.\cite{Kess4} and \cite{GK4} is presented. The  TMT model including also the terms contained in (\ref{non invar terms Ed})
is described by the following primordial action
\begin{eqnarray}
&S=&\int d^{4}x e^{\alpha\phi /M_{p}}
\Bigl[-\frac{M_P^2}{2}R(\Gamma ,g)(\Upsilon +b_{g}\sqrt{-g})
\nonumber
\\
&&+(\Upsilon+b_{\phi}\sqrt{-g})\frac{1}{2}g^{\mu\nu}\phi_{,\mu}\phi_{,\nu}-e^{\alpha\phi
/M_{p}}\left(\Upsilon V_{1} +\sqrt{-g}V_{2}\right)\Bigr],
 \label{totaction k-ess}
\end{eqnarray}
where parameters $b_g$ and $b_{\phi}$ are introduced based on the discussion in sectiom \ref{Some additional mathematical aspects taken into account in TMT};
 $\alpha$ is a real parameter that is assumed to be positive. Note that the notation used for the constants $V_{1}$ and $V_{2}$
 differs from that used in  sections \ref{toy model} and \ref{quint with additional symm}.
This action is invariant under the global scale transformations 
\begin{eqnarray}
   && g_{\mu\nu}\rightarrow e^{\theta }g_{\mu\nu}, \quad 
\Gamma^{\mu}_{\alpha\beta}\rightarrow \Gamma^{\mu}_{\alpha\beta}, \quad
\phi\rightarrow \phi-\frac{M_{p}}{\alpha}\theta,
\label{st}
\\
&& \varphi_{a}\rightarrow \lambda_{ab}\varphi_{b}, \quad
\text{where} \quad \det(\lambda_{ab})=e^{2\theta} \quad \text{and, therefore,} \quad \Upsilon\rightarrow e^{2\theta }\Upsilon. 
\nonumber
\end{eqnarray}
Note that scalar $\zeta\equiv\frac{\Upsilon}{\sqrt{-g}}$ is  invariant under these transformations.

Performing the TMT procedure with action (\ref{totaction k-ess}) produces the following results:

a) similar to the model described in section \ref{Guend mod}, variation of the primordial action (\ref{totaction k-ess})
with respect to $\varphi_a$ subject to the condition $\Upsilon\neq 0$ leads to an
equation where the appearance of an arbitrary integration constant ${\mathcal M}$ also  breaks the scale invariance;

b) the constraint defining scalar $\zeta$ reads now
\begin{equation}
\zeta(\phi, X;{\mathcal M})=b_{g}\frac{{\mathcal M}e^{-2\alpha\phi /M_P}+V_1-2V_2-\delta\cdot b_g^2X}
{{\mathcal M}e^{-2\alpha\phi /M_P}+V_1+\delta\cdot b_gX}
\label{constraint in k-ess}
\end{equation}
where
\begin{equation}
X\equiv\frac{1}{2}\tilde{g}^{\alpha\beta}\phi_{,\alpha}\phi_{,\beta} \, ,
 \qquad \delta =\frac{b_{g}-b_{\phi}}{b_{g}}
\label{delta}
\end{equation}
and the metric tensor in the Einstein frame is defined by
\begin{equation}
\tilde{g}_{\mu\nu}=e^{\alpha\phi/M_{p}}(\zeta +b_{g})g_{\mu\nu};
\label{ct}
\end{equation}

c) Einstein equations have canonical GR form (\ref{Grav.eq toy}) with 
the TMT-effective energy-momentum tensor
\begin{equation}
T_{\mu\nu}^{eff}=\frac{\zeta +b_{\phi}}{\zeta +b_{g}}
\phi_{,\mu}\phi_{,\nu}-X
\tilde{g}_{\mu\nu}
+\tilde{g}_{\mu\nu}U_{eff}(\phi;\zeta,{\mathcal M}),
 \label{Tmn k-ess}
\end{equation}
where the function $U_{eff}(\phi;\zeta,{\mathcal M})$ is defined as
following
\begin{equation}
U_{eff}(\phi;\zeta ,{\mathcal M})=
\frac{b_{g}\left[{\mathcal M}e^{-2\alpha\phi/M_{p}}+V_{1}\right]
-V_{2}}{(\zeta +b_{g})^{2}};
 \label{Ueff zeta k-ess}
\end{equation}

d) the $\zeta$-dependent form of the field $\phi$ equation
\begin{equation}
\frac{1}{\sqrt{-\tilde{g}}}\partial_{\mu}\left[\frac{\zeta
+b_{\phi}}{\zeta
+b_{g}}\sqrt{-\tilde{g}}\tilde{g}^{\mu\nu}\partial_{\nu}\phi\right]-\frac{2\alpha\zeta}{(\zeta
+b_{g})^{2}M_{p}}{\mathcal M}e^{-2\alpha\phi/M_{p}} =0.
\label{phi-after-con}
\end{equation}

Similar to what was discussed at the end of the previous subsection, since $\tilde{g}_{\mu\nu}$ is invariant under scale transformations (\ref{st}), spontaneous breaking of scale symmetry in the Einstein frame reduces to spontaneous breaking of shift symmetry
$\phi\rightarrow\phi +const.$. The discussion of the inapplicability
of Goldstone's theorem in this theory, \cite{G2}, is relevant for this model as well.

If $X>0$, the TMT-effective energy-momentum tensor (\ref{Tmn k-ess}) can be
represented in a form of that of  a perfect fluid
\begin{equation}
T_{\mu\nu}^{eff}=(\rho +p)u_{\mu}u_{\nu}-p\tilde{g}_{\mu\nu},
\qquad \text{where} \qquad
u_{\mu}=\frac{\phi_{,\mu}}{(2X)^{1/2}}\label{Tmnfluid}
\end{equation}
with the following energy and pressure densities resulting from
Eqs.(\ref{Tmn k-ess}) and (\ref{Ueff zeta k-ess}) after substituting $\zeta(\phi, X;{\mathcal M})$ given by  the constraint (\ref{constraint in k-ess}):
\begin{equation}
\rho(\phi,X;{\mathcal M}) =X+ \frac{({\mathcal M}e^{-2\alpha\phi/M_p}+V_1)^2-
2\delta b_{g}({\mathcal M}e^{-2\alpha\phi/M_p}+V_1)X -3\delta^{2}b_{g}^2X^2}
{4[b_{g}({\mathcal M}e^{-2\alpha\phi/M_{p}}+V_{1})-V_{2}]},
\label{rho1}
\end{equation}
\begin{equation}
p(\phi,X;{\mathcal M}) =X- \frac{\left({\mathcal M}e^{-2\alpha\phi/M_{p}}+V_{1}+
\delta b_{g}X\right)^2}
{4[b_{g}({\mathcal M}e^{-2\alpha\phi/M_{p}}+V_{1})-V_{2}]}. 
\label{p1}
\end{equation}

Just as the TMT-effective action (\ref{Seff toy}), (\ref{Leff toy})  was constructed for the toy model, in the model under consideration the TMT-effective action, the variation of which leads to the equations written out above, has the form of
the k-essence type  action \cite{Chiba Kess}. \cite{Mukh Kess}
\begin{equation}
S_{eff}=\int\sqrt{-\tilde{g}}d^{4}x\left[-\frac{1}{\kappa}R(\tilde{g})
+p\left(\phi,X;{\mathcal M}\right)\right] \label{k-eff},
\end{equation}
where $p(\phi,X;{\mathcal M})$ is given by Eq.(\ref{p1}).  The TMT-effective Lagrangian $p\left(\phi,X;{\mathcal M}\right)$ can be
represented in the form
\begin{equation}
p\left(\phi,X;{\mathcal M}\right)=K(\phi)X+
L(\phi)X^2-\frac{[V_{1}+{\mathcal M}e^{-2\alpha\phi/M_{p}}]^{2}}
{4[b_{g}\left(V_{1}+{\mathcal M}e^{-2\alpha\phi/M_{p}}\right)-V_{2}]}
\label{eff-L-ala-Mukhanov}
\end{equation}
where $K(\phi)$ and $L(\phi)$ depend on $\phi$ only via
${\mathcal M}e^{-2\alpha\phi/M_{p}}$.

A detailed analysis taking into account the dependence on the model parameters shows that in the initial state of the classical evolution of the universe, the curvature may not have an initial singularity, but its time derivative is singular. This is the type of sudden singularity studied by Barrow \cite{Barrow} on purely kinematic grounds (in the classification of future singularities);

Numerical solutions were found and the phase portrait was studied in detail
for different initial conditions.
Depending on the choice of regions in the parameter space, the model exhibits various possible outcomes
 of cosmological dynamics. Among them, it is especially worth noting power law inflation, which,  depending on the region in
the parameter space (but without fine tuning)  ends
with a  graceful exit either into the state with zero
cosmological constant  or into the state driven by both a
small CC and the field $\phi$ with a quintessence-like potential.

a) In the model with $\delta =0$, the following TMT-effective potential of the $\phi$ field  results
from eqs.(\ref{Ueff zeta k-ess}) and (\ref{constraint in k-ess})
\begin{equation}
U_{eff}(\phi;{\mathcal M})|_{\delta =0}
=\frac{[V_{1}+{\mathcal M}e^{-2\alpha\phi/M_{p}}]^{2}}
{4[b_{g}\left(V_{1}+{\mathcal M}e^{-2\alpha\phi/M_{p}}\right)-V_{2}]}
\label{Veffvac-delta=0}
\end{equation}
If ${\mathcal M}>0$, $V_{1}<0$, and $V_{2}<0$, then the old cosmological constant problem is solved without any fine-tuning in a manner very similar to that used in refs. \cite{G1}, \cite{G2}, \cite{G4} and described in section \ref{Guend mod}.
$\phi =\phi_{0}$, where
$V_{1}+ {\mathcal M}e^{-2\alpha\phi_{0}/M_{p}}=0$, is the minimum of the effective potential, and $U_{eff}(\phi_{0};{\mathcal M})|_{\delta =0}=0$ without any further tuning of parameters and initial conditions. However, as shown in ref. \cite{GK4}, in the model under consideration, the measure density $\Upsilon$ and the primordial metric $g_{\mu\nu}$ oscillate around zero during transition to  the vacuum state $\phi_{0}$. The latter means that it can only happen with very non-trivial modifications in the understanding of the mathematical foundations of TMT, described in sections \ref{Some additional mathematical aspects taken into account in TMT} and \ref{toy model}.
The papers \cite{Kess4} and \cite{GK4} present a detailed analysis that clarifies how and due to what, {\bf contrary to Weinberg's no-go theorem \cite{Weinberg}, the old CC problem can be solved in TMT}.
Namely,  in footnote 8 of review \cite{Weinberg} S. Weinberg indicates the only exception when the conditions required by the theorem are not satisfied.
This is precisely the exception that is implemented in TMT, as shown in section VIII.B of paper \cite{Kess4}.

   b) There is a wide range of parameters for which the late time  Universe undergoes superacceleration: the equation of state is $w=p/\rho <-1$; $w$ asymptotically, as $t\rightarrow\infty$, tends to $-1$ from below; $\rho$ asymptotically approaches from below a positive cosmological constant, the smallness of which does not require fine-tuning of the dimensional parameters.  In ref.\cite{GK4}  it is shown that the crossing of the phantom divide $w=-1$ occurs simultaneously with the change of sign of the total volume measure density $(\Upsilon+b_{\phi}\sqrt{-g})$ in the field $\phi$ kinetic term in the primordial action (\ref{totaction k-ess}), see Fig. 1.

\begin{figure}[htb]
\includegraphics[width=10.0cm,height=8.0cm]{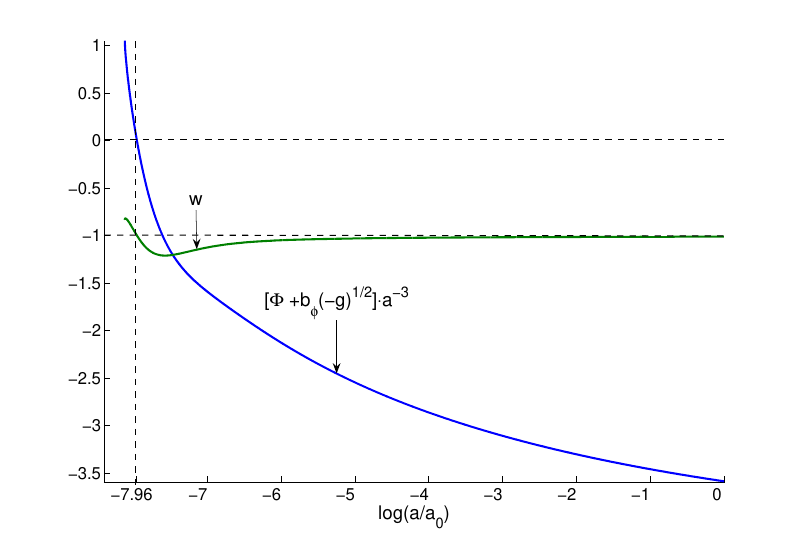}
\caption{For the model with $\alpha =0.2$, $\delta =0.1$,
$V_{1}=10M^{4}$ and $V_{2}=9.9b_{g}M^{4}$ and with the  the initial conditions
$\phi_{in}=M_{p}$, $\dot\phi_{in} =5.7M^2/\sqrt{b_g}$: crossing the phantom divide $w=-1$ and changing sign
of the total volume measure $(\Upsilon +b_{\phi}\sqrt{-g})$ in the
scalar field $\phi$ kinetic term (in the primordial action
(\ref{totaction k-ess})) occur simultaneously. The figure is copied from ref.\cite{GK4}, where $\Phi$ was used instead of $\Upsilon$.}
\label{fig5}
\end{figure}

In paper \cite{Emergent} the described model was applied to the closed FRW cosmology and it was shown that it admits stable emerging universe solutions. The transition from the emerging phase to inflation and then to the phase with zero cosmological constant was studied. The spectrum of density perturbations and constraints imposed on the model parameters were also examined.

\subsection{Reproducing Einstein's General Relativity without the fifth force problem
\\
in a scale-invariant TMT model containing the dilaton $\phi$ and dust.}
\label{dilaton and dust}

The study announced in the title was published in ref.\cite{GK5}.
The model is invariant under the same scale transformations (\ref{st}) as the model described in section \ref{K-Essence and inflation in the scale invariant}.
The primordial action of the model differs from the primordial action (\ref{totaction k-ess}) only by adding the scale invariant action for dust (as a model for matter)
\begin{equation}
S_{m}=\int d^4x(\Upsilon +b_{m}\sqrt{-g}) \Biggl[-m\sum_{i}\int e^{\frac{1}{2}\alpha\phi/M_{p}}
\sqrt{g_{\alpha\beta}\frac{dx_i^{\alpha}}{d\lambda}\frac{dx_i^{\beta}}{d\lambda}}\,
\frac{\delta^4(x-x_i(\lambda))}{\sqrt{-g}}d\lambda\Biggr],
\label{S dust}
\end{equation}
where $\lambda$ is an arbitrary parameter. For simplicity we
consider the collection of the particles with the same mass
parameter $m$. We assume in addition that $x_i(\lambda)$ do not
participate in the scale transformations (\ref{st}).

In this model we restrict ourselves to a zero
temperature gas of particles, i.e. we will assume that
$d\vec{x}_i/d\lambda \equiv 0$  for all particles. It is
convenient to proceed in the frame where $g_{0l}=0$, \, $l=1,2,3$.
Then the particle density is defined by
\begin{equation}
n(\vec{x})=\sum_{i}\frac{1}{\sqrt{-g_{(3)}}}\delta^{(3)}(\vec{x}-\vec{x}_i(\lambda))
\label{n(x)}
\end{equation}
where $g_{(3)}=\det(g_{kl})$ and
\begin{equation}
S_{m}=-m\int d^{4}x(\Phi
+b_{m}\sqrt{-g})\,n(\vec{x})\,e^{\frac{1}{2}\alpha\phi/M_{p}}
\label{S-n(x)}
\end{equation}
  We will  assume that  the dimensionless real parameters
$b_g$, $b_{\phi}$ and $b_m$ are
positive, have the same or very close orders of magnitude
\begin{equation}
 b_g\sim b_{\phi}\sim b_m \quad \text{and besides} \quad  b_m>b_g.
\label{sim-bg-bm-bphi}
\end{equation}

In the absence of matter (dust), the model coincides with the model of section \ref{K-Essence and inflation in the scale invariant} and, hence, all results concerning dark energy coincide with the results discussed in section \ref{K-Essence and inflation in the scale invariant}. Therefore,  we will focus on the effects caused by dust. The first one concerns the important effect observed  when
varying $S_m$ with respect to $g^{\mu\nu}$:
\begin{equation}
\frac{\delta S_{m}}{\delta g^{00}}=\frac{b_m}{2}\,\sqrt{-g} \, m
\, n(\vec{x}) \, e^{\frac{1}{2}\alpha\phi/M_{p}}\, g_{00},
\label{00var}
\end{equation}
\begin{equation}
\frac{\delta S_{m}}{\delta g^{kl}}=-\frac{1}{2}\,\Upsilon\, m \,
n(\vec{x}) \, e^{\frac{1}{2}\alpha\phi/M_{p}}\, g_{kl}.
\label{klvar}
\end{equation}
The latter equation shows that due to the measure density $\Upsilon$, {\em the
zero temperature gas generically possesses a pressure}. As we will
see this pressure  disappears automatically together with the
fifth force as the matter energy density is many orders of
magnitude larger then the dark energy density, which is evidently
true in all physical phenomena tested experimentally.

 The transformation to the Einstein frame 
(\ref{ct}) causes the transformation of the particle density
\begin{equation}
\tilde{n}(\vec{x})=(\zeta +b_g)^{-3/2}\,
e^{-\frac{3}{2}\alpha\phi/M_{p}}\, n(\vec{x}).
 \label{ntilde}
\end{equation}
Einstein equations  (\ref{Grav.eq toy}) appear now with 
the TMT-effective energy-momentum tensor with the following components
\begin{equation}
T_{00}^{eff}=\frac{\zeta +b_{\phi}}{\zeta +b_{g}}
\dot{\phi}^2-\tilde{g}_{00}X
+\tilde{g}_{00}\Biggl[U_{eff}(\phi;\zeta,{\mathcal M}) +\frac{3\zeta +b_m +2b_g}{2\sqrt{\zeta
+b_{g}}}\, m\, \tilde{n}\Biggr],
 \label{T00 dust}
\end{equation}
\begin{equation}
T_{jk}^{eff}=\frac{\zeta +b_{\phi}}{\zeta +b_{g}}
\phi_{,j}\phi_{,k}-\tilde{g}_{jk}X
+\tilde{g}_{jk}\Biggl[U_{eff}(\phi;\zeta,{\mathcal M}) +\frac{\zeta -b_m +2b_g}{2\sqrt{\zeta
+b_{g}}}\, m\, \tilde{n}\Biggr],
 \label{Tjk dust}
\end{equation}
where the function $U_{eff}(\phi;\zeta,{\mathcal M})$ is given by the formula (\ref{Ueff zeta k-ess}) and notations (\ref{delta}) are used.
Function $\zeta(\phi,X,\tilde{n})$ is defined now by the following constraint:
\begin{equation}
\frac{(b_{g}-\zeta)\left({\mathcal M}e^{-2\alpha\phi/M_{p}}+V_{1}\right)-2V_{2}}{(\zeta
+b_g)^2} -\frac{\delta\cdot b_{g}X}{\zeta +b_g}=
\frac{\zeta -b_m +2b_g}{2\sqrt{\zeta +b_{g}}}\, m\, \tilde{n}
\label{constraint dust}
\end{equation}
The dilaton $\phi$ field equation in the Einstein frame is as
follows
\begin{eqnarray}
\frac{1}{\sqrt{-\tilde{g}}}\partial_{\mu}\left[\frac{\zeta
+b_{\phi}}{\zeta
+b_{g}}\sqrt{-\tilde{g}}\tilde{g}^{\mu\nu}\partial_{\nu}\phi\right]&-&\frac{\alpha}{M_{p}}\,\frac{(\zeta
+b_{g}){\mathcal M}e^{-2\alpha\phi/M_{p}}-(\zeta -b_{g})V_{1}
-2V_{2}-\delta b_{g}(\zeta
+b_{g})X}{(\zeta +b_{g})^{2}}\nonumber\\
 &&=\frac{\alpha}{M_{p}}\,\frac{\zeta
-b_m +2b_g}{2\sqrt{\zeta +b_{g}}}\, m\,\tilde{n}
 \label{phief}
\end{eqnarray}

One should now pay attention to the interesting result that  the
explicit $\tilde{n}$ dependence involving {\bf the  same form of
$\zeta$ dependence}
\begin{equation}
 \frac{\zeta -b_m +2b_g}{2\sqrt{\zeta +b_{g}}}\, m\,
\tilde{n} \label{universality}
\end{equation}
 appears simultaneously in the dust contribution to the pressure (through
the last term in Eq. (\ref{Tjk dust})),  in the
r.h.s. of the constraint (\ref{constraint dust}). and in the  effective dilaton to
dust coupling (see the r.h.s. of Eq. (\ref{phief})).
Let us analyze consequences of this wonderful coincidence in the
case when the matter energy density (modeled by
 dust) is much larger than the dilaton contribution to the dark
 energy density in the space region occupied by this matter. Evidently
this is the condition under which all tests of Einstein's GR,
including the question of the fifth force, are fulfilled.
Therefore, if this condition is satisfied we will say that {\bf the
matter is in  normal conditions}. In conventional (non-TMT) models, 
the existence of the fifth force becomes a problem precisely under normal conditions. The opposite
situation may be realized  if the
matter is diluted up to a magnitude of the macroscopic energy
density comparable with the dilaton contribution to the dark
energy density. In this case we
 say that the matter is in
 the  state of cosmo-low energy physics ({\bf CLEP}). It is
 evident that the fifth force acting on the matter in the CLEP
 state cannot be detected now and in the near future, and therefore
 does not appear to be a problem. But effects of the CLEP may be
 important in cosmology, see ref. \cite{GKneutr DE}  and the next subsection.

  The last terms in eqs. (\ref{T00 dust}) and (\ref{Tjk dust}),
 being the matter contributions to the energy density ($\rho_m$) and the
 pressure ($-p_m$) respectively, generally speaking have the same
 order of magnitude. But if the dust is in the normal conditions
 there is a possibility to provide the desirable feature of the dust in GR: it
 must be pressureless. A detailed analysis of the equations of motion together
 with the constraint (\ref{constraint dust}) in Appendix A of ref. \cite{GK5} showed
 that dust under normal conditions has no pressure, provided that under normal conditions (n.c.)
 the following equality is satisfied with extremely high accuracy:
\begin{equation}
 \zeta^{(n.c.)}\approx b_m-2b_g
\label{decoupling-cond}
\end{equation}
Remind that we have assumed  $b_m >b_g$. Then $\zeta^{(n.c.)}+b_g
>0$, and the transformation (\ref{ct}) and the subsequent
equations in the Einstein frame are well defined.
 Inserting
(\ref{decoupling-cond}) in the last term of eq.(\ref{T00 dust}) we
obtain the TMT-effective dust energy density in normal conditions
\begin{equation}
 \rho_m^{(n.c.)}=2\sqrt{b_m-b_g} \, m\tilde{n}
\label{rho-m-n.c.}
\end{equation}
Substitution of (\ref{decoupling-cond}) into the rest of the terms
of the components of the energy-momentum tensor (\ref{T00 dust}) and
(\ref{Tjk dust}) gives the dilaton contribution to the energy density
and pressure of the dark energy which have the orders of magnitude
close to those in the absence of matter case, Eqs. (\ref{rho1})
and (\ref{p1}). The latter statement may be easily checked by
using eqs.
(\ref{constraint in k-ess}), (\ref{decoupling-cond}) and
(\ref{sim-bg-bm-bphi}).

Detailed analysis shows that the constraint (\ref{constraint dust}) describes a balance between the
pressure of the dust in normal conditions on the one hand and the
vacuum energy density on the other hand. This balance is realized
due to the condition (\ref{decoupling-cond}). The most obvious way to verify that there 
is no fifth force problem under normal matter conditions is to consider the $\phi$-equation (\ref{phief})
 and estimate the Yukawa type coupling constant in
the r.h.s. of this equation. In fact, using the constraint
(\ref{constraint dust}) and representing the particle density in the
form $\tilde{n}\approx N/\upsilon$ where $N$ is the number of
particles in a volume $\upsilon$, one can make the following
estimation for the effective dilaton to matter coupling "constant"
$f$ defined by the Yukawa type interaction term $f\tilde{n}\phi$
(if we were to invent an effective action whose variation with
respect to $\phi$ would result in Eq. (\ref{phief})):
\begin{equation}
f \equiv\alpha\frac{m}{M_{p}}\,\frac{\zeta -b_m
+2b_g}{2\sqrt{\zeta +b_{g}}}\approx
\alpha\frac{m}{M_{p}}\,\frac{\zeta -b_m
+2b_g}{2\sqrt{b_m-b_{g}}}\sim
\frac{\alpha}{M_{p}}\,\frac{\rho_{vac}}{\tilde{n}} \approx
\alpha\frac{\rho_{vac}\upsilon}{NM_{p}} \label{Archimed}
\end{equation}
 Thus we conclude that {\it the
effective coupling "constant" of the dilaton to cold matter (dust) in the
normal conditions is of the order of the ratio of  the "mass of
the vacuum" in the volume occupied by the matter to   the Planck
mass taken $N$ times}. In some sense this result resembles the
{\it Archimedes law}. At the same time Eq. (\ref{Archimed}) gives
us an estimation of the exactness of the condition
(\ref{decoupling-cond}).

\subsection{Scale invariant TMT model including  fermions. Neutrino dark energy.}
\label{scale with ferm}

Considering  the model of ref.\cite{Kess4} simplified by fitting $b_g=b_{\phi}=b$ and adding fermions, in paper \cite{GKneutr DE} the scale 
invariant model including  fermions was studied. In this section we present the main results of \cite{GKneutr DE} with some additional analysis. Omitting  the gauge field contribution, the corresponding primordial action is $S=\int\mathcal{L}d^{4}x$ 
with the Lagrangian density
\begin{eqnarray}
&\mathcal{L}&=e^{\alpha\phi /M_{p}}(\Phi +b\sqrt{-g})
\left(-\frac{1}{\kappa}R(\omega ,e) +
\frac{1}{2}g^{\mu\nu}\phi_{,\mu}\phi_{,\nu}\right)-e^{2\alpha\phi /M_{p}}[\Phi V_{1} +\sqrt{-g}V_{2}]
\nonumber\\
&&+ e^{\alpha\phi /M_{p}}(\Phi +k\sqrt{-g})
\frac{i}{2}\sum_{i}\overline{\Psi}_{i}
\left(\gamma^{a}e_{a}^{\mu}\overrightarrow{\nabla}_{\mu}-
\overleftarrow{\nabla}_{\mu}\gamma^{a}e_{a}^{\mu}\right)\Psi_{i}
\nonumber\\
     &&-e^{\frac{3}{2}\alpha\phi /M_{p}}
\left[(\Phi +h_{E}\sqrt{-g})\mu_{E}\overline{E}E +(\Phi
+h_{N}\sqrt{-g})\mu_{N}\overline{N}N\right],
\label{totaction wth ferm}
\end{eqnarray}
where $\Psi_{i}$ ($i=N,E$) is the general notation for the
primordial fermion fields $N$ and $E$ (say, the primordial neutrino and the primordial electron);
$\mu_{N}$ and $\mu_{E}$ are the mass parameters;
$\overrightarrow{\nabla}_{\mu}=\vec{\partial}+
\frac{1}{2}\omega_{\mu}^{cd}\sigma_{cd}$,
 $R(\omega ,e)
=e^{a\mu}e^{b\nu}R_{\mu\nu ab}(\omega)$ is the scalar curvature;
$e_{a}^{\mu}$ and $\omega_{\mu}^{ab}$ are the vierbein and
spin-connection; $g^{\mu\nu}=e^{\mu}_{a}e^{\nu}_{b}\eta^{ab}$ and
$R_{\mu\nu ab}(\omega)=\partial _{mu}\omega_{\nu
ab}+\omega^{c}_{\mu a}\omega_{\nu cb}-(\mu \leftrightarrow\nu)$.
 Constants $b, k, h_{N}, h_{E}$  are non specified
dimensionless real parameters and we will only assume that they
 have close orders of magnitude
\begin{equation}
b\sim k\sim h_{N}\sim h_{E}.
 \label{sim-parameters}
\end{equation}
The formulation of the primordial action(\ref{totaction wth ferm}) in the Palatini formalism means that
vierbein and spin-connection are considered as independent degrees of freedom.

The action (\ref{totaction wth ferm}) is invariant under the global scale
transformations:
\begin{eqnarray}
    &&e_{\mu}^{a}\rightarrow e^{\theta /2}e_{\mu}^{a}, \quad  g_{\mu\nu}\rightarrow e^{\theta }g_{\mu\nu}, \quad 
\omega^{\mu}_{ab}\rightarrow \omega^{\mu}_{ab}, \quad
\varphi_{a}\rightarrow \lambda_{ab}\varphi_{b}, \quad  \Upsilon\rightarrow e^{2\theta}\Upsilon,
 \nonumber
\\
 &&\phi\rightarrow
\phi-\frac{M_{p}}{\alpha}\theta ,\quad \Psi_{i}\rightarrow
e^{-\theta /4}\Psi_{i}, \quad \overline{\Psi}_{i}\rightarrow
e^{-\theta /4} \overline{\Psi}_{i}, 
\label{stferm}
\end{eqnarray}
where $\theta =const.$, $\lambda_{ab}=const.$ and
 $\det(\lambda_{ab})=e^{2\theta}$; scalar $\zeta\equiv\frac{\Upsilon}{\sqrt{-g}}$ is  invariant under these transformations.

Similar to the models of the previous subsections, variation of the primordial action (\ref{totaction wth ferm})
with respect to $\varphi_a$ subject to the condition $\Upsilon\neq 0$ leads to an
equation where the appearance of an arbitrary integration constant ${\mathcal M}$ also  breaks the scale invariance.
All other equations of motion
resulting from (\ref{totaction wth ferm}) in the Palatini formalism
contain terms proportional to $\partial_{\mu}\zeta$ that makes the
space-time non-Riemannian and equations of motion - non canonical.
However, with the new set of variables ($\phi$ 
remains unchanged)
\begin{eqnarray}
&&\tilde{e}_{a\mu}=e^{\frac{1}{2}\alpha\phi/M_{p}}(\zeta
+b)^{1/2}e_{a\mu}, \quad
\tilde{g}_{\mu\nu}=e^{\alpha\phi/M_{p}}(\zeta +b)g_{\mu\nu},
\nonumber\\
&&\Psi^{\prime}_{i}=e^{-\frac{1}{4}\alpha\phi/M_{p}} \frac{(\zeta
+k)^{1/2}}{(\zeta +b)^{3/4}}\Psi_{i} , \quad i=N,E 
\label{ctferm-1}
\end{eqnarray}
which we call the Einstein frame,
 the spin-connections become those of the
Einstein-Cartan space-time. The fermion equations in the Einstein frame take canonical
general-coordinate invariant form with $\zeta$-dependent mass
\begin{equation}
m_{i}(\zeta)= \frac{\mu_{i}(\zeta +h_{i})}{(\zeta +k)(\zeta
+b)^{1/2}}.
 \label{m Ein}
\end{equation}
Since $\tilde{e}_{a\mu}$,
$\tilde{g}_{\mu\nu}$, $N^{\prime}$ and $E^{\prime}$ are invariant
under the scale transformations (\ref{stferm}), spontaneous
breaking of the scale symmetry  is
reduced in the new variables to the  spontaneous breaking of
the shift symmetry $\phi\rightarrow\phi +const.$

 After the change of
variables (\ref{ctferm-1}) to the Einstein frame and some simple
algebra, the gravitational equations  take the standard GR form
(\ref{grav eq Ein toy}) with the TMT-effective energy-momentum tensor
\begin{equation}
T_{\mu\nu}^{eff}=\phi_{,\mu}\phi_{,\nu}-\frac{1}{2}
\tilde{g}_{\mu\nu}\tilde{g}^{\alpha\beta}\phi_{,\alpha}\phi_{,\beta}
+\tilde{g}_{\mu\nu}U_{eff}(\phi;\zeta)+T_{\mu\nu}^{(ferm,can)}+T_{\mu\nu}^{(ferm,noncan)};
 \label{Tmn-1}
\end{equation}
\begin{equation}
U_{eff}(\phi;\zeta)=
\frac{b\left(\mathcal{M}e^{-2\alpha\phi/M_{p}}+V_{1}\right)
-V_{2}}{(\zeta +b)^{2}}; 
\label{Veff1-1}
\end{equation}
 $T_{\mu\nu}^{(ferm,can)}$ is the canonical
energy-momentum tensor for fermions   in curved space-time \cite{Birrel Devies}. 
$T_{\mu\nu}^{(ferm,noncan)}$ is the
{\em noncanonical} contribution of the fermions into the TMT-effective energy
momentum tensor
\begin{equation}
 T_{\mu\nu}^{(ferm,noncan)}=-\tilde{g}_{\mu\nu}\Lambda_{dyn}^{(ferm)}
 \label{Tmn-noncan-1}
\end{equation}
where
\begin{equation}
\Lambda_{dyn}^{(ferm)}\equiv Z_{N}(\zeta)m_{N}(\zeta)
\overline{N^{\prime}}N^{\prime}+
Z_{E}(\zeta)m_{E}(\zeta)\overline{E^{\prime}}E^{\prime},
\label{Lambda-ferm-1}
\end{equation}
$m_{i}(\zeta)$ are given by the formula (\ref{m Ein}) and
\begin{equation}
Z_{i}(\zeta)\equiv \frac{(\zeta -\zeta^{(i)}_{1})(\zeta
-\zeta^{(i)}_{2})}{2(\zeta +k)(\zeta +h_{i})}, \qquad i=N^{\prime},E^{\prime},
 \label{Zeta-1}
\end{equation}
\begin{equation}
\zeta_{1,2}^{(i)}=\frac{1}{2}\left[k-3h_{i}\pm\sqrt{(k-3h_{i})^{2}+
8b(k-h_{i}) -4kh_{i}}\,\right].
 \label{zeta12-1}
\end{equation}
The noncanonical contribution $T_{\mu\nu}^{(ferm,noncan)}$ of the
fermions into the energy momentum tensor has the transformation
properties of a cosmological constant term but it is proportional
to fermion densities
$\overline{\Psi}^{\prime}_{i}\Psi^{\prime}_{i}$ \,
($i=N^{\prime},E^{\prime}$). This is why we will refer to it as
"dynamical fermionic $\Lambda$ term". This fact is displayed
explicitly in Eqs.(\ref{Tmn-noncan}),(\ref{Lambda-ferm}) by
defining $\Lambda_{dyn}^{(ferm)}$.

The "dilaton" $\phi$ field equation  in the Einstein frame reads
\begin{equation}
\Box\phi -\frac{\alpha}{M_{p}(\zeta +b)}
\left[\mathcal{M}e^{-2\alpha\phi/M_{p}}-\frac{(\zeta -b)V_{1}
+2V_{2}}{\zeta +b}\right]= -\frac{\alpha
}{M_{p}}\Lambda_{dyn}^{(ferm)},
 \label{phief+ferm1}
\end{equation}
where $\Box\phi =(-\tilde{g})^{-1/2}\partial_{\mu}
(\sqrt{-\tilde{g}}\tilde{g}^{\mu\nu}\partial_{\nu}\phi)$.

The  scalar  $\zeta$ 
is determined by the following constraint that has the same origin as in models of sections \ref{Two-Measure Theory in empty space},
\ref{toy model}, \ref{quint with additional symm}, \ref{Guend mod}-\ref{dilaton and dust}:
\begin{equation}
\frac{1}{(\zeta
+b)^{2}}\Bigl[(b-\zeta)\left(\mathcal{M}e^{-2\alpha\phi/M_{p}}+
V_{1}\right)-2V_{2})\Bigr]=
\Lambda_{dyn}^{(ferm)}
 \label{constraint ferm}
\end{equation}
The novelty is that now $\zeta$ depends not only on the scalar field $\phi$, but also on
$\overline{\Psi}^{\prime}_{i}\Psi^{\prime}_{i}$ \,
($i=N^{\prime},E^{\prime}$).

Similar to section \ref{dilaton and dust}, we notice an unexpected and very important fact, namely, that the same function $\Lambda_{dyn}^{(ferm)}$,
Eq.(\ref{Lambda-ferm-1}), occurs
in three different places:

(i) in the form of the noncanonical fermion contribution $T_{\mu\nu}^{(ferm,noncan)}$
 to the TMT-effective energy-momentum tensor, eq. (\ref{Tmn-noncan});

(ii)  in the
 TMT-effective Yukawa coupling of the dilaton $\phi$ to fermions
 (see the right hand side of eq.(\ref{phief+ferm1}));

(iii)  as the right hand side of the constraint (\ref{constraint ferm}).

Applying the constraint (\ref{constraint ferm}) to
Eq.(\ref{phief+ferm1}) one can reduce the latter to the form
\begin{equation}
\Box\phi-\frac{2\alpha\zeta}{(\zeta
+b)^{2}M_{p}}\mathcal{M}e^{-2\alpha\phi/M_{p}} =0, \label{phi-constr}
\end{equation}
where $\zeta$  is a solution of the constraint (\ref{constraint ferm}).
 This result is true both in the presence of
fermions and in their absence.

In general, the fermions in the model under consideration are very different
 from those we are accustomed to in conventional field theory.
For example, the dependence of the fermion mass on $\zeta$, formula (\ref{m Ein}), 
together with constraint (\ref{constraint ferm}) and definition (\ref{Lambda-ferm-1}), 
mean that in general the mass of a non-relativistic fermion depends on its
 density (see Appendix \ref{Nonrelativistic fermions}).
Only if the local
energy density of the fermion is much larger than the vacuum
energy density, the fermion can have a constant mass. However this
is exactly the case of atomic, nuclear and particle physics,
including accelerator physics and high density objects of
astrophysics. This is why to such "high density" (in comparison
with the vacuum energy density) phenomena we will refer as "{\it
normal particle physics conditions}" and the appropriate fermion
states in TMT  we will call "{\it regular fermions}". For generic
fermion states in TMT we will use the term "{\it primordial
fermions}" in order to distinguish them from regular fermions.

Before moving on to the study of the properties of fermions in the TMT model under consideration, it is worth noting that the results described below do not require the introduction of a specially selected interaction of fermions with a scalar field into the primordial action, except for the imposition of the scale invariance condition. This observation is particularly important when comparing the model under study with models known in the literature as the variable-mass neutrino models \cite{Massive n 1,Massive n 2,Massive n 3,Massive n 4,Massive n 5,Massive n 6}.

\vspace{0.4cm}
{\bf 1. Dark energy in the absence of massive fermions}

 In the limiting case where massive fermions are absent, the model reduces to a special case (where $b_g=b_{\phi}$) of the model presented in section \ref{K-Essence and inflation in the scale invariant}: the field $\phi$ equation  reduces to 
\begin{equation}
\Box\phi +U^{(0)\prime}_{eff}(\phi)=0
\label{eq-phief-without-ferm}
\end{equation}
 with the TMT-effective potential
\begin{equation}
U_{eff}^{(0)}(\phi)\equiv
U_{eff}(\phi;\zeta_{0})|_{\overline{\psi^{\prime}}\psi^{\prime}=0}
=\frac{[V_{1}+\mathcal{M}e^{-2\alpha\phi/M_{p}}]^{2}}
{4[b\left(V_{1}+\mathcal{M}e^{-2\alpha\phi/M_{p}}\right)-V_{2}]}
\label{Veffvac}
\end{equation}
Applying this to the cosmology of the late FLRW Universe  and assuming that the scalar field
$\phi\rightarrow\infty$ as $t\rightarrow\infty$, we find that the cosmological evolution  is driven by the dark energy density
\begin{equation}
\rho^{(0)}_{d.e}=\frac{1}{2}\dot{\phi}^{2}+\Lambda^{(0)}+U^{(0)}_{q-like}(\phi).
\label{rho-without-ferm}
\end{equation}
where $\Lambda^{(0)}$ is the cosmological constant
\begin{equation}
\Lambda^{(0)} =\frac{V_{1}^{2}} {4[bV_{1}-V_{2}]}
\label{lambda-without-ferm}
\end{equation}
and $U^{(0)}_{q-like}(\phi)$ is the quintessence-like scalar field
 potential
\begin{equation}
U^{(0)}_{q-like}(\phi) =\frac{V_{1}(bV_{1}-2V_{2})+
(bV_{1}-V_{2})\mathcal{M}e^{-2\alpha\phi/M_{p}}}
{4[bV_{1}-V_{2}][b(V_{1}+
\mathcal{M}e^{-2\alpha\phi/M_{p}})-V_{2}]}\cdot \,
\mathcal{M}e^{-2\alpha\phi/M_{p}}. 
\label{V-quint-without-ferm}
\end{equation}
In what follows we will assume
\begin{equation}
\mathcal{M}>0, \quad V_{1}>0 \quad V_2>0 \quad  \quad b>0
\label{param-vac}
\end{equation}
and
\begin{equation}
 bV_{1}> V_2 \quad \text{which means that} \quad \Lambda^{(0)}>0.
\label{bV1>V2}
\end{equation}
In the absence of massive fermions, the constraint (\ref{constraint ferm}) is reduced to
\begin{equation}
\zeta=\zeta^{(0)}(\phi)=\frac{b\left(V_1+\mathcal{M}e^{-2\alpha\phi/M_{p}}\right)-2V_2}
{V_1+\mathcal{M}e^{-2\alpha\phi/M_{p}}}.
 \label{constraint absence ferm}
\end{equation}
 It follows from (\ref{constraint absence ferm}) that
in the late universe,  that is when $\mathcal{M}e^{-2\alpha\phi/M_{p}}\ll V_1$,
\newline
(i)  $\zeta^{(0)}>0$ \, if \, $bV_{1}>2V_{2}$,
\newline
(ii) $\zeta^{(0)}<0$ \, if \, $bV_{1}<2V_{2}$.

In section \ref{toy model}, the pregeometry effect was formulated as a key conclusion of TMT: "only those cosmological solutions   are valid for which 
$\zeta(x)$ does not change the sign throughout the evolution of the universe". Thus, results (i) and (ii) for the sign of  $\zeta$ are valid 
throughout the evolution of the universe.
In addition, depending on the relation between $bV_1$ and $V_2$, two different scenarios for the late  Universe
 in the absence of massive fermions are possible:
\newline
(i) $U_{eff}^{(0)}(\phi)$ asymptotically monotonically approaches $\Lambda^{(0)}>0$ from above if $bV_{1}>2V_{2}$ (that is, this is the cosmological scenario with $\zeta>0$),
\newline
(ii) $U_{eff}^{(0)}(\phi)$ asymptotically monotonically approaches $\Lambda^{(0)}>0$ from below if $bV_{1}<2V_{2}$ (that is, this is the cosmological scenario with $\zeta<0$).
Case (ii) is analogous to the late universe phantom dark energy scenario discussed  in point b) of section \ref{K-Essence and inflation in the scale invariant}.

\vspace{0.4cm}
{\bf 2. Reproducing Einstein equations with regular fermions
and resolution of the fifth force problem.}

When analyzing the field theory model (\ref{totaction wth ferm}) in more general cases, 
we would first like to make sure that it is able to correctly reproduce the model of 
General Relativity with Einstein's equations, where the scalar field $\phi$ and massive fermions are the sources of gravity.
It is easy to see that eqs.(\ref{grav eq Ein toy}) and
(\ref{Tmn-1}) reduce to the Einstein equations in this
field theory model  if $\zeta$ is constant and
\begin{equation}
\Lambda_{dyn}^{ferm}=0 \qquad \text{or at least} \qquad
|T_{\mu\nu}^{(ferm,noncan)}|\ll |T_{\mu\nu}^{(ferm,can)}|.
\label{noncan-ll-can}
\end{equation}
According to eqs.(\ref{Tmn-noncan-1})-(\ref{zeta12-1}), in the case
when a single massive fermion is a source of gravity, the
condition (\ref{noncan-ll-can}) is realized  if
\begin{equation}
Z_{i}(\zeta)\approx 0 \quad \Longrightarrow \quad
\zeta=\zeta_{1}^{i} \quad \text{or} \quad \zeta=\zeta_{2}^{i}, \quad
i=N^{\prime},E^{\prime}, \label{Z-0}
\end{equation}
where $\zeta_{1,2}^{i}$ are defined in Eqs.(\ref{zeta12-1}).
The only component of the non-relativistic fermion energy-momentum tensor $T_{\mu\nu}^{(ferm,can)}+T_{\mu\nu}^{(ferm,noncan)}$
 which survive in this approximation is the energy density $T_{00}^{(f,non-rel)}=\rho_f^{(can)}=m(\zeta) n$ (see Appendix \ref{Nonrelativistic fermions}). Thus, the form of Einstein's equations describing gravity created by a gas of nonrelativistic fermions with density n coincides with the canonical one.

Reproducing Einstein equations in the studied TMT model when the primordial fermions are in
the states of the regular fermions is not enough in order to
assert that GR is reproduced. The reason is that at the late
universe, as $\phi\gg M_{p}$, the TMT-effective
potential of scalar field $\phi$ is very flat and therefore due to the Yukawa-type
coupling of massive fermions to $\phi$, (the r.h.s. of
Eq.(\ref{phief+ferm1})),  the long range scalar force appears to
be possible in general. The Yukawa coupling "constant" is
$\alpha\frac{m_{i}(\zeta)}{M_{p}}Z_{i}(\zeta)$. A detailed analysis 
of the meaning of the constraint (\ref{constraint ferm}), carried out in paper \cite{GKneutr DE}, shows that 
for regular fermions with $\zeta^{(i)} =\zeta_{1,2}^{(i)}$ 
the factor $Z_{i}(\zeta)$ is of the order of the ratio of the vacuum energy density 
to the energy density of a regular fermion.  Thus we conclude
that {\em the 5-th force is extremely suppressed for the fermionic
matter observable in classical tests of GR. It is very important
that this result is obtained automatically, without tuning of the
parameters.}

\vspace{0.4cm}
{\bf 3. Nonrelativistic neutrinos and dark energy.} 

It turns out that besides the normal particle physics situations, TMT
predicts possibility of so far unknown
 states which can be realized,  for example,
in astrophysics and cosmology. Roughly speaking such exotic states
may be created if the degree of  localization of the fermion  is
very small.

\vspace{0.3cm}

\underline{ 1. Cosmo-Particle Phenomena in a toy model}

 Let us start from a simplest (but idealized) model
describing the following self-consistent system: the spatially
flat FRW universe filled with the homogeneous scalar field $\phi$
and a homogeneous primordial neutrino field $N^{\prime}(t)$. The
non-canonical contribution of the primordial neutrino $N^{\prime}$
into the TMT-effective energy-momentum tensor reads
\begin{equation}
 T_{\mu\nu}^{(N,noncan)}=-\tilde{g}_{\mu\nu}\Lambda_{dyn}^{(N)}=-\tilde{g}_{\mu\nu}Z_{N}(\zeta)m_{N}(\zeta)
\overline{N}^{\prime}{N}^{\prime}
 \label{Tmn-N-noncan}
\end{equation}
where $Z_{N}(\zeta)$ and $m_{N}(\zeta)$ are defined by eqs.(\ref{Zeta-1}) and (\ref{m Ein}).
The neutrino has zero momenta and therefore one can rewrite
$\overline{N}^{\prime}N^{\prime}$ in the form of density
$\overline{N}^{\prime}N^{\prime}= u^{\dagger}u$ where $u$ is the
large component of the Dirac spinor $N^{\prime}$. The space
components of the 4-current
$\tilde{e}_{a}^{\mu}\overline{N}^{\prime}\gamma^{a}N^{\prime}$
equal zero. It follows from the 4-current conservation  that
$\overline{N}^{\prime}N^{\prime}= u^{\dagger}u
=\frac{const}{a^{3}}$ where $a=a(t)$ is the scale factor.
 Then the constraint (\ref{constraint ferm}) reduces the following form:
\begin{equation}
\frac{(b-\zeta )\left[\mathcal{M}e^{-2\alpha\phi/M_{p}}+
V_{1}\right]-2V_{2}}{(\zeta +b)^{3/2}}=\frac{(\zeta
-\zeta^{(N)}_{1})(\zeta -\zeta^{(N)}_{2})}{(\zeta +k)^{2}}\mu_{N}
\frac{const}{a^{3}}.
 \label{constraint-toy}
\end{equation}
In addition to the solution where the contribution of massive fermions is negligible, there is another solution
 where the decaying fermion
contribution $u^{\dagger}u\sim \frac{const}{a^{3}}$ to the
constraint is compensated by the appropriate behavior of the
scalar  $\zeta$. Namely if expansion of the universe is
accompanied by approaching $\zeta\rightarrow -k$ in such a way
that $(\zeta +k)^{-2} \propto a^{3}$, then the r.h.s. of
(\ref{constraint-toy}) approaches a constant. Note that the l.h.s.
of (\ref{constraint-toy}) also approaches a constant if
$\phi\rightarrow\infty$ as $a(t)\rightarrow\infty$ (recall we
assume $\alpha >0$). The described regime corresponds to a very
unexpected state of the primordial neutrino with the following
exotic features:

(i) This state is different from the regular
neutrino states since  $-k\neq\zeta_{1,2}$ unless a fine tuning is
made.

(ii)  The effective mass of the neutrino in this state increases
like $(\zeta +k)^{-1}\propto a^{3/2}$ and therefore the canonical neutrino density $\rho_{(N,canon)}$ decreases as
$\rho_{(N,canon)}=T_{00}^{(N,canon)}=
m_{N}(\zeta)u^{\dagger}u\propto a^{-3/2}$.
 At the same time the dynamical neutrino $\Lambda_{dyn}^{(N)}$ term
approaches a constant: $\Lambda_{dyn}^{(N)}\propto(\zeta
+k)^{-2}u^{\dagger}u\rightarrow constant$. This means that at the
late time universe, the canonical neutrino energy density
$\rho_{(N,canon)}$
 becomes negligible in comparison with the non-canonical neutrino
 energy density $\rho_{(N,noncanon)}=T_{00}^{(N,noncanon)}=
-\Lambda_{dyn}^{(N)}$.

(iii) It follows immediately from eq.(\ref{Tmn-noncan-1}) and item (ii)
that such cold neutrino matter possesses a pressure $p_{N}$ and
its equation of state  in the late time universe approaches the
form $p_{N}=-\rho_{N}=-\rho_{(N,noncanon)}=\Lambda_{dyn}^{(N)}=const$ typical for a
cosmological constant. Therefore {\it the primordial neutrino in
the described regime behaves as a sort of dark energy}.

(iv) In the regime $\zeta\rightarrow -k$, the scalar field $\phi$
effective potential $U_{eff}(\phi;\zeta)$, eq.(\ref{Veff1-1}), and
the l.h.s. of the constraint (\ref{constraint-toy}) have the same
order of magnitude while the r.h.s. of the constraint is
$\Lambda_{dyn}^{(N)}$. Therefore in this toy model, contributions
of the scalar field $\phi$ and the primordial neutrino into the
dark energy density are of the same order of magnitude if the
 regime $\zeta\rightarrow -k$ is realized.

   Thus TMT
predicts a possibility of new type of states which we  refer
as {\it Cosmo-Low-Energy-Physics  (CLEP) }states.

\vspace{0.3cm}
\underline{ 2. Gas of uniformly distributed non-relativistic neutrinos in the CLEP state.}

Consider a model where the spatially flat
FRW universe filled with the homogeneous scalar field $\phi$ and
 a cold gas of uniformly distributed stable non-relativistic
neutrinos. After averaging over spins of
the neutrinos, the cosmological averaging of the microscopic
non-canonical contribution to the energy-momentum tensor
$T_{\mu\nu}^{(N,can)}$ results in
\begin{equation}
<T_{\mu\nu}^{(N,can)}>_{cosm.av.}=\delta_{\mu}^{0}\delta_{\nu}^{0}
\frac{h_{N}-k}{(b-k)^{{1/2}}}\mu_{N}\frac{n}{(\zeta +k)a^{3}}+
{\cal O}\left(\frac{1}{a^{3}}\right), 
\label{Tmn-can-aver}
\end{equation}
where $n$ is a constant determined by the total number of the cold
neutrinos entering the CLEP regime, i.e. in the regime with $\zeta\approx
-k$. Note that the formula (\ref{Tmn-can-aver}), as well as $m_N|_{(\zeta\to -k)}$ (see eq.(\ref{m Ein})), are defined if
\begin{equation}
b>k
 \label{b>k}
\end{equation}
Similarly, averaging of the $\Lambda_{dyn}^{(N)}$ term gives
\begin{equation}
<\Lambda_{dyn}^{(N)}>_{cosm.av.}=(b-k)^{1/2}(h_{N}-k)\mu_{N}\frac{n}{(\zeta
+k)^{2}a^{3}}+ {\cal O}\left(\frac{1}{(\zeta +k)a^{3}}\right).
 \label{L-N-aver}
\end{equation}
Hence the appropriate averaged expression of the microscopic
non-canonical contribution to the energy-momentum tensor
$T_{\mu\nu}^{(N,noncan)}$, Eq.(\ref{Tmn-N-noncan}), is then
\begin{equation}
<T_{\mu\nu}^{(N,noncan)}>_{cosm.av.}=
-\tilde{g}_{\mu\nu}<\Lambda_{dyn}^{(N)}>_{cosm.av.}
\label{Tmn-noncan-aver}
\end{equation}

Taking into account the homogeneity of the scalar field $\phi$ and
Eq.(\ref{L-N-aver}), the result of the cosmological averaging of
the constraint (\ref{constraint-toy}) can be represented in the
form
\begin{equation}
\frac{(b+k)\left(\mathcal{M}e^{-2\alpha\phi/M_{p}}+V_{1}\right)
-2V_{2}}{(b-k)^{2}}=(b-k)^{1/2}(h_{N}-k)\cdot\mu_{N}\frac{n}{(\zeta +k)^{2}a^{3}}+{\cal
O}\left(\zeta +k\right)
\label{constr-k-av}
\end{equation}
As with the above toy model designated 1, the constraint
(\ref{constr-k-av}) allows a "CLEP solution" where the decay of
the neutrino density with the expansion of the universe is
accompanied by approaching $\zeta\rightarrow -k$ in such a way
that $(\zeta +k)^{-2} \propto a^{3}$, and then the r.h.s. of
(\ref{constr-k-av}) as well as $<\Lambda_{dyn}^{(N)}>_{cosm.av.}$
in  Eq.(\ref{Tmn-noncan-aver}) approach  constants. At the same
time, $<T_{\mu\nu}^{(N,can)}>_{cosm.av.}$,
Eq.(\ref{Tmn-can-aver}), approaches zero. Therefore in the CLEP
regime, the neutrino energy-momentum tensor
$<T_{\mu\nu}^{(N)}>_{cosm.av.}$ is reduced to the approaching
constant (as $a(t)\rightarrow\infty$) non-canonical part of the
neutrino energy-momentum tensor
\begin{equation}
<T_{\mu\nu}^{(N)}>_{cosm.av.}= -\tilde{g}_{\mu\nu}
(b-k)^{1/2}(h_{N}-k)\mu_{N}\frac{n}{(\zeta +k)^{2}a^{3}}+{\cal
O}\left(\zeta +k\right) \label{Tmn-noncan-aver-asympt}
\end{equation}
The presence of the
cold neutrino gas in the CLEP regime essentially changes
$U_{eff}(\phi;\zeta)$, Eq.(\ref{Veff1-1}). In fact, instead of
$U_{eff}^{(0)}(\phi)$ as $\zeta =\zeta_{0}$, eq.(\ref{Veffvac}),
we obtain in the case $\zeta\rightarrow -k$
\begin{equation}
U_{eff}(\phi;\zeta)|_{\zeta\rightarrow -k}=
\frac{b\left(\mathcal{M}e^{-2\alpha\phi/M_{p}}+V_{1}\right)
-V_{2}}{(b-k)^{2}}+{\cal O}\left(\zeta +k\right). \label{Veff3}
\end{equation}
Using the
constraint (\ref{constr-k-av})  one can represent the
neutrino contribution $<T_{\mu\nu}^{(N)}>_{cosm.av.}$  into
$<T_{\mu\nu}^{(tot)}>_{cosm.av.}$ {\it in terms of the scalar
field} $\phi$ {\it alone} and thus the total energy and pressure
in the CLEP state can be written in an equivalent form where they
are only $\phi$-dependent:
\begin{equation}
\rho_{tot}\equiv <T_{00}^{(tot)}>_{cosm.av.}=
\frac{1}{2}\dot{\phi}^{2}+U_{eff}^{(tot)}(\phi) \label{rho-tot}
\end{equation}
\begin{equation}
p_{tot}\equiv <T_{ii}^{(tot)}>_{cosm.av.}=
\frac{1}{2}\dot{\phi}^{2}-U_{eff}^{(tot)}(\phi). \label{p-tot}
\end{equation}
where the TMT-effective potential $U_{eff}^{(tot)}(\phi)$ is the sum
\begin{equation}
U_{eff}^{(tot)}(\phi)\equiv \Lambda +U_{q-like}(\phi),
\label{Ueff-phi}
\end{equation}
of the TMT-effective cosmological constant
\begin{equation}
\Lambda = \frac{V_{2}-kV_{1}}{(b-k)^{2}} \label{Lambda-nu}
\end{equation}
and the quintessence-like potential
\begin{equation}
U_{q-like}(\phi)=
-\frac{k}{(b-k)^{2}}\mathcal{M}e^{-2\alpha\phi/M_{p}}+{\cal
O}\left(\zeta +k\right). \label{pot-nu}
\end{equation}
$\Lambda$ is  positive if
$V_{2}>kV_{1}$ 
that we will assume in what follows. From inequalities (\ref{bV1>V2}) and (\ref{b>k}) it follows that $kV_{1}<V_2<bV_1$.

Since  the sign of $\zeta(x)$ does not change throughout the evolution of the universe, in accordance with the results obtained
 at the end of the paragraph A devoted to scenarios for the late  Universe in the absence of massive fermions  case, we conclude:
\newline
(i) if $bV_{1}>2V_{2}$ (that is, this is the cosmological scenario with $\zeta\approx -k>0$, and therefore $k<0$), then $U_{eff}^{(tot)}(\phi)$ asymptotically monotonically approaches $\Lambda>0$ from above; 
\newline
(ii) if $bV_{1}<2V_{2}$ (that is, this is the cosmological scenario with $\zeta\approx -k<0$, and therefore $k>0$), then $U_{eff}^{(tot)}(\phi)$ asymptotically monotonically approaches $\Lambda>0$ from below.
Again, the case (ii) is analogous to the late universe phantom dark energy scenario discussed at the end of section \ref{K-Essence and inflation in the scale invariant} in point b).

The {\it remarkable result} consists
in the fact that
\begin{equation}
U_{eff}^{(0)}(\phi)-U_{eff}^{(tot)}(\phi)\equiv
\frac{\left[(b+k)\left(V_{1}+\mathcal{M}e^{-2\alpha\phi /M_{p}}\right)
-2V_{2}\right]^{2}}
{4(b-k)^{2}\left[b\left(V_{1}+\mathcal{M}e^{-2\alpha\phi /M_{p}}\right)
-V_{2}\right]}>0. \label{L-L0}
\end{equation}
where $U_{eff}^{(0)}(\phi)$ is defined by the formula (\ref{Veffvac}) and the conditions (\ref{param-vac})  have
been used. This means that {\it the universe with the gas of
uniformly distributed non-relativistic neutrinos in the CLEP state
is energetically more preferable than the one in the absence of
fermions case}.

For illustration of what kind of solutions one can expect, let us
take the {\em particular value} for the parameter $\alpha$, namely
$\alpha =\sqrt{3/8}$. Then for the late time universe in the CLEP
regime, when $\zeta$ is close enough to $-k$,
 the cosmological equations allow the following analytic solution:
 \begin{equation}
\phi(t)=\frac{M_{p}}{2\alpha}\varphi_{0}+
\frac{M_{p}}{2\alpha}\ln(M_{p}t), \qquad a(t)\propto
t^{1/3}e^{\lambda t}, \label{a-sol-nu}
\end{equation}
where
\begin{equation}
\lambda =\frac{1}{M_{p}}\sqrt{\frac{\Lambda}{3}}, \qquad
e^{-\varphi_{0}}=
\frac{2(b-k)^{2}M_{p}^{2}}{\sqrt{3}|k|M^{4}}\sqrt{\Lambda}.
\label{phi-0}
\end{equation}
and $\Lambda$ is determined by Eq.(\ref{Lambda-nu}). The mass of
the neutrino in such CLEP state increases exponentially in time
and its $\phi$ dependence is double-exponential:
\begin{equation}
m_{N}|_{clep}\sim (\zeta +k)^{-1}\sim a^{3/2}(t)\sim
t^{1/2}e^{\frac{3}{2}\lambda t}\sim \exp\left[\frac{3\lambda
e^{-\varphi_{0}}}{2M_{p}} \exp\left(\frac{2\alpha}{M_{p}}\phi
\right)\right]. \label{m-t-phi}
\end{equation}
It is important to note that {\em the exponential growth of the effective neutrino mass in the CLEP regime is consistent with the model assumption of a gas of non-relativistic neutrinos}.

\section{Model of dark energy generated by neutrinos without a scalar field}
\label{Neutrino genereted dark energy}

It turns out that in TMT, unlike the variable-mass neutrino models \cite{Massive n 1,Massive n 2,Massive n 3,Massive n 4,Massive n 5,Massive n 6},  the effect of the neutrino dark energy can be realized even without a scalar field and therefore without the requirement of scale invariance. This possibility was considered in paper \cite{GK6}, where the TMT model describing a self-consistent system including four-dimensional gravity, a primordial cosmological constant (CC), and fermions was studied.
The primordial action has the form  $S =\int\mathcal{L}d^{4}x$ with  the following Lagrangian density:
\begin{eqnarray}
\mathcal{L}&=&-\frac{M_P^2}{2} (\Upsilon +b\sqrt{-g}) R(\omega ,e)
 - \sqrt{-g}V_0
\label{tot Lagr density neutr de}
\\
&& + (\Upsilon +k\sqrt{-g})
\frac{i}{2}\sum_{i}\overline{\Psi}_{i}
\left(\gamma^{a}e_{a}^{\mu}\overrightarrow{\nabla}^{(i)}_{\mu}-
\overleftarrow{\nabla}^{(i)}_{\mu}\gamma^{a}e_{a}^{\mu}\right)\Psi_{i}
     -(\Upsilon +h\sqrt{-g})\sum_{i}\mu_i\overline{\Psi}_i\Psi_i,
 \nonumber
\end{eqnarray}
where $V_0=const$, which would be a cosmological constant in conventional gravity models. 
Note that, as in the model of section \ref{quint with additional symm} and all models in  section \ref{Scale invar models}, there is no term (\ref{V2 term}) in the primordial action.
$\Psi_{i}$  are the fermion fields  of  species labeled by the index $i$;
$\mu_i$ are the primordial mass parameters; all other notations are essentially the same as in section \ref{scale with ferm}.

Variation of the primordial action 
with respect to $\varphi_a$ subject to the condition $\Upsilon\neq 0$ leads to an
equation $L_2=\mathcal{M}$  where  an arbitrary integration constant ${\mathcal M}$ appears.
Equations of motion for fermion fields as well as the equations for the spin-connection
 in the first order formalism
contain terms proportional to $\partial_{\mu}\zeta$ that makes the
space-time non-Riemannian and equations of motion - non canonical.
However, with the new set of variables
\begin{eqnarray}
&&\tilde{e}_{a\mu}=(\zeta
+b)^{1/2}e_{a\mu}, \quad
\tilde{g}_{\mu\nu}=(\zeta +b)g_{\mu\nu},
\nonumber\\
&&\Psi^{\prime}_{i}=\frac{(\zeta
+k)^{1/2}}{(\zeta +b)^{3/4}}\Psi_{i} ,
\label{ctferm}
\end{eqnarray}
which we call the Einstein frame,
 the spin-connections become those of the
Einstein-Cartan space-time with the metric $\tilde{g}_{\mu\nu}$
The fermion equations also take the standard form of the Dirac equation in
the Einstein-Cartan space-time where now the fermion masses become $\zeta$ dependent
\begin{equation}
m_{i}(\zeta)= \mu_{i}F(\zeta), \qquad \text{where} \qquad F(\zeta)=\frac{(\zeta +h)}{(\zeta +k)(\zeta
+b)^{1/2}} \label{m}
\end{equation}
From here on, we assume that
\begin{equation}
\zeta +b>0 \qquad \text{and} \qquad \zeta+k>0.
\label{zb-zk-positive}
\end{equation}

 It is easy to check that  the gravitational equations in the Einstein frame take the standard GR form
(\ref{grav eq Ein toy}) with the TMT-effective energy-momentum tensor
\begin{equation}
T_{\mu\nu}^{eff}=T_{\mu\nu}^{(f,can)}+T_{\mu\nu}^{(\Lambda)}
 \label{Tmn}
\end{equation}
where
$T_{\mu\nu}^{(f,can)}$ is the canonical
energy momentum tensor for fermions in curved space-time\cite{Birrel Devies}; $T_{\mu\nu}^{(\Lambda)}$
is the total variable CC term in the presence of fermions
\begin{equation}
T_{\mu\nu}^{(\Lambda)}=\tilde{g}_{\mu\nu}\Lambda_{tot}; \qquad \Lambda_{tot}=\Lambda_{eff}(\zeta)
-\Lambda_{dyn}^{(f)},
\label{Tmn_Lambda}
\end{equation}
where
\begin{equation}
\Lambda_{eff}(\zeta)=
\frac{b{\mathcal M}
-V_0}{(\zeta +b)^{2}},
\label{Veff1}
\end{equation}
and
\begin{equation}
\Lambda_{dyn}^{(f)}\equiv  Z(\zeta)F(\zeta)\sum_i \mu_{i}
\overline{\Psi^{\prime}}_i\Psi^{\prime}_i
=Z(\zeta)\sum_i m_{i}(\zeta)
\overline{\Psi^{\prime}}_i\Psi^{\prime}_i
\label{Lambda-ferm}
\end{equation}
\begin{equation}
Z(\zeta)\equiv \frac{(\zeta -\zeta_1)(\zeta
-\zeta_2)}{2(\zeta +k)(\zeta +h)}
\label{Zeta}
\end{equation}
\begin{equation}
\zeta_{1,2}=\frac{1}{2}\left[k-3h\pm\sqrt{(k-3h)^{2}+
8b(k-h) -4kh}\,\right].
 \label{zeta12}
\end{equation}
Taking into account our assumption (\ref{sim-parameters}) made in section (\ref{scale with ferm}), we see that
\begin{equation}
\zeta_1\sim\zeta_2\sim b\sim k\sim h.
 \label{sim-parameters_1}
\end{equation}
if no a special fine tuning of the parameters is assumed.

As in all TMT models, the constraint, describing the selfconsistency condition of equations of motion, 
has the form of an algebraic equation.
In the model under study, the constraint, rewritten in the Einstein frame, has the following form
\begin{equation}
\frac{(b-\zeta){\mathcal M}-2V_0}{(\zeta
+b)^2}=\Lambda_{dyn}^{(f)} =Z(\zeta)\sum_i m_{i}(\zeta)
\overline{\Psi^{\prime}}_i\Psi^{\prime}_i.
\label{constraint}
\end{equation}
To simplify the presentation of the cosmological results of the model, in what follows, instead of the sum over $i$, we will explicitly write down the contribution of only one fermion:
$m(\zeta)\overline{\Psi}^{\prime}\Psi^{\prime}$.

We are going to study the application of this field theory model to the case of {\it nonrelativistic fermions} in the cold fermions (dust) approximation. This means that {\it we neglect the effect of fermion 3-momenta}.  The only component of the canonical fermion energy-momentum tensor $T_{\mu\nu}^{(f,can)}$ which survive in this approximation is the energy density $T_{00}^{(f,can)}=\rho_f^{(can)}$. Making use of the Dirac equation in the same approximation we obtain (see also Eq.(\ref{m}))
\begin{equation}
\rho_f^{(can)}=
 m(\zeta)
\overline{\Psi^{\prime}}\Psi^{\prime}; \qquad  m(\zeta)=\frac{\mu(\zeta +h)}{(\zeta +k)(\zeta
+b)^{1/2}},
\label{rhofcan}
\end{equation}
where $\mu$ is the primordial fermion mass parameter. 
Based on the analysis in  Appendix \ref{Nonrelativistic fermions}, in the semiclassical approximation we can replace the field operator 
$\overline{\Psi}^{\prime}\Psi^{\prime}$ by the fermion number density $n$. Then
\begin{equation}
\rho_f^{(can)}=
 m(\zeta)n
\label{rhofcan n}
\end{equation}
 and the constraint (\ref{constraint}) takes the form
\begin{equation}
\frac{(b-\zeta){\mathcal M}-2V_0}{(\zeta
+b)^2}= Z(\zeta) m(\zeta)\cdot n,
\label{constraint via n}
\end{equation} 
\underline{The constraint (\ref{constraint via n}) determines the  scalar  $\zeta$ as a function of the fermion  number density $n$:}
 $\zeta=\zeta(n)$.

The TMT-effective CC $\Lambda_{eff}(\zeta)$, eq.(\ref{Veff1}). is generated by the integration constant ${\mathcal M}$ and the $V_0$-term in
 the Lagrangian density (\ref{tot Lagr density neutr de}). \textit{The first new, very unusual effect} that we observe here is {\bf the functional dependence of $\Lambda_{eff}(\zeta)$ on $n$}, arising via $\zeta(n)$: $\Lambda_{eff}=\Lambda_{eff}(\zeta(n))$.
$\Lambda_{eff}(\zeta)$  is positive provided
 \begin{equation}
b{\mathcal M}>V_0.
 \label{bsM_bigger_L0}
\end{equation}
\

\textit{The second novelty} consists in  generating the
{\bf fermion noncanonical  TMT-effective energy
momentum tensor}
\begin{equation}
 T_{\mu\nu}^{(f,noncan)}=-\tilde{g}_{\mu\nu}\Lambda_{dyn}^{(f)}=-\tilde{g}_{\mu\nu}Z(\zeta) m(\zeta)\cdot n,
 \label{Tmn-noncan}
\end{equation}
which has the transformation
properties of a CC term but it is proportional to $n$. This is why we refered to it in \cite{GK6} as
{\bf "dynamical fermionic $\Lambda$ term"}.
The appearance of $T_{\mu\nu}^{(f,noncan)}$ means that {\bf even cold fermions generically
possess pressure} $P_f^{(noncan)}$ and
\begin{equation}
P_f^{(noncan)}=-\rho_f^{(noncan)}=\Lambda_{dyn}^{(f)} =Z(\zeta) m(\zeta)\cdot n,
\label{Pfnoncan}
\end{equation}
where $\rho_f^{(noncan)}$ is the noncanonical contribution of fermions into the energy density.
It is intereting that $P_f^{(noncan)}$ coincides with  the r.h.s. of the constraint. 
   Using the constraint (\ref{constraint via n}) and eq.(\ref{Tmn_Lambda}), the total CC term $T_{\mu\nu}^{(\Lambda)}$ may be represented in the form
\begin{equation}
T_{\mu\nu}^{(\Lambda)}=\tilde{g}_{\mu\nu}\Lambda_{tot}(\zeta), \qquad \Lambda_{tot}(\zeta)=\frac{{\mathcal M}\zeta +V_0}{(\zeta +b)^2}.
\label{T_DE}
\end{equation}
where only the $\zeta$ dependence remains explicitly.

 In the case of the absence of massive fermions, it follows from
the constraint  that in such a case $\zeta$ is a constant
\begin{equation}
\zeta=\zeta_0 =b-\frac{2V_0}{{\mathcal M}}.
  \label{zete-vacuum}
  \end{equation}
Therefore, the TMT-effective cosmological constant {\em in the fermion vacuum} reads
\begin{equation}
\rho_{vac} =\Lambda_{eff}(\zeta_0)=
\frac{{\mathcal M}}{2(\zeta_0 +b)}=\frac{{\mathcal M}^2}{4(b{\mathcal M}-V_0)},
\label{Veff-vac}
\end{equation}
which is positive due to the condition (\ref{bsM_bigger_L0}). 
  Then $\zeta_0\sim b$ and Eq.(\ref{sim-parameters_1}) may be completed by adding $\zeta_0$:
  \begin{equation}
\zeta_0\sim\zeta_{1}\sim\zeta_{2}\sim b\sim k\sim h.
 \label{sim-all_parameters_0}
\end{equation}
The observed tiny value of the present-day vacuum energy density $\rho_{vac}\sim {\mathcal M}/4b$ can be achieved by choosing a huge value of the dimensionless primordial parameter $b$ and hence $k$ and $h$. Unlike the usual way of achieving a tiny CC by {\em mutual fine-tuning} of the coupling constants and masses, the way in which large dimensionless numbers appear in the primordial TMT action is completely different: 1) the parameters $b$, $k$, and $h$ are huge numbers compared to the dimensionless parameters in particle field theories, but {\em there is no need for mutual fine-tuning} of $b$, $k$, and $h$; 2) the huge values of $b$, $k$, and $h$ have no observable consequences except for the small value of the vacuum energy density compared to the typical local matter energy density.

Using eqs.(\ref{zete-vacuum}) and (\ref{rhofcan n}) the constraint (\ref{constraint via n})
can be rewritten as follows
\begin{equation}
\frac{(\zeta_0-\zeta){\mathcal M}}{(\zeta
+b)^2}=\Lambda_{dyn}^{(f)}= Z(\zeta)\rho_f^{(can)}.
\label{constraint via n-1}
\end{equation} 
It is very important that due to   our basic assumption concerning the parameters of the model, eq.(\ref{sim-all_parameters_0}),  the l.h.s. of the constraint (\ref{constraint via n}) has generically the same order of magnitude as $\rho_{vac}$, eq.(\ref{Veff-vac}), if no a special fine tuning of the parameters is assumed. In the opposite to the fermion vacuum case, i.e. for the high fermion density case (that is {\em when the local fermion energy density  is many tens orders of magnitude bigger than the vacuum energy density}), the only possible way to realize the ballance between two sides of the constraint is to allow for $\zeta$ to be very close either to $\zeta_1$ or to $\zeta_2$. Then the factor  $Z(\zeta)$ in the r.h.s. becomes very small and it is able to compensate the large value of $\rho_f^{(can)}$, {\em while the l.h.s. of the constraint remains of order $\sim\rho_{vac}$}. Moreover, the detailed analysis in ref.\cite{GK6}  shows that the $\zeta$ dependence of the fermion mass, eq.(\ref{rhofcan n}), has no observable effects under laboratory conditions (i.e. at high fermion density).

By virtue of the constraint (\ref{constraint via n-1}), the total energy density as well as the total pressure  of cold fermions
 may be expressed in the form where they are functions of $\zeta$ alone:
\begin{equation}
\rho =\frac{{\mathcal M}}{(\zeta +b)^2}\left[\frac{1}{2}(b-\zeta_0)+\zeta+\frac{\zeta_0 -\zeta}{Z(\zeta)}\right]
\label{rho-zeta}
\end{equation}
\begin{equation}
p=-\frac{{\mathcal M}}{(\zeta +b)^2}\left[\frac{1}{2}(b-\zeta_0)+\zeta\right]
\label{p-zeta}
\end{equation}
Further analysis is based on the following  \textit{natural assumptions}:
 There should be possible transitions from  the high density of cold fermions  (when $\zeta$  is very close to  $\zeta_1$ or $\zeta_2$)  to the low fermion density ending with asymptotic transition $\rho\rightarrow \rho_{vac}$ as $\zeta\rightarrow \zeta_0$. We suppose that  $\zeta$, as the solution of the constraint (\ref{constraint via n-1}), is a continuous  function of  $n$ evolving  from $\zeta \approx\zeta_1$ (or $\zeta\approx\zeta_2$) to the regime $\zeta\rightarrow \zeta_0$.  Besides, in the course of the monotonic decay of  $n$, the total energy density $\rho$ must be continuous and positive function of $\zeta$.   This means that $\zeta$ cannot cross over values $\zeta_1$ and $\zeta_2$, where $Z(\zeta)$ equals zero and $\rho$ is singular. As it was already supposed by Eq.(\ref{zb-zk-positive}), $\zeta$ cannot cross also over the values $-b$ and $-k$  because the transformation to the Einstein frame, Eq.(\ref{ctferm}), as well as the fermion mass, eq.(\ref{rhofcan n}), become singular and the constraint turns out to be senseless (it looks then as an equality  of finite and infinite quantities). Adding to this that the fermion number density $n$ cannot be negative,  it can be shown that there are only two  regimes for evolution of $\zeta$, where 
\begin{equation}
\text{The case (A):}  \qquad \zeta_2<\zeta_1<\zeta<\zeta_0;
\label{interval_zeta_A}
\end{equation}

\begin{equation}
\text{The case (B):} \qquad \zeta_2<\zeta_0<\zeta<\zeta_1;
\label{interval_zeta_B}
\end{equation}
and the wide region in the parameter space exists where such regimes are possible.

 In the course of the afore-mentioned monotonic decay of  $n$,  $\zeta$ remains of the order of the parameters $b\sim k\sim h$ which means that  the l.h.s. of the constraint (\ref{constraint via n}) has  the same order (or very close to) as $\rho_{vac}$,
and hence the same is valid for $P_f^{(noncan)}$ and $\rho_f^{(noncan)}$:
\begin{equation}
 |\Lambda^{(f)}_{dyn}|=|P_f^{(noncan)}|=|\rho_f^{(noncan)}|\sim \Lambda_{tot}\sim\rho_{vac}
\label{Pf-sim_vac}
\end{equation}
Only in the fermion vacuum $\Lambda^{(f)}_{dyn}=0$ and $\Lambda_{tot}=\rho_{vac}$.
Such the narrow intervals for values of  $\zeta$,  $\Lambda^{(f)}_{dyn}$ and $\Lambda_{tot}$ explain why the noncanonical fermion energy density and pressure are unobservable under regular physics conditions, but in the cosmology of the late time universe they may be important.
It is worth noting that this result follows from the structure of the constraint, in which the function $Z(\zeta)$ plays the role of a “self-locking clamp”, allowing very narrow intervals for the values of the listed quantities in the entire gigantic range of fermion densities, from very high to zero.

Applying the model to the late Universe cosmology, it is necessary to add the first Friedmann equation, 
which in the spatially flat FLRW Universe, as usual, has the form
\begin{equation}
\left(\frac{\dot{a}}{a}\right)^{2}=\frac{1}{3M_{p}^{2}}\bar{\rho}.
 \label{FRW}
\end{equation}
where $\bar{\rho}$ should be the cosmologically averaged total energy density.
The formulation of a cosmological model requires an answer to the question of whether the constraint, being highly nonlinear in $\zeta$  algebraic equation , retains its structure after cosmological averaging, that is it
also has the form as in Eq.(\ref{constraint via n-1})
\begin{equation}
\frac{{\mathcal M}(\zeta_0-\bar{\zeta})}{(\bar{\zeta} +b)^2}=\overline{\Lambda_{dyn}^{(f)}}; \qquad \overline{\Lambda_{dyn}^{(f)}}=\overline{P_f^{(noncan)}},
\label{constraint_av}
\end{equation}
where now $\overline{\Lambda_{dyn}^{(f)}}$ and $\overline{P_f^{(noncan)}}$  are  the cosmologically averaged values of $\Lambda_{dyn}^{(f)}$ and $P_f^{(noncan)}$;
 $\bar{\zeta}$ is the cosmological averaged of its local space-time values $\zeta(x)$ and  now $\bar{\zeta}$  is a function only of the cosmic time.
Solution to this averaging problem turns out to be simple if we take into account the observed inhomogeneity of the Universe: the existence of regions with  clumped matter and domains where matter is very diluted. In the large scale structure this is manifested in the existence of filaments and voids. The characteristic feature of such a structure is that, on each level of scales, the volume of the low density domains is tens orders of magnitude larger than the volume of regions with clumped matter. On the other hand, due to the self-locking  effect,  the relative differences both in values of $\zeta$ and in values of $P_f^{(noncan)}$ in these two types of regions do not exceed one order of magnitude. Hence the cosmological average values   $\bar{\zeta}$ and $\bar{P}_f^{(noncan)}$ with very high accuracy coincide with the values of $\zeta$ and $P_f^{(noncan)}$  in the maximal volume domains of low fermion density. Therefore the cosmologically  averaged constraint indeed has the form of Eq.(\ref{constraint_av}), where $\bar{\zeta}$ and  $\bar{P}_f^{(noncan)}$ practically equal to the corresponding values in the maximal volume domains of low fermion density.

  In the cold matter-dominated epoch, in the course of the cosmological expansion, the Universe enters a  stage when
 in the maximal volume domains of low fermion density, $\rho_f^{(can)}$ becomes smaller than $\Lambda_{tot}(\zeta)$. Then the regions of the Universe populated with the maximal volume domains of low fermion density start to expand with acceleration which is, properly speaking, \textit{the transition from the cold matter dominated epoch to the DE dominated epoch}. We come to the quite natural conclusion that the \textit{accelerated} cosmological expansion of the Universe can proceed due to \textit{accelerated} expansion of voids and other low matter density domains.
 It is evident  that such a manner of the accelerated expansion  becomes more and more dominant in the course of evolution of the late time Universe. If the described mechanism of the accelerated  expansion is right, it means that the transition to it could happen only after the voids are formed. But the latter must be accompanied by simultaneous formation of clustered matter.
 The cosmological averaged energy density of the clustered matter scales as $1/a^3$ and we will describe it using the only
 phenomenological parameter $M_{cl}$ (which has the dimensionality of mass)
\begin{equation}
\bar{\rho}_{cl}=\frac{M_{cl}^4}{a^3}.
\label{rho-clust}
\end{equation}
Averaging  eq.(\ref{Pfnoncan}) yields
 \begin{equation}
 \overline{\Lambda_{dyn}^{(f)}}=Z(\bar{\zeta})\overline{\rho_f^{(can)}},
 \label{Lambda-canon-av}
\end{equation}
where the cosmological averaged of the canonical energy density of the cold fermion matter disposed in the  maximal volume domains of low fermion density are obtained by averaging eq.(\ref{rhofcan n}):
\begin{equation}
\overline{\rho_f^{(can)}}=m(\bar{\zeta})\bar{n}=\frac{(\bar{\zeta} +h)}{(\bar{\zeta} +k)(\bar{\zeta}
+b)^{1/2}}\mu \bar{n},
 \label{rho-canon-av}
\end{equation}
where $\bar{n}=n_0/a^3$. Therefore $\overline{\rho_f^{(can)}}$ scales in a way different from $1/a^3$.

In the  FLRW universe, the averaged components of $T_{\mu\nu}^{(\Lambda)}$, eq.(\ref{Tmn_Lambda}), are  governed by $\bar{\zeta}$ and they read
 \begin{equation}
\bar{p}_{(\Lambda)}=-\bar{\rho}_{(\Lambda)}=-\Lambda_{tot}(\bar{\zeta}),
\label{p_rho_av}
\end{equation}
where
\begin{equation}
\Lambda_{tot}(\bar{\zeta})=\Lambda_{eff}(\bar{\zeta})
-\overline{\Lambda_{dyn}^{(f)}}=\frac{{\mathcal M}}{(\bar{\zeta} +b)^2}\left[\frac{1}{2}(b-\zeta_0)+\bar{\zeta}\right],
\label{Leff_av}
\end{equation}
and $\Lambda_{eff}$ is determined by eq.(\ref{Veff1}).
We would like to stress here that solving the constraint (\ref{constraint_av}) for $\bar{\zeta}$ we obtain it as a function of $\bar{n}$: $\bar{\zeta}=\bar{\zeta}(\bar{n})$. We come to \textit{the crucial role result of the model}:   $\Lambda_{tot}$ is a function of the number density $\bar{n}$ of the cold fermion matter disposed in the  maximal volume domains  of low fermion density  (\underline{presumably cold neutrinos in cosmic voids}):
\begin{equation}
\Lambda_{tot}=\Lambda_{tot}(\bar{n})
\label{Ltot-func-n}
\end{equation}
and  {\bf the dynamical DE effect is driven by fermion degrees of freedom without any specially added DE degrees of freedom}.

\begin{figure}[htb]
\begin{center}
\includegraphics[width=8.0cm,height=6.0cm]{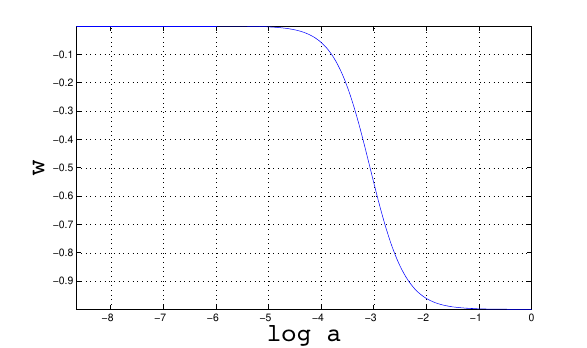}
\end{center}
\label{fig1}
\caption{Total EoS $w=\bar{p}/\bar{\rho}$ vs $\log a$.}
\end{figure}
\begin{figure}[htb]
\begin{center}
\includegraphics[width=8.0cm,height=6.0cm]{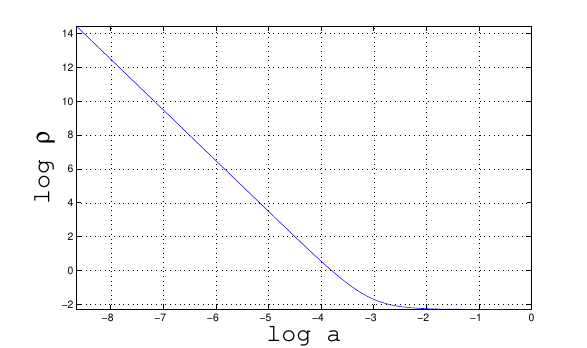}
\end{center}
\label{fig2}
\caption{Total energy density $\log\bar{\rho}$ vs $\log a$.}
\end{figure}
\begin{figure}[htb]
\begin{center}
\includegraphics[width=8.0cm,height=6.0cm]{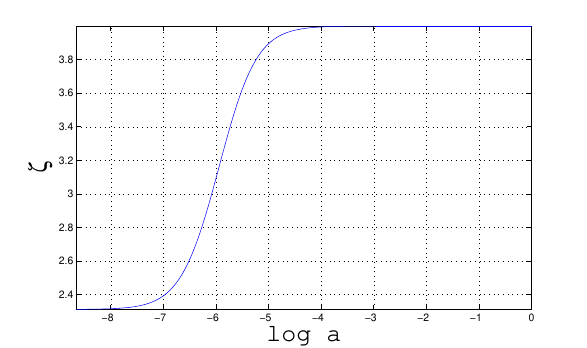}
\end{center}
\label{fig3}
\caption{ Evolution of the averaged value  $\bar{\zeta}$ that with very high accuracy coincides with the value of $\zeta$ in the maximal volume domains of low fermion density. The result of this numerical solution confirms the analytical estimates, the results of which are formulated in the paragraph after 
 eq. (\ref{Pf-sim_vac}) under the name of the “self-locking” effect: as the total energy density decays from the value $\sim 10^{14}$ at the cold matter dominated epoch with $w\approx 0$ up to the value $\sim 10^{-2}$ at the DE dominated epoch with $w\approx -1$ (see fig.2), $\bar{\zeta}$ changes only from $\bar{\zeta}\approx 2.3$ to the fermion vacuum value $\bar{\zeta}=\zeta_0 =4$ defined by eq.(\ref{zete-vacuum}).}
\end{figure}
\begin{figure}[htb]
\begin{center}
\includegraphics[width=8.0cm,height=6.0cm]{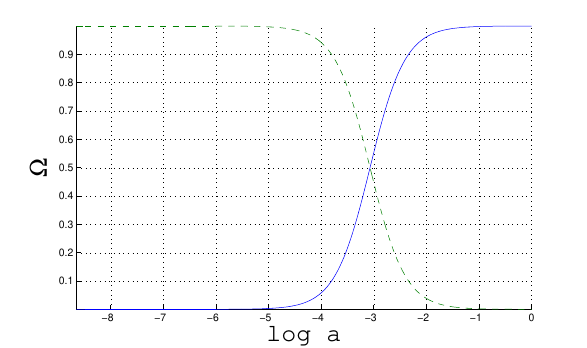}
\end{center}
\label{fig4}
\caption{ $\Omega$ vs $\log a$ where fractions of clustered (dark) matter $\Omega_m$ (the black dash line) and effective DE $\Omega_{DE}$ (the blue solid line).}
\end{figure}
\begin{figure}[htb]
\begin{center}
\includegraphics[width=14.0cm,height=10.0cm]{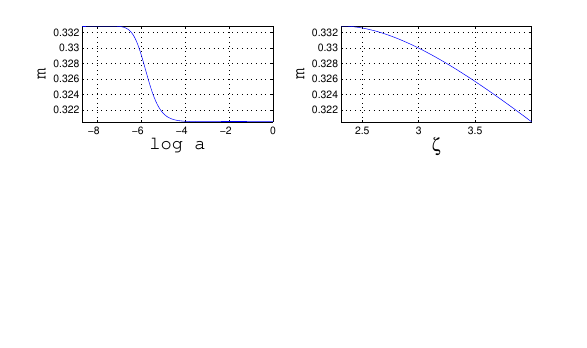}
\end{center}
\caption{ Mass of the cold neutrinos in the maximal volume domains of low density: $m$ vs $\ln a$ and $m$ vs $\bar{\zeta}$.}
\label{fig5}
\end{figure}
Ignoring the clumped matter pressure  in the late Universe, we have that the total averaged pressure
\begin{equation}
\bar{p}=\bar{p}_{(\Lambda)}=-\Lambda_{tot}=-\bar{\rho}_{(\Lambda)}.
\label{p-av}
\end{equation}
The total canonical matter energy density consists of contributions of the clumped (dark) matter and the canonical energy density 
$\overline{\rho_f^{(can)}}$ of the cold fermions  disposed in the  maximal volume domains:
\begin{equation}
\bar{\rho}_m=\bar{\rho}_{cl}+ \overline{\rho_f^{(can)}}
\label{rho-tot-normal-av}
\end{equation}
The total averaged energy density, which in the flat Universe is the critical one, is then
\begin{equation}
\bar{\rho} =\bar{\rho}_m+\bar{\rho}_{(\Lambda)}=\bar{\rho}_m+\Lambda_{tot}=\bar{\rho}_m-\bar{p},
\label{rho-av}
\end{equation}
and fractions of the normal matter  energy density and of the one which mimics the DE density are then defined as usual:
\begin{equation}
\Omega_m =\frac{\bar{\rho}_m}{\bar{\rho}} \qquad \text{and} \qquad \Omega_{DE} =\frac{\bar{\rho}_{(\Lambda)}}{\bar{\rho}}
\label{Omega-definitions}
\end{equation}

Making use of the constraint, eq.(\ref{constraint_av}), one can rewrite $\bar{\rho}$ and $\bar{p}$ in the form where  the contributions from the  maximal volume domains  depend only on the averaged scalar field $\bar{\zeta}$:
\begin{equation}
\bar{\rho} =\bar{\rho}_{cl}+\frac{M^4}{(\bar{\zeta} +b)^2}\left[\frac{1}{2}(b-\zeta_0)+\bar{\zeta}+\frac{\zeta_0 -\bar{\zeta}}{Z(\bar{\zeta})}\right]
\label{rho-zeta-av}
\end{equation}
\begin{equation}
\bar{p}=-\frac{M^4}{(\bar{\zeta} +b)^2}\left[\frac{1}{2}(b-\zeta_0)+\bar{\zeta}\right]
\label{p-zeta-av}
\end{equation}
Using  the total averaged energy-momentum tensor  in the perfect fluid form $\bar{T}^{\mu\nu}=(\bar{\rho}+\bar{p}) u^{\mu}u^{\nu}
-\bar{p}\tilde{g}^{\mu\nu}$ and the last equality in (\ref{rho-av}) we get
\begin{equation}
\bar{T}^{\mu\nu}=\bar{\rho}_m u^{\mu}u^{\nu}+\Lambda_{tot}\tilde{g}^{\mu\nu}
\label{perfect-Tmunu}
\end{equation}
where $\tilde{g}_{\mu\nu}=diag(1,-a^2,-a^2,-a^2)$, the four-velocity $u^{\mu}$ of the cold fermion gas in the co-moving frame is $u^{\mu}=(1,0,0,0)$ and $\Lambda_{tot}$ is determined by Eq.(\ref{Leff_av}). The structure of the cosmological model under consideration is very unusual. For example,  the fermion masses and  $\Lambda_{tot}$ are $\zeta$ dependent and evolve in the course of cosmological expansion. But, despite this, with the help of fairly cumbersome calculations, one can verify that  
the standard GR equation of the energy-momentum conservation
 \begin{equation}
 \dot{\bar{\rho}}=-3\frac{\dot{a}}{a}(\bar{\rho} +\bar{p})
 \label{en-mom-conserv}
\end{equation}
is satisfied.

The results of numerical calculations for a scenario, when the EoS $w=\bar{p}/\bar{\rho}$ decreases monotonically from $w\approx 0$ in the cold matter-dominated epoch to $w\to -1$ in the very late Universe,  are presented in figures 1-5.
 In this case, dimensionless units are used, obtained by the following redefinitions:
 \begin{equation}
 \frac{\zeta}{b}\rightarrow\zeta; \quad \frac{k}{b}\rightarrow k; \quad  \frac{h}{b}\rightarrow h; \quad
 \frac{V_0}{b{\mathcal M}}\rightarrow V_0; \quad  \frac{b}{{\mathcal M}}M_{cl}^4\rightarrow M_{cl}^4; \quad
\frac{b}{{\mathcal M}}\bar{\rho}\rightarrow\bar{\rho}.
 \nonumber
 \end{equation}
Numerical calculations were carried out with the following parameters: $V_0 =-1.5$, $k=2$, $h=0.3$,
 $M_{cl}^4=10^{-5}$. Then $\zeta_1=2.3114$, $\zeta_2 =-1.2114$. The initial value of $\bar{\zeta}$ is chosen
to be $\bar{\zeta}_{in}=2.312$.

\section{A possible connection between the Borde-Guth-Vilenkin theorem and initial conditions for inflation as a TMT effect}
\label{BGV and initial conditions for inflation}

The inflationary stage of the Universe's evolution is one of the most important and complex problems, and its intensity of study has not waned for many years. This is explained by the fact that inflation appears to be a natural laboratory providing invaluable information not only for studying the very early universe but also for particle physics and astrophysics. For recent studies see, for example, \cite{Vilenkin,KLY,Odintsov}.

Conditions that the initial kinetic and gradient energy densities of the canonically normalised scalar field  should not exceed the potential energy density 
\begin{equation}
\rho_{kin}^{(in)}= \frac{1}{2}\dot{\varphi}_{in}^2\lesssim V(\varphi_{in}) \quad \text{and} \quad  \rho_{grad}^{(in)}=\frac{1}{2}|(\partial^k\varphi)_{in}(\partial_{k}\varphi)_{in}|\lesssim V(\varphi_{in})
\label{kin and grad less V} 
\end{equation}
are well known as the constraints neeeded for the onset of inflation.  According to the understanding developed in the first models of chaotic inflation \cite{Linde 1983}, \cite{Linde 1985}, when the classical space-time domain first appears after the Planck quantum era, the total energy density is of the order of $M_P^4$, and inflation begins with $V(\varphi_{in})\sim M_P^4$. Then all admissible values $\varphi_{in}$, $\dot{\varphi}_{in}$, $(\partial_{k}\varphi)_{in}$ of the classical scalar field $\varphi$ satisfying (\ref{kin and grad less V} ) can serve as initial values for inflation. At first glance, such an idea of the beginning of inflation cannot contradict {\em the constraint} $V(\varphi)\sim 10^{-10}M_P^4$ {\em in the last stages of inflation}. However, the situation has changed dramatically in light of recent cosmological observations data \cite{Planck}, \cite{BICEP} which favor inflationary models with plateau-like potentials, and with the height of the plateau $V_{pl}\sim 10^{-10}M_P^4$.
There exist a number of field theory models in which the plateau-like potentials arise due to the implementation of various original ideas, and these potentials satisfy the CMB constraints.
These include  
 the Starobinsky model \cite{Star},
the Goncharov-Linde model \cite{GL1}, the Higgs inflation models \cite{Bardeen}-\cite{MetricHI-5}.
 Of particular interest are $\alpha$-attractor models, which were initiated by the pioneering works \cite{KL1}-\cite{KL3} and which have been intensively studied in recent years. To date, there is  the broad class of the cosmological attractor models, which generalize most of the previously proposed models with plateau potentials.

Despite such an impressive success of plateau-like models compared to all other models, a lively discussion 
 ensued, during which even the very idea of inflation has been called into question \cite{Stein}-\cite{Stein1}.
An obvious disadvantage of the models with plateau potentials mentioned above is the infinite length of the potential energy density plateau. In such a theory, all initial values of the homogeneous component of the scalar field $\varphi$ are equally probable. Therefore, an excessively long duration of inflation is possible.
The main problem formulated in paper\cite{Stein} is also related to the height of the plateau  $V_{pl}\sim 10^{-10}M_P^4$, since in this case there is a huge range of possible values of the initial kinetic and gradient energy densities greater than $V_{pl}$, up to the Planck density. 
Therefore, in contrast to the understanding developed in  first models of chaotic inflation \cite{Linde 1983}, \cite{Linde 1985} of how initial conditions for inflation arise, there is no reason to believe that conditions (\ref{kin and grad less V} ) necessary for the onset of inflation are satisfied. A possible response to this challenge may be to modify the model in such a way that the potential has a plateau of finite length, after which, at very large $\varphi$, the potential rapidly increases. An example of this type of model is the "singular $\alpha$-attractor" model proposed by Linde in \cite{singular alpha}, in which the simplest $\alpha$-attractor potential takes an exponentially steep form for very large $\varphi $. This makes it possible to provide conditions for power-law inflation, which starts at the Planck density, and thus makes it possible to solve the problem of initial conditions in the spirit of \cite{Linde 1985}. 

Another, completely independent aspect of the problem, related to the onset of inflation is predicted by the Borde--Guth--Vilenkin 
(BGV) theorem. It was initially proved in refs. \cite{BV1,BV2}, and then strengthened  in paper \cite{BGV}. According to the BGV theorem\cite{BGV} which strengthens earlier proofs of singularity theorems \cite{BV1}, \cite{BV2},  in inflationary cosmology
  almost all past-directed timelike and null geodesics cannot be extended to the past beyond some boundary  spacelike hypersurface $\mathcal{B}$.
 The statement of the BGV theorem is quite general because it is based on a kinematic argument. The main problem that inevitably follows  from the statement of the BGV theorem is that the inflationary universe must have had some kind of beginning, and, therefore, some new physics is necessary in order
to determine the correct conditions at the boundary $\mathcal{B}$. However, the BGV theorem says nothing about the boundary conditions on 
$\mathcal{B}$, or even about its location.

This section is based on the TMT-model of paper \cite{my JCAP 2023}, where it is shown that:  (A) there is an exact upper  bound $\varphi_0$ of the interval of possible values of the inflaton field $\varphi$ in which the onset of inflation is guaranteed; (B) this maximum allowed value $\varphi_0$ determines the location of the boundary  spacelike hypersurface that can be identified  with the  boundary  spacelike hypersurface $\mathcal{B}$ in the BGV theorem.
The primordial action is chosen as follows
\begin{eqnarray}
S&=&\int d^4x\left[-\frac{M_P^2}{2}(\sqrt{-g}+\Upsilon)\left(1+\xi \frac{\phi^2}{M_P^2}\right)R(\Gamma,g)
-\sqrt{-g}V_1+\frac{\Upsilon^2}{\sqrt{-g}}V_2\right]
\nonumber
\\
&+&\int d^4x\left[(b_k\sqrt{-g}+\Upsilon) \frac{1}{2}g^{\alpha\beta}\phi_{,\alpha}\phi_{,\beta} 
- (b_p\sqrt{-g}+\Upsilon)V(\phi) \right],
\label{S without fine tun}
\end{eqnarray}
where
\begin{equation}
V(\phi)=\frac{1}{2}m^2\phi^2+\frac{\lambda}{4}\phi^4
\label{V in BGV}
\end{equation}
Here $b_k$ and $b_p$ are primordial model parameters, introduced in accordance with the general structure  of the TMT primordial Lagrangian density 
discussed in  section \ref{Some additional mathematical aspects taken into account in TMT}.
But unlike the models studied in sections \ref{quint with additional symm}, \ref{Scale invar models} and \ref{Neutrino genereted dark energy}, the action (\ref{S without fine tun}) contains, firstly, a $V_2$-term (\ref{V2 term}) and, secondly, a non-minimal coupling of the scalar field to curvature  with a non-minimal coupling constant $\xi$.

Following the prescription of the TMT procedure, we must consider the equations of motion that follow from the primordial action (\ref{S without fine tun}). Under the condition (\ref{Phi neq 0}), i.e. everywhere $\Upsilon(x)\neq 0$, the consistency of the equations imposes a constraint, and, as usual, a transition to the Einstein frame is necessary. The latter is described by the formula
\begin{equation}
\tilde{g}_{\mu\nu}=(1+\zeta)\left(1+\xi\frac{\phi^2(x)}{M_P^2}\right)g_{\mu\nu}.
 \label{gmunuEin}
\end{equation}
Then the constraint in the Einstein frame reads as follows 
\begin{equation}
\zeta(\phi,X_{\phi})=\frac{{\mathcal M}-2V_1-(2b_p-1)V(\phi)
-(1-b_k)\left(1+\xi\frac{\phi^2}{M_P^2}\right)\cdot X_{\phi}}
{{\mathcal M}-2V_2+V(\phi)+(1-b_k)\left(1+\xi\frac{\phi^2}{M_P^2}\right)\cdot X_{\phi}},  \qquad
 X_{\phi}=\frac{1}{2}{\tilde g}^{\alpha\beta}\phi_{,\alpha}\phi_{,\beta}
\label{1+ zeta no fine tun phi M-1}
\end{equation}

Similar to what was in section \ref{K-Essence and inflation in the scale invariant}, the TMT-effective energy-momentum tensor 
and $\phi$-equation in the
Einstein frame have the  structure of the $k$-essence type model.
The minimum of the TMT-effective potential is reached at $\phi =0$, and from the constraint  (\ref{1+ zeta no fine tun phi M-1}) it follows
 that the scalar $\zeta$ in this vacuum is equal to
\begin{equation}
\zeta|_{\phi =0}=\zeta_v=\frac{{\mathcal M}-2V_1}
{{\mathcal M}-2V_2}.
\label{zeta at phi =0}
\end{equation}
We choose the value of the TMT effective potential in the vacuum, i.e. the TMT-effective CC, to be zero.  This can be achieved by choosing the integration constant either ${\mathcal M}_{+}=2\sqrt{V_1V_2}$ or ${\mathcal M}_{-}=-2\sqrt{V_1V_2}$. This is very similar to the toy model 
from section \ref{toy model}, and  the  corresponding values of $\zeta$ coincide with $\zeta_{v+}=\sqrt{\frac{V_1}{V_2}}$ and $\zeta_{v-}=-\sqrt{\frac{V_1}{V_2}}$, as in eqs.(\ref{zeta toy vac Mplus}) and (\ref{zeta toy vac Mminus}). We choose  a solution with ${\mathcal M}_{+}$ and the corresponding 
$\zeta_{v+}$, omitting the $+$ sign further. Recall that this also means choosing $\Upsilon >0$. To indicate that we are considering a solution with a zero cosmological constant, we will use the symbol $(0)$ in the notation of all relevant quantities.  We will also assume that $V_1<0$ and $V_2<0$. After using the redifinition of the scalar field $\phi$ to $\varphi$
\begin{equation}
\frac{\phi}{M_P}=\frac{1}{\sqrt{\xi}}{\sinh}\left(\sqrt{\xi}\frac{\varphi}{M_P}\right),
\label{phi via canonical varphi}
\end{equation}
which is more appropriate when studying inflation, the constraint (\ref{1+ zeta no fine tun phi M-1}) reduces to the following
\begin{equation}
\zeta(\varphi,X_{\varphi})=\frac{2|V_2|\zeta_v(1+\zeta_v)-(2b_p-1)\left[\frac{m^2M_P^2}{2\xi}\sinh^2z+\frac{\lambda}{4\xi^2}M_P^4\sinh^4z\right]
-(1-b_k)\cosh^4z\cdot X_{\varphi}}
{2|V_2|(1+\zeta_v)+\frac{m^2M_P^2}{2\xi}\sinh^2z+\frac{\lambda}{4\xi^2}M_P^4\sinh^4z+(1-b_k)\cosh^4z\cdot X_{\varphi}}
\label{1+ zeta no fine tun}
\end{equation}

Omitting the explicitly written Einstein equations and the $\phi$-field equation, in complete analogy with sections \ref{toy model} and \ref{K-Essence and inflation in the scale invariant}, we use the pressure density, which plays the role of the matter Lagrangian in the TMT-effective action; a variation of the latter yields these equations and is much more convenient to work with. In terms of the $\varphi$ field, and after inserting $\zeta$ given by the constraint (\ref{1+ zeta no fine tun}), the TMT-effective action takes the form
\begin{equation}
S_{eff}^{(0)}=\int \left[-\frac{M_P^2}{2}R(\tilde{g})+
L_{eff}^{(0)}\left(\varphi,\zeta(\varphi,X_{\varphi})\right)\right]
\sqrt{-\tilde{g}}d^4x, 
\label{Seff in BGV}
\end{equation}
where the TMT-effective Lagrangian for the scalar field $\varphi$ appears as following
\begin{equation}
L_{eff}^{(0)}(\varphi,X_{\varphi})=X_{\varphi}-V_{eff}^{(0)}(\varphi)-K_1(\varphi)X_{\varphi}-K_2(\varphi)\frac{X_{\varphi}^2}{M_P^4},
\label{Leff final no fine tun}
\end{equation}
where
\begin{equation}
V_{eff}^{(0)}(\varphi) =\frac{M_P^4}{4\xi^2}{\tanh}^4z\cdot F(z),
\label{Veff varphi tanh no fine tun 1}
\end{equation}
\begin{equation}
F(z)=\frac{\lambda q^4(\zeta_v+b_p)+
\frac{m^4}{4M_P^4}+
\frac{\lambda}{4\xi}\frac{m^2}{M_P^2}{\sinh}^2z+\frac{\lambda^2}{16\xi^2}\sinh^4z+
2\xi q^4(\zeta_v+b_p)\frac{m^2}{M_P^2}\cdot \sinh^{-2}z }
{(1+\zeta_v)^2q^4+(1-b_p)\left(\frac{1}{2\xi}\frac{m^2}{M_P^2}\sinh^2z+\frac{\lambda}{4\xi^2}{\sinh}^4z\right)},
\label{Veff varphi tanh no fine tun 2}
\end{equation}

\begin{equation}
K_1(\varphi)=\frac{1-b_k}{2\cosh^2z}\, \cdot\frac{8q^4\xi^2(1+\zeta_v)+2\xi\frac{m^2}{M_P^2}\cdot \sinh^{2}z+\lambda\sinh^4z}
{4q^4\xi^2(1+\zeta_v)^2+2(1-b_p)\xi\frac{m^2}{M_P^2}\cdot \sinh^{2}z+\lambda(1-b_p)\sinh^4z},
\label{K1}
\end{equation}
\begin{equation}
K_2(\varphi)=\frac{(1-b_k)^2}{4\left[q^4(1+\zeta_v)^2+\frac{1-b_p}{2\xi}\frac{m^2}{M_P^2}\cdot \sinh^{2}z
+\frac{\lambda(1-b_p)}{4\xi^2}\sinh^4z\right]},
\label{K2}
\end{equation}
\begin{equation}
z=\sqrt{\xi}\frac{\varphi}{M_P}; 
 \qquad X_{\varphi}=\frac{1}{2}\tilde{g}^{\alpha\beta}\varphi_{,\alpha}\varphi_{,\beta}
\label{z and X varphi}
\end{equation}
and for the partameter $V_2<0$ the parametrization $|V_2|=q^4M_P^4$ was used. The appearance in the TMT-effective
potential  $V_{eff}^{(0)}(\varphi)$ the hyperbolic functions like ${\tanh}\frac{\varphi}{\sqrt{6}M_P}$ is typical for the simplest T-Model obtained in the conformal theory \cite{KL1}. An example of the shape of the TMT-effective
potential  $V_{eff}^{(0)}(\varphi)$ is shown in figure 7 for the  following choice of parameters:
  $q^4=10^{-6}$, \, $\xi=0.18=\frac{1}{6\alpha}$, that is $\alpha=0.926$, \, $\zeta_v=0.2$, \, $b_p=0.5+10^{-6}$, \, $b_k=0.5$, \, $\lambda= 2.9\cdot 10^{-11}$, \,  $m\approx 2\cdot 10^{13}GeV$. 
It is similar to a case of  
the single field $\alpha$ attractor models \cite{KL3}, where   the hyperbolic functions depend on the  combination $\frac{\varphi}{\sqrt{6\alpha}M_P}$.

A detailed analysis in \cite{my JCAP 2023} shows that in the region of $\varphi$ corresponding to the plateau of $V_{eff}^{(0)}(\varphi)$,
the contribution of the last two terms to the TMT-effective Lagrangian (\ref{Leff final no fine tun}) is negligible. Consequently, the presence of these terms in (\ref{Leff final no fine tun}) does not change the conditions (\ref{kin and grad less V}) for the onset of inflation. The standard, non-TMT, formulation of the initial conditions for inflation driven by a scalar field includes specifying or at least estimating the initial value of the field $\varphi_{in}$ and its first derivatives or, equivalently, the initial kinetic energy density  $\rho_{kin,in}$ and the gradient energy density $\rho_{grad,in}$.
It is fundamentally important and taken for granted that there is no dependence in any form between  $\varphi_{in}$ and $\rho_{kin,in}$, as well as between   $\varphi_{in}$ and $\rho_{grad,in}$. 
 But in the studied TMT model {\em the scalar $\zeta$ given by the constraint (\ref{1+ zeta no fine tun}) turns out to depend not only on $\varphi$, but also on $X_{\varphi}=\frac{1}{2}\tilde{g}^{\alpha\beta}\varphi_{,\alpha}\varphi_{,\beta}$}. After choosing the integration constant ${\mathcal M}=2\sqrt{V_1V_2}>0$ and,  correspondingly,  $\zeta_{v}=\sqrt{\frac{V_1}{V_2}}>0$, throughout the entire process of cosmological evolution, $\zeta$ must be positive.
As a result, the condition $\zeta>0$ imposes restrictions in the form of inequalities on the admissible ranges of $\varphi$ {\bf and} $X_{\varphi}$. We are interested in the constraints on the initial values $\varphi_{in}$ and $X_{\varphi}^{(in)}$ imposed  {\em both by  the condition $\zeta(\varphi_{in}, X_{\varphi }^{(in )})>0$ and by the upper bounds on $\rho_{kin,in}$ {\em and} $\rho_{grad,in}$}, eq.(\ref{kin and grad less V}).
 Inflation can only begin if all of these conditions are met simultaneously.

The results of the requirement that $\zeta(\varphi, X_{\varphi})$ be positive can be obtained from a detailed study of the inequality
\begin{equation}
\zeta=\frac{2|V_2|\zeta_v(1+\zeta_v)-(2b_p-1)\left[\frac{m^2M_P^2}{2\xi}\sinh^2z+\frac{\lambda}{4\xi^2}M_P^4\sinh^4z\right]
-(1-b_k)\cosh^4z\cdot X_{\varphi}}
{2|V_2|(1+\zeta_v)+\frac{m^2M_P^2}{2\xi}\sinh^2z+\frac{\lambda}{4\xi^2}M_P^4\sinh^4z+(1-b_k)\cosh^4z\cdot X_{\varphi}}>0,
\label{1+ zeta no fine tun larger 0}
\end{equation}
 taking into account the significant difference between how kinetic $\rho_{kin}$ and gradient $\rho_{grad}$ energy densities enter into the Einstein equations and into the left-hand side of the inequality (\ref{1+ zeta no fine tun larger 0}). Indeed, while 
the kinetic and gradient energy densities enter the Einstein equations in the form of a sum, they enter the constraint  in the form 
\begin{equation}
X_{\varphi}=\frac{1}{2}\tilde{g}^{\alpha\beta}\varphi_{,\alpha}\varphi_{,\beta}=\frac{1}{2}\left(\dot{\varphi}^2-
\frac{1}{a^2}(\nabla\varphi)^2\right)
=\rho_{kin}-\rho_{grad},
\label{X as difference}
\end{equation}
 that is, in fact, in the form of a difference. Therefore, 
$X_{\varphi}^{(in)}$ can be positive or negative depending on how inhomogeneous and anisotropic at the beginning of inflation was the space domain  whose expansion generates our Universe. Since the sign of $X_{\varphi}^{(in)}$ can significantly affect the results obtained from the condition $\zeta>0$, the cases $X_{\varphi}^{(in)}>0$ and $X_{ \varphi}^{(in)}<0$ should be considered separately. The main results are summarized below.

Let us start fron considering the constraint (\ref{1+ zeta no fine tun}) at $X_{\varphi }= 0$. 
We see that there exists the value $\varphi_0$ such that $\zeta\rightarrow 0^+$ when $\varphi\rightarrow \varphi_0^{\,\,\,-}$ 
and $\varphi_0$ is defined by the relation 
\begin{equation}
\sinh^4\sqrt{\xi}\frac{\varphi_0}{M_P}=\frac{8q^4\xi^2\zeta_v(1+\zeta_v)}{\lambda (2b_p-1)}
\label{sinh 4 max no fine tuned case}
\end{equation}
obtained after neglecting a very small correction from the  term $\propto \sinh^2\sqrt{\xi}\frac{\varphi_0}{M_P}$.
In other words, in the case $X_{\varphi }= 0$ there exists a maximum allowed value $\varphi_0$ such that values $\varphi>\varphi_0$ are prohibited by the condition $\zeta >0$, equivalent to $\Upsilon >0$, i.e. the infinite interval $\varphi\geq\varphi_0$ is an artifact. For the parameters used in the graph in Fig. 7, the relation (\ref{sinh 4 max no fine tuned case}) gives $\varphi_0\approx 13.7M_P$. The corresponding point on the curve is marked with a dot.

\begin{figure}
\includegraphics[width=13.0cm,height=8cm]{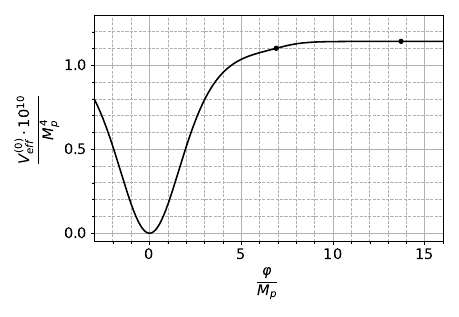}
\caption{Plot of $V_{eff}^{(0)}(\varphi)$ for the case of the following choice of parameters:
  $q^4=10^{-6}$, \, $\xi=0.18=\frac{1}{6\alpha}$, that is $\alpha=0.926$, \, $\zeta_v=0.2$, \, $b_p=0.5+10^{-6}$, \, $b_k=0.5$, \, $\lambda= 2.9\cdot 10^{-11}$, \,  $m\approx 2\cdot 10^{13}GeV$. The two dots on the plot correspond to $\varphi_0\approx 13.7 M_P$ 
defined by relation (\ref{sinh 4 max no fine tuned case}) and
$\varphi_{in}^{(min)}\approx 6.8 M_P$ defined by Eq.(\ref{interval phi for X positive}).}
\label{figBGV.eps}
\end{figure}

The analysis of the case $X_{\varphi}^{(in)}<0$ in \cite{my JCAP 2023} shows that it is impossible to specify precise conditions that guarantee the onset of inflation. It is only worth noting that  $X_{\varphi}^{(in)}<0$ means the dominance of the spatial gradients of the  field $\varphi(x)$, which usually arise due to quantum fluctuations. The interpretation and possible cosmological effect of this result may be of interest, but its study was beyond the scope of the paper
 \cite{my JCAP 2023}.

 \underline{The main results in the case $X_{\varphi}^{(in)}>0$:}

1) There exists the minimal initial value  $\varphi_{in}^{(min)}$ defined by the formula
\begin{equation}
\varphi_{in}^{(min)}=\varphi_0-  \frac{M_P}{4\sqrt{\xi}}\ln\left(1+\frac{k_1(1-b_k)(\zeta_v+b_p)}{(2b_p-1)(1+\zeta_v)^2}\right),
\quad (\text{where} \quad 1\leq k_1\lesssim 1.03)
\label{interval phi for X positive}
\end{equation}
such that
{\bf on the interval}
\begin{equation}
\varphi_{in}^{(min)}\leq  \varphi_{in}
<\varphi_0
\label{phi 0 and phi min}
\end{equation}
{\bf the condition $\zeta(\varphi_{in} , X_{\varphi}^{(in)})>0$  and the necessary conditions (\ref{kin and grad less V}) for the beginning of inflation are guaranteed to be fulfilled.} 
For the set of parameters used in Fig. 6, we get  $\varphi_{in}^{(min)}\approx 6.8 M_P\leq\varphi_{in}<\varphi_0\approx 13.7M_P$. With the chosen value of $\xi =0.18$  we get  that $\varphi(N)$ correspoding to 60 $e$-foldings is $\varphi(60)\approx 6.9M_P>\varphi_{in}^{(min)}\approx 6.8 M_P$, i.e. is inside the  interval (\ref{phi 0 and phi min}).

2) It turns out that  the upper bound of admissible  values of  $X_{\varphi}^{(in)}$  tends to zero as $\varphi_{in}\rightarrow \varphi_0^{\,\, -}$:
\begin{equation}
\lim_{\varphi_{in} \to \varphi_0^{\,-}}\zeta(\varphi_{in},X_{\varphi}^{(in)})=0 \quad {\text {\bf and}} \quad 
\lim_{\varphi_{in} \to \varphi_0^{\,-}}X_{\varphi}^{(in)}=0.
\label{X>0 lim phi0}
\end{equation}
Thus, for the case $X_{\varphi}^{(in)}>0$ in the plane $(X_{\varphi},\varphi)$ the point $(0,\varphi_0)$ is the boundary point such that
$\varphi_{in}$ cannot be extended to  $\varphi_{in}>\varphi_0$ without changing the sign of $\zeta$ from positive to negative. 

3) In the  interval (\ref{phi 0 and phi min}) of $\varphi_{in}$, the TMT effective  potential has a plateau-like shape and $X_{\varphi}^{(in)}$ is less than the plateau height. Therefore, one can expect that the Universe  inflates in the slow-roll regime.

4) For $\varphi_{in}<\varphi_{in}^{(min)}$ the condition $\zeta(\varphi, X_{\varphi})>0$ is satisfied, but the conditions
 (\ref{kin and grad less V}) required for the onset of inflation  may or may not hold.

\vspace{0.3cm}

 \underline{\em{ Initial conditions for inflation and the spacelike character of the hypersurface $\Upsilon(x)= 0$.}}

In the model under study (as in all models in the previous sections), we came to the key conclusion: the dynamics of the model dictates that nontrivial solutions exist only when the measure density  $\Upsilon(x)\neq 0$ and, therefore, $\Upsilon(x)$ must be sign-definite. However, the condition (\ref{X>0 lim phi0}) can be represented as following
\begin{equation}
\Upsilon\equiv\zeta(\varphi,X_{\varphi})\cdot\sqrt{-g}\rightarrow 0^+  \quad \text{and} \quad X_{\varphi}\rightarrow 0^+\quad \text{as} \quad  \varphi\rightarrow\varphi_0^{\,\,\,-}.
\label{upsilon 0}
\end{equation} 
This means that the hypersurface $\Upsilon(x)=0$ does indeed exist and it separates the space-time manifold $M_4$ into two regions with $\Upsilon(x)>0$ and $\Upsilon(x)<0$, and we must study equations and their solutions in these regions independently of each other. These two regions can be considered as submanifolds  of  ${M_4}$ with a common boundary $\Upsilon(x)=0$. In what follows we will use the notations ${M_4^{(+)}}$ and ${M_4^{(-)}}$ for the submanifolds with $\Upsilon(x)>0$ and $\Upsilon(x)<0$ respectively.
The condition $X_{\varphi}>0$ means that the normal vectors $\partial_{\alpha}\varphi$ 
to all hypersurfaces described by the equations $\varphi(x)=constant$ are timelike and consequently, the hypersurfaces $\varphi(x)=constant$ are spacelike.
All of the above remains true for $\varphi(x)$ arbitrarily close to $\varphi_0$. To find out the geometric properties of the hypersurface $\Upsilon(x)=0$, we must, given the definition $\Upsilon(x)\equiv\zeta(x)\sqrt{-g(x)}$, study what happens to $\zeta(x)\equiv\zeta(\varphi(x),X_{\varphi})$ when $\varphi(x)\to \varphi_0$. In doing this analysis, we must not forget that $\varphi(x)$ and $X_{\varphi}$ are considered independent of each other and serve as initial values for inflation. Therefore, it would be wrong to conclude that $X_{\varphi}\to 0$ because $\varphi(x)\to\varphi_0=const$.
But we can  take into account the very nontrivial result containing in Eq.(\ref{upsilon 0}):    $X_{\varphi}> 0$  decreases  to 0 when $\varphi(x) \to \varphi_0^{\,\, -}$.
Therefore, for values of $\varphi(x)$ very close to  $\varphi_0$ the function $\zeta(\varphi(x),X_{\varphi})$ can be considered approximately as a function depending only on $\varphi(x)$: $\zeta(\varphi(x),X_{\varphi})\approx \zeta(\varphi(x))$. Hence, for  $\varphi(x)\approx \varphi_0$ 
the equation $\zeta(\varphi(x))=constant$ describes a spacelike  hypersurface. The latter remains true when $\varphi(x) \to \varphi_0^{\,\, -}$ (which is accompanied by $constant \to 0$).
Therefore, {\bf the hypersurface  $\Upsilon(x)\equiv\zeta(x)\sqrt{-g(x)}=0$ is a  space-like boundary of the submanifold  ${M_4^{(+)}}$.}

It is worth remembering that by applying the principle of least action to the primordial action (\ref{S without fine tun}), 
we mean, as usual, that all primordial variables (scalar field $\phi$, non-degenerate metric tensor $g_{ \mu\nu}$, 
the affine connection $\Gamma^{\lambda}_{\mu\nu}$, the functions $\varphi_a$ 
and hence $\Upsilon$) are defined globally {\em on an orientable space-time manifold $M_4$} and are smooth functions.
 In particular, this means that: 1) they are all continuous on $M_4$; 2) one can choose the  orientation of $M_4$. 
The latter should be expressed in the possibility of choosing a certain sign of $\Upsilon(x)$. As we have seen, in order to describe our Universe, when solving equations in the model under consideration, we had to put $\Upsilon(x)>0$, i.e. choose a positive orientation. Based on our experience in field theory, we might naively expect this condition to still hold globally on $M_4$. However, it turns out that this is not the case due to the dynamics of the model. The reason for this phenomenon is a very profound change introduced by TMT into field theory: as a result of the inclusion of the variables $\varphi_a$ (from which $\Upsilon(x)$ is built) into the principle of least action, the possibility arises of {\em mutual influence of the  matter field dynamics and the continuous function $\Upsilon(x)$ or, in other words, between matter and such a fundamental property of the space-time manifold as orientability.} This fundamentally new dynamical effect can lead to the creation of a hypersurface $\Upsilon(x)= 0$ in $M_4$ , which, as we have seen, can in turn impose significant restrictions on the dynamics of mater fields.

The creation of the hypersurface $\Upsilon(x)= 0$ (despite the fact that originally $\Upsilon(x)$ was nonvanishing on $M_4$) and the splitting of $M_4$ into two submanifolds $M_4^{(+)} $ and $M_4^ {(-)}$ with the opposite orientation means that $M_4$ is not oriented anymore. Since this effect is the result of solving dynamical equations, it can be interpreted as {\bf spontaneous violation of the orientability of $M_4$}.
 Moreover, we found that {\bf the dynamically arisen spacelike hypersurface $\Upsilon(x)=0$ does not allow extension of solutions for $\varphi(t)$ from $\varphi(t)<\varphi_0$ to $\varphi(t) \geq\varphi_0$.}

The totality of the obtained results, namely

- impossibility of extension of solutions for $\varphi(t)$ from $\varphi(t)<\varphi_0$ to $\varphi(t) \geq\varphi_0$

- guaranteed onset of inflation  subject to (\ref{phi 0 and phi min})

- the spacelike character of the hypersurface $\Upsilon(x)=0$ and the timelike character of vectors $\partial_{\alpha}\varphi$,

allows us to assume that in the model under study, the spacelike hypersurface $\Upsilon(x)=0$ is a concrete realization of the spacelike hypersurface 
${\mathcal B}$ from the BGV theorem.

Considering the existence of two  submanifolds $M_4^{(+)} $ and $M_4^ {(-)}$ with the opposite orientation, the paper \cite{my JCAP 2023}  also studies question  what is hidden behind the space-like hypersurface $\Upsilon(x)=0$
   and boundary conditions on this hypersurface.

\section{Discussion}

The wide range of applications of TMT to gravity and cosmology is evident from the section headings of the paper. The most important advantage of TMT as an alternative theory is that, under the conditions under which all classical tests of general relativity are performed, TMT accurately reproduces Einstein's GR.
 At the same time, TMT offers a number of very interesting models of dark energy. In this paper, we also discussed the new possibilities TMT opens for describing inflation and its initial conditions. In this regard, it is worth noting that refs. \cite{My Higgs Infl,2-nd paper}, not included in this review, propose a new approach to implementing the Higgs inflation model based on the TMT modification of the $SU(2)\times U(1)$ gauge-invariant electroweak Standard Model (TMSM). The homogeneous scalar field $\phi(t)$ obtained by cosmological averaging of the local Higgs doublet  plays the role of inflaton. TMSM provides a new type of connection between particle physics at accelerator energies and at the inflation energy scale. This allows to realize the Higgs inflation model with small  constant $\xi$ of non-minimal coupling to the scalar curvature, for example, with $\xi =1/6$. 
 This result is of  fundamental importance when compared with  the Higgs inflation models known in the literature \cite{MetricHI-1,MetricHI-2,MetricHI-3,MetricHI-4,MetricHI-5,MetricHI-6,MetricHI-7,MetricHI-8,MetricHI-9,MetricHI-10,Palatini-H-1,Palatini-H-2,Palatini-H-3,Palatini-H-4,Palatini-H-5,Palatini-H-6,Palatini-H-7,Palatini-H-8}, where
 suitable description of  inflation in accordance with the CMB data \cite{Planck,BICEP} is only possible with an unnaturally  large non-minimal coupling constant to the scalar curvature: depending on the type of model, metric or Palatini, the coupling constant varies in the range $10^4\lesssim\xi \lesssim 10^8$.
The TMSM  overcomes  this problem due to the presence of $\zeta(\phi(t))$ in all equations of motion. As a result, all TMSM coupling constants are kind of running (classical) TMT-effective parameters. During the evolution of 
the cosmological background, changing these parameters yields new results that are important for both particle physics and inflation:

\begin{enumerate}
\item[(1)]	The classical running TMT-effective Higgs self-coupling increases from $\lambda\sim 10^{-11}$ at the stage of slow-roll inflation
 (which ensures consistency with the Planck's CMB data at $\xi=\frac{1}{6}$)
 to $\lambda_{SM}\sim 0.1$ at the stage close to the vacuum; 
\item[(2)]	The mass term in the TMT-effective Higgs potential changes sign from positive to negative,
which provides spontaneous symmetry breaking in the standard way of the Glashow--Weinberg--Salam (GWS) theory; 
\item[(3)]	The classical running constants of the gauge and Yukawa couplings change by several orders of magnitude;
\item[(4)] The GWS theory is reproduced in such a way that the observed  hierarchy of fermion masses is obtained quite naturally;
\item[(5)] It is clearly shown that the theory, considered on the cosmological background at the the slow-roll inflation regime, is renormalizable.
  Taking into account quantum corrections 
 in the one-loop approximation preserves the slow-roll inflation regime and does not violate the vacuum stability;
\item[(6)] All known models of preheating after inflation, such as \cite{KLS 1,KLS 2,KLS 3}, assume the interaction of the inflaton with the quantum fields of particles in the original  action, despite the fact that this may, for example, disrupt the harmony of the GWS theory, in particular with respect to its renormalizability. Further developments, not included in the preprints \cite{My Higgs Infl,2-nd paper}, show that after the end of inflation, but long before reaching the vacuum state, the effective TMT self-interaction parameter of the quantum Higgs field acquires a value very close to that in GWS theory. Moreover, a fundamentally new TMSM effect is discovered, consisting of the emergence of interactions between the quantum Higgs  and fermion fields with the classical field $\phi(t)$ without any additional assumptions. By applying the approaches to studying preheating developed in the works of refs. \cite{Turner,KLS 3,Brandenberger} for the case of oscillating solutions in the conventional $\frac{\lambda}{4}\phi^4$ model, it can be shown that a similar preheating mechanism can be effective in TMSM as well.
\end{enumerate}

 In the final discussion, it makes sense to focus on the fundamentally new ideas underlying TMT and leading to the results presented in the article. First of all, it is necessary to pay attention to the validity of the statement that {\bf the volume elements $dV_g=\sqrt{-g}d^4x$ and $dV_{\Upsilon}=\Upsilon d^4x$ are equally natural}. This conclusion,  which we reached in the Introduction, section \ref{Introduction}, in the course of our analysis of Wald's argumentation, is a concept of fundamental importance and there is no longer any reason to continue to ignore the volume measure $dV_{\Upsilon}$  when constructing field theory models.

In this paper, while demonstrating various TMT models and their highly interesting results, we paid special attention to the effects arising from
 including the $\varphi_a$ functions, from which $\Upsilon$ is constructed, as degrees of freedom in the primordial action. As we have seen, in all the models studied, these effects are described by a scalar function $\zeta=\Upsilon/\sqrt{-g}$.  Bearing in mind that gravity in GR
 has a geometric nature described by the metric tensor, we  used the term pregeometry for effects caused by the scalar $\zeta(x)$.

One of the pregeometry effects is related to sign-indefinitness of $\Upsilon$. Therefore, the use of the volume element   $dV_{\Upsilon}$ in the primordial action means that a field theory model is formulated on a  space-time manifold orientation of which is initially unfixed. 
Only after applying the least action principle and choosing a solution to the equations of motion does a specific sign of $\zeta$ appear, and hence a specific sign of orientation. Moreover, as we have seen, there always exists another solution corresponding to the opposite orientation  sign.
Thus, firstly, we are dealing with the pregeometry effect, which consists of {\bf a spontaneous restoration  of the space-time orientation}; secondly there is a strict correlation between the physical results predicted by the TMT (via solutions to the equations) and the sign of the orientation of spacetime.

Representing the primordial action as $S = \int\left[ L_{1}\sqrt{-g}+L_{2}\Upsilon\right]d^{4}x$ and varying it wih respect to $\varphi_a$, we find  that the solution $L_{2}=\mathcal{M}$ exists under the condition $\Upsilon(x)\neq 0$, that is, only if spacetime is orientable. In this case, only those cosmological solutions (along with their initial conditions and  vacuum states)  are valid for which 
$\zeta(x)$ does not change  sign throughout the evolution of the Universe. This is another pregeometry effect that has played a significant role in all the models considered in this article.

However, when discussing the models under consideration, it is worth noting three special cases where the condition $\Upsilon(x)\neq 0$ may be violated, which means the orientability of the space-time manifold is lost. 

1) In the model of refs.\cite{Kess4} and \cite{GK4} briefly discussed in section 
\ref{K-Essence and inflation in the scale invariant}, the measure density $\Upsilon$ and the primordial metric $g_{\mu\nu}$ oscillate around their zeros during transition to  the vacuum state $\phi_{0}$ with zero vacuum energy. This means that transition to a state with zero cosmological constant, which 
occurs  in this model without fine tuning,  must be accompanied not only by a periodic reversal of the orientation sign but also by passage through a singularity in the primordial metric. Therefore, the assumption of regularity of the function $\zeta=\Upsilon/\sqrt{-g}$ may be violated. This pregeometry effect apparently indicates that attempting to implement TMT models in which zero CC is achieved without fine-tuning requires a more careful analysis of possible solutions in TMT.

2)  In ref.\cite{Kess4}, studying possible  dark energy scenarios for the late time Universe, it was demonstrated that there is a wide range of the primordial model parameters  where the superaccelerating cosmological expansion is obtained without introducing an explicit phantom scalar field into the primordial action (\ref{totaction k-ess}). 
The numerical solution of the cosmological equations and its detailed analysis in the paper \cite{GK4} show that during the evolution, the volume measure density $\Upsilon$ changes sign from positive to negative, so that $(\Upsilon+b_{\phi}\sqrt{-g})$ becomes negative. Returning to the primordial action (\ref{totaction k-ess}), we see that the kinetic term of the $\phi$ field changes sign, that is, the standard scalar  field is dynamically transformed into a phantom field. Some features of this scenario are briefly described in section \ref{K-Essence and inflation in the scale invariant} of this paper, from which it is clear that the sign change of  the field $\phi$ kinetic term in the primordial action (\ref{totaction k-ess}) occurs simultaneously with crossing of the phantom divide $w=-1$, see Fig. 1.

3) An investigation of a possible connection between the BGV theorem and initial conditions for inflation in the paper \cite{my JCAP 2023}, briefly discussed in section \ref{BGV and initial conditions for inflation}, reveals that the boundary spacelike hypersurface predicted by the BGV theorem does indeed exist, and in the TMT model under consideration this hypersurface is described by the equation $\Upsilon(x) =0$. 
Assuming the continuity of $\Upsilon(x)$, we showed that, despite the initial assumption $\Upsilon(x)\neq 0$, a hypersurface $\Upsilon(x)= 0$ arises, partitioning the spacetime manifold $M_4$ into two submanifolds $M_4^{(+)}$ and $M_4^ {(-)}$ with opposite signs of $\Upsilon(x)$ and hence with opposite orientations. Consequently, the equations of motion must be solved separately for $\Upsilon(x)>0$ and for $\Upsilon(x)<0$, where, generally speaking, different integration constants must be chosen. It then turns out that, in the general case, on the hypersurface $\Upsilon(x)=0$, there is a discontinuity of the metric tensor $\tilde{g}_{\mu\nu}$ used in the Einstein frame.

\appendix

\section{Nonrelativistic fermions in sections \ref{scale with ferm} and \ref{Neutrino genereted dark energy}}
\label{Nonrelativistic fermions}

In sections \ref{scale with ferm} and \ref{Neutrino genereted dark energy} we study the application of the field theory models to the case of {\it nonrelativistic fermions}. This means that  we neglect the effect of fermions 3-momenta.  The only component of the canonical fermion energy-momentum tensor $T_{\mu\nu}^{(f,can)}$ which survive in this approximation is the energy density $T_{00}^{(f,can)}=\rho_f^{(can)}$. Making use of the Dirac equation in the same approximation we obtain (see also eqs.(\ref{m}) and (\ref{rhofcan}))
\begin{equation}
\rho_f^{(can)}=
 m(\zeta)
\overline{\Psi^{\prime}}\Psi^{\prime}; \qquad  m(\zeta)=\frac{\mu(\zeta +h)}{(\zeta +k)(\zeta
+b)^{1/2}}.
\label{rhofcan_1}
\end{equation}
To provide the consistent description of the fermion energy density and pressure we have to regard
$\overline{\Psi}^{\prime}\Psi^{\prime}$ as the field operator. Then the standard semiclassical approach to the Einstein eqs.(\ref{grav eq Ein toy}) (when the gravity is treated as the classical field) implies the need to evaluate the  matrix element of $\overline{\Psi^{\prime}}\Psi^{\prime}$ between appropriate states of the Fock space resulting in the fermion (particles and antiparticles) number density $n$ \cite{Massive n 1,Peccei,Goldman}. Then
 \begin{equation}
\rho_f^{(can)}=
 m(\zeta) n
\label{rhofcan_n}
\end{equation}
 In the case of nonrelativistic fermions, corrections to the cold fermion approximation reduce to the appearance of the Fermi momentum depending factors  in front of $n$  in Eq.(\ref{rhofcan_n}). For the goals of the present paper it is enough to notice that these factors are positive.


\begin{thebibliography}{99}

\bibliography{}

\bibitem{Hawking}
Hawking, S.W.; Ellis, G.F.R. \emph{The Large Scale Structure of Space-Time}; 
Cambridge University Press: Cambridge, UK, 1973.

\bibitem{Wald}
Wald, R.M. 
\emph{General Relativity}; The University of Chicago Press:  Chicago, 
 1984.


\bibitem{GK2}
Guendelman, E.I.; Kaganovich, A.B.
Dynamical measure and field theory models free of the cosmological constant problem.
  \emph{Phys. Rev. D} \textbf{1999}, {\em 60}, 065004. arXiv:9905029. 


\bibitem{Lee}
Lee, J.M.
\emph{Introduction to Smooth Manifolds}, 2nd ed.; Springer Science+Business Media: New York, NY, USA, 2013.

\bibitem{more than one non-canonical form}
Benisty, D.; Guendelman, E.; Kaganovich, A.; Nissimov, E.; Pacheva, S.
Modified Gravity Theories Based on the Non-canonical Volume-Form Formalism.
 \emph{Springer Proc. Math. Stat.} \textbf{2019}, {\em 335}, 239--252. arXiv:1905.09933



\bibitem{G1}
 {Guendelman, E.I. 
Scale invariance and vacuum energy.} 
 \emph{Mod. Phys. Lett.} \textbf{1999}, {\em A14}, 1043. arXiv:0106084.

\bibitem{G2}
 Guendelman, E.I. 
Scale symmetry spontaneously broken by asymptotic behavior.
 \emph{Class. Quant. Grav.} \textbf{2000}, {\em 17}, 361. arXiv:9906025.


\bibitem{K}
Kaganovich, A.B.
Field theory model giving rise to ‘quintessential inflation’ without the cosmological constant and other fine tuning problems.
 \emph{Phys. Rev. D} \textbf{2000}, {\em 63},  025022. arXiv:0007144.

\bibitem{G4}
Guendelman, E.I.
The Volume element of space-time and scale invariance.
  \emph{Found. Phys.} \textbf{2001}, {\em 31}, 1019.
 arXiv:hep-th/0011049. 






\bibitem{PV}
Peebles, P.J.E.; Vilenkin, A. Quintessential Inflation.
\emph{Phys. Rev.} \textbf{1999}, {\em D59}, 063505. arXiv:9810509.

\bibitem{Planck}
Akrami, Y.; Arroja, F.; Ashdown, M.; Aumont, J.; Baccigalupi, C.; Ballardini, M.; Banday, A.J.; Barreiro, R.B.; Bartolo, N.; Basak, S.; et al. {Planck 2018 results. X. Constraints on inflation}. \emph{Astron.
Astrophys.} \textbf{2020}, {\em 641}, A10. arXiv:1807.06211. 

\bibitem{BICEP}
 Ade1, P.; Ahmed, Z.; Amiri, M.; Barkats, D.; Thakur, R.B.; Bischoff, C.; Beck, D.; Bock, J.; Boenish, H.;  E. Bullock; et al. 
 {Improved Constraints on Primordial Gravitational
Waves using Planck, WMAP, and BICEP/Keck Observations through the 2018 Observing
Season}. \emph{Phys. Rev. Lett.} \textbf{2021}, {\em 127}, 151301. arXiv:2110.00483. 


\bibitem{Kess4}
Guendelman, E.I.; Kaganovich, A.B.
Fine tuning free paradigm of Two Measures Theory: K-essence, absence of initial singularity of the curvature and inflation with graceful exit to zero cosmological constant state.
   \emph{Phys. Rev. D} \textbf{2007}, {\em 75}, 083505. arXiv:0607111. 

\bibitem{GK4}
Guendelman, E.I.; Kaganovich, A.B.
Transition to zero cosmological constant and phantom dark energy as solutions involving change of orientation of space-time manifold.
 	\emph{Class. Quant. Grav.} \textbf{2008}, {\em 25}, 235015.
arXiv:0804.1278. 

\bibitem{Chiba Kess}
Chiba, T.; Okabe, T.; Yamaguchi, M.
Kinetically driven quintessence. \emph{Phys. Rev. D} \textbf{2000}, {\em 62}, 023511. arXiv:9912463.

\bibitem{Mukh Kess}
Armendariz-Picon, C.; Mukhanov, V.F.; Steinhardt, P.J.
Essentials of k essence. \emph{Phys. Rev. D}  \textbf{2001}, {\em 63}, 103510. arXiv:0006373.

\bibitem{Barrow}
Barrow, J.D. More general sudden singularities.
\emph{Class Quant. Grav.} \textbf{2004}, {\em 21},  5619. arXiv:0409062.

\bibitem{Weinberg} 
Weinberg, S. The Cosmological Constant Problem. \emph{Rev. Mod. Phys.} \textbf{1989}, {\em 61}, 1.

\bibitem{Emergent}	
Campo, S.; Guendelman, E.I.; Kaganovich, A.B.; Herrera, R.
	Emergent universe from scale invariant Two Measures Theory. \emph{
 Phys. Lett. B} \textbf{2011}, {\em  699}, 211.
arXiv:1105.0651.

\bibitem{GK5}
Guendelman, E.I.; Kaganovich, A.B.
Absence of the fifth force problem in a model with spontaneously broken dilatation symmetry. \emph{   Ann. Phys.} \textbf{2008}, {\em 323},  866.   arXiv:0704.1998.

\bibitem{GKneutr DE}
Guendelman, E.I.; Kaganovich, A.B.
Exotic low density fermion states in the two measures field theory: Neutrino dark energy. \emph{
 Int. J. Mod. Phys. A} \textbf{2006}, {\em 21}, 4373.

\bibitem{Birrel Devies}
Birrell, N.D.; Davies, P.C.W. 
\emph{Quantum Fields in Curved Space};
Cambridge University Press, 1982.
 

\bibitem{Massive n 1}
{Fardon, R.; Nelson, A.E.; Weiner, N.
Dark energy from mass varying neutrino.} 
 \emph{ JCAP} \textbf{2004},  {\em 10}, 5. arXiv:0309800.


\bibitem{Massive n 2}
Brookfield, A.W.; Bruck, C.v.; Mota, D.F.; Tocchini-Valentini, D. 
Cosmology with massive neutrinos coupled to dark energy. \emph{
Phys. Rev. Lett.} \textbf{2006}, {\em 96}, 061301. arXiv:0503349.

\bibitem{Massive n 3}
Wetterich, C. 
Growing neutrinos and cosmological selection. \emph{ Phys. Lett. B} \textbf{2007}, {\em 655}, 201. arXiv:0706.4427.

\bibitem{Massive n 4}
Amendola, L.; Baldi, M.; Wetterich, C.
Quintessence cosmologies with a growing matter component. \emph{ Phys. Rev. D} \textbf{2008},
{\em 78}, 023015. arXiv:0706.3064.

\bibitem{Massive n 5}
Pettorino, V.; Wintergerst, N.; Wetterich, C.
Neutrino lumps and the Cosmic Microwave Background. \emph{ Phys. Rev. D} \textbf{2010},
{\em 82}, 123001. arXiv:1009.2461.

\bibitem{Massive n 6}
Wintergerst, N.; Pettorino, V.; Mota, D.F.; Wetterich, C.
Very large scale structures in growing neutrino quintessence. \emph{ Phys. Rev. D} \textbf{2010},
{\em 81}, 063525. arXiv:0910.4985.

\bibitem{GK6} 
Guendelman, E.I.; Kaganovich, A.B.
Neutrino generated dynamical dark energy with no dark energy field. 	\emph{
Phys. Rev. D}  \textbf{2013}, {\em 87},   044021.
arXiv:1208.2132.

\bibitem{Vilenkin}
Blanco-Pillado, J.J.; Deng, H.; Vilenkin, A.
Eternal Inflation in Swampy Landscapes. \emph{JCAP} \textbf{2020},  {\em 5}, 014. arXiv:1909.00068.

\bibitem{KLY}
Kallosh, R.; Linde, A.; Wrase, T.; Yamada, Y.
Sequestered Inflation.  \emph{ Fortsch. Phys.} \textbf{2021}, {\em 69}, 11--12. arXiv:2108.08491.

\bibitem{Odintsov}
Odintsov, S.D.; Oikonomou, V.K.; Giannakoudi, I.; Fronimos, F.P.; Lymperiadou, E.C.
Recent Advances in Inflation. \emph{ Symmetry} \textbf{2023}, {\em 15}, 1701. arXiv:2307.16308.

\bibitem{Linde 1983}
Linde, A.D. Chaotic inflation. \emph{Phys. Lett. B} \textbf{1983}, {\em  129}, 177--181. 

\bibitem{Linde 1985}
Linde, A.D.  Initial conditions for inflation.
\emph{Phys. Lett. B} \textbf{1985}, {\em  162},  281--286.

\bibitem{Star}
 Starobinsky, A.A. A New Type of Isotropic Cosmological Models Without Singularity. \emph{
Phys. Lett. B} \textbf{1980}, {\em  91}, 99.

\bibitem{GL1}
 Goncharov, A.B.; Linde, A.D. Chaotic Inflation in Supergravity.  \emph{Phys. Lett. B} \textbf{1984}, {\em  139},
  27.


\bibitem{Bardeen}
 Salopek, D.S.; Bond, J.R.; Bardeen, J.M. Designing density fluctuation spectra in inflation.  \emph{
Phys. Rev. D} \textbf{1989}, {\em 40},  1753.

\bibitem{MetricHI-1}
Bezrukov, F.; Shaposhnikov, M. The Standard Model Higgs boson as the inflaton. \emph{
Phys. Lett. B} \textbf{2008}, {\em 659}, 703. arXiv:0710.3755.

\bibitem{MetricHI-2}
Barvinsky, A.O.; Kamenshchik, A.Y.; Starobinsky, A.A. Inflation scenario via the Standard Model Higgs boson and LHC. \emph{ JCAP} \textbf{2008}, {\em 11}. arXiv:0809.2104.

\bibitem{MetricHI-3}
Bezrukov, F.L.; Magnin, A.; Shaposhnikov, M. Standard Model Higgs boson mass from
inflation. \emph{Phys. Lett. B} \textbf{2009}, {\em 675}, 88. arXiv:0812.4950.

\bibitem{MetricHI-4}
 Bezrukov, F.; Shaposhnikov, M. Standard Model Higgs boson mass from inflation: Two
loop analysis. \emph{ JHEP}  \textbf{2009}, {\em 0907}, 089. arXiv:0904.1537.

\bibitem{MetricHI-5}
Bezrukov, F.; Magnin, A.; Shaposhnikov, M.; Sibiryakov, S. Higgs inflation: consistency and generalisations. \emph{ JHEP} \textbf{2011},
 {\em 01}, 016. arXiv:1008.5157.


\bibitem{KL1}
Kallosh, R.; Linde, A. Universality Class in Conformal Inflation. \emph{ JCAP} \textbf{2013}, {\em 1307}, 002. arXiv:1306.5220.

\bibitem{KL2}
 Ferrara, S.; Kallosh, R.; Linde, A.; Porrati, M. 
Minimal Supergravity Models of Inflation. \emph{Phys. Rev. D} \textbf{2013}, {\em 88}, 085038. arXiv:1307.7696.

\bibitem{KL3}
 Kallosh, R.; Linde, A.; Roest, D. Superconformal Inflationary $\alpha$-Attractors. \emph{JHEP}
\textbf{2013}, \emph{1311}, 198. arXiv:1311.0472.


\bibitem{Stein}
Ijjas, A.; Steinhardt, P.J.; Loeb, A. Inflationary paradigm in trouble after Planck2013. \emph{Phys. Lett. B} \textbf{2013}, {\em723},  261--266. arXiv:1304.2785.

\bibitem{Resp to Steinh-1}
Guth, A.H.; Kaiser, D.I.; Nomura, Y. Inflationary paradigm after Planck 2013. \emph{ 
Phys. Lett. B} \textbf{2014}, {\em  733}, 112--119. 
arXiv:1312.7619.


\bibitem{Resp to Steinh-2} 
Linde, A. Inflationary Cosmology after Planck 2013.
Contribution to: Post-Planck Cosmology: Lecture Notes of the Les Houches Summer School. \emph{arXiv} \textbf{2013},  \emph{100}, 231--316. arXiv:1402.0526.

\bibitem{Stein1}
Ijjas, A.; Steinhardt, P.J.; Loeb, A.  Inflationary schism. \emph{
 Phys. Lett. B} \textbf{2014}, {\em  736}, 142--146. arXiv:1402.6980.

\bibitem{singular alpha} 
Linde, A.  {On the problem of initial conditions for inflation}.
\emph{Found. Phys.} \textbf{2018}, {\em 48}, 1246--1260. arXiv:1710.04278.

\bibitem{BV1}
Borde, A.; Vilenkin, A.
Eternal inflation and the initial singularity.
 \emph{Phys. Rev. Lett.} \textbf{1994}, {\em 72}, 3305--3309. 

\bibitem{BV2}
Borde, A.; Vilenkin, A.
Violations of the weak energy condition in inflating space-times. \emph{
 Phys. Rev. D} \textbf{1997}, {\em 56}, 717--723. 

\bibitem{BGV}
Borde, A.; Guth, A.H.; Vilenkin, A.
Inflationary space-times are incomplete in past directions.
 \emph{Phys. Rev. Lett.} \textbf{2003}, {\em90}, 151301. arXiv:0110012.

\bibitem{my JCAP 2023}
Kaganovich, A.B.
Possible relationship between initial conditions for inflation and past geodesic incompleteness of the inflationary spacetime.
  \emph{JCAP} \textbf{2023}, {\em 05}, 007. arXiv:2209.00378.


\bibitem{My Higgs Infl}
Kaganovich, A.B.
Higgs inflation model with small non-minimal coupling constant. \textbf{2025}. arXiv:2501.15597.
 
\bibitem{2-nd paper}
Kaganovich, A.B.
Two-Measure Electroweak Standard Model.
 Some aspects of cosmological evolution and vacuum stability.  \textbf{2025}. arXiv:2501.15623.

\bibitem{MetricHI-6}
 Barvinsky, A.O.; Kamenshchik, A.Y.; Kiefer, C.; Starobinsky, A.A.;
Steinwachs, C.F. 
Higgs boson, renormalization group, and naturalness in
cosmology.  \emph{ Eur. Phys. J. C} \textbf{2012}, {\em 72}, 2219. arXiv:0910.1041.

\bibitem{MetricHI-7}
Bezrukov, F.; Shaposhnikov, M. 
Higgs inflation at the critical point.
 \emph{Phys. Lett. B } \textbf{2014}, {\em 734}, 249--254. arXiv:1403.6078.

\bibitem{MetricHI-8}
Bezrukov, F.; Rubio, J.; Shaposhnikov, M.
{Living beyond the edge: Higgs inflation and vacuum metastability}.
\emph{Phys. Rev. D} \textbf{2015}, {\em 92}, 083512. arXiv:1412.3811.

\bibitem{MetricHI-9}
Rubio, J. Higgs inflation. \emph{ Front. Astron. Space Sci.} \textbf{2019}, {\em 5}, 50. arXiv:1807.02376.

\bibitem{MetricHI-10}
Barvinsky, A.O.; Kamenshchik, A.Y.
 {Nonminimal Higgs Inflation and Initial Conditions in Cosmology}.  \textbf{2022}. 
 	arXiv:2212.13077.




\bibitem{Palatini-H-1}
Bauer, F.; Demir, D.A. Inflation with Non-Minimal Coupling: Metric versus Palatini
Formulations. \emph{ Phys. Lett. B} \textbf{2008}, {\em 665}, 222. arXiv:0803.2664.



\bibitem{Palatini-H-2}
Rasanen, S.; Wahlman, P.
Higgs inflation with loop corrections in the Palatini formulation. \emph{  JCAP} \textbf{2017}, {\em 11}, 047. arXiv:1709.07853.

\bibitem{Palatini-H-3}
Markkanen, T.; Tenkanen, T.; Vaskonen, V.; Veermäe, H. 
Quantum corrections to quartic inflation with a non-minimal coupling: metric vs. Palatini. \emph{
JCAP} \textbf{2018}, {\em 03}, 029. arXiv:1712.04874.

\bibitem{Palatini-H-4}
Rubio, J.; Tomberg, E.S.
Preheating in Palatini Higgs inflation. \emph{  JCAP} \textbf{2019}, {\em 04}, 021. arXiv:1902.10148.

\bibitem{Palatini-H-5}
Shaposhnikov, M.; Shkerin, A.; Zell, S.
Quantum Effects in Palatini Higgs Inflation. \emph{
 JCAP} \textbf{2020}, {\em 07}, 064. arXiv:2002.07105.


\bibitem{Palatini-H-6}
Tenkanen, T.
Tracing the high energy theory of gravity: an introduction to Palatini inflation. \emph{
 Gen. Rel. Grav.} \textbf{2020}, {\em 52}, 33. arXiv:2001.10135.

\bibitem{Palatini-H-7}
Enckell, Ve.; Nurmi, S.; Räsänen, S.; Tomberg, E.
Critical point Higgs inflation in the Palatini formulation. \emph{
 JHEP} \textbf{2021}, {\em 04}, 059. arXiv:2012.03660.

\bibitem{Palatini-H-8}
Gialamas, I.D.; Karam, A.; Pappas, T.D.; Tomberg, E.
Implications of Palatini gravity for inflation and beyond. \emph{ Int. J. Geom. Meth. Mod. Phys.} \textbf{2023}, {\em 20}, 13. arXiv:2330007.

\bibitem{KLS 1}
Kofman, L.; Linde, A.D.; Starobinsky, A.A.
Reheating after inflation. \emph{ Phys. Rev. Lett.} \textbf{1994}, {\em 73}, 3195. arXiv:9405187.

\bibitem{KLS 2}
Kofman, L.; Linde, A.D.; Starobinsky, A.A.
Towards the theory of reheating after inflation.
\emph{Phys. Rev. D} \textbf{1997}, {\em  56}, 3258. arXiv:9704452.

\bibitem{KLS 3}
Greene, P.B.; Kofman, L.; Linde, A.D.; Starobinsky, A.A.
Structure of resonance in preheating after inflation. \emph{
Phys. Rev. D} \textbf{1997}, {\em  56}, 6175. arXiv:9705347.



\bibitem{Turner}
Turner, M.S.
Coherent Scalar Field Oscillations in an Expanding Universe. \emph{Phys. Rev. D} \textbf{1983}, {\em28}, 1243. 


\bibitem{Brandenberger}
Shtanov, Y.; Traschen, J.H.; Brandenberger, R.H.
Universe reheating after inflation. \emph{ Phys. Rev. D}  \textbf{1995}, {\em 51}, 5438. arXiv:9407247.








\bibitem{Peccei}
 Peccei, R.D. Neutrino models of dark energy. \emph{Phys. Rev. D} \textbf{2005}, {\em 71}, 023527.


\bibitem{Goldman}
Goldman, T.; Stephenson, G.J., Jr.; Alsing, P.M.; McKellar, B.H.J.
A Possible Connection Between Massive Fermions and Dark Energy. \emph{arXiv}
\textbf{2009}.  	arXiv:0905.4308.

\end{thebibliography}
\end{document}